\documentclass[a4paper,11pt]{article}
\pdfoutput=1 % if your are submitting a pdflatex (i.e. if you have
\usepackage{jcappub} % for details on the use of the package, please
\usepackage[T1]{fontenc} % if needed

\def\l{\left}
\def\r{\right}
\def\be{\begin{equation}}
\def\ee{\end{equation}}
\def\ba{\begin{eqnarray}}
\def\ea{\end{eqnarray}}
\def\bl#1\el{\begin{align}#1\end{align}}
\def\nn{\nonumber}

\newcommand{\gsim}{\mathrel{\hbox{\rlap{\lower.55ex \hbox {$\sim$}}
			\kern-.3em \raise.4ex \hbox{$>$}}}}
\newcommand{\lsim}{\mathrel{\hbox{\rlap{\lower.55ex \hbox {$\sim$}}
			\kern-.3em \raise.4ex \hbox{$<$}}}}

\title{\boldmath Impacts of Hawking Radiation from Primordial Black Holes in Critical Collapse Model on the Light Element Abundances}

\author[a,b]{Yudong Luo}
\author[c,d,e]{Chao Chen}
\author[f]{Motohiko Kusakabe}
\author[a,b,f]{and Toshitaka Kajino}

\affiliation[a]{National Astronomical Observatory of Japan,2-21-1 Osawa, Mitaka, Tokyo 181-8588, Japan}
\affiliation[b]{Department of Astronomy, Graduate School of Science,University of Tokyo, 7-3-1 Hongo, Bunkyo-ku, Tokyo 113-0033, Japan}

\affiliation[c]{Department of Astronomy, School of Physical Sciences, University of Science and Technology of China, Hefei, Anhui 230026, China}
\affiliation[d]{CAS Key Laboratory for Researches in Galaxies and Cosmology, University of Science and Technology of China, Hefei, Anhui 230026, China}
\affiliation[e]{School of Astronomy and Space Science, University of Science and Technology of China, Hefei, Anhui 230026, China}
\affiliation[f]{School of Physics, International Research Center for Big-Bang Cosmology and Element Genesis, Beihang University 37, Xueyuan Rd., Haidian-qu, Beijing 100083 China}
\emailAdd{ydong.luo@grad.nao.ac.jp}
\emailAdd{cchao012@mail.ustc.edu.cn}
\emailAdd{kusakabe@buaa.edu.cn}
\emailAdd{kajino@nao.ac.jp}

\abstract{

  We study the photodisintegration process triggered by the nonthermal electromagnetic Hawking radiation from primordial black holes (PBHs) in critical collapse model. We consider the simplest case that all PBHs formed at a single epoch stemming from an inflationary spectrum with a narrow peak, and an extended mass distribution is obtained due to critical phenomena of gravitational collapse. The presence of a low-mass tail of critical collapse mass function could lead to an enhancement of energetic photon emissions from Hawking radiation of PBHs.
  Nuclear photodisintegration rates are calculated with a nonthermal photon spectrum derived by solving the Boltzmann equation iteratively. The exact spectrum is much different than that based on an often-used analytical bended power-law spectrum and it is found to significantly depend on the adopted PBH mass functions. With the newest observational limit on the $^3$He abundance in Galactic H II  regions, the updated $^3$He constraints on PBH mass spectrum in the horizon mass range $10^{12} - 10^{13}$ g are derived.
  Our results for the first time show that $^3$He constraints on the critical mass function are about one order of magnitude severer than the monochromatic one although the fraction of PBHs in the low-mass tail region is relatively small. The $^6$Li elemental abundance is also enhanced significantly for the critical mass function. More precise measurement of $^6$Li abundance is highly desirable to provide a promising constraint on PBHs in the future.
   For monochromatic mass function, we provide the analytical bounds for photodisintegration and hadrodissociation from PBH radiation, and we report discrepancies between our updated $^3$He constraints and the previous results.
}


\arxivnumber{2011.10937}

\begin{document}
\maketitle
\flushbottom

\section{Introduction}
\label{sec:intro}

Primordial black holes (PBHs) may be formed from density fluctuations in the very early Universe and have been studied over more than $50$ years \cite{Zeldovich:1966, Hawking:1971ei, Carr:1974nx, Carr:1975qj}, which have been attracting a lot of attentions even though there is still no definite evidence for their existence. There are several major motivations for studying PBHs from the theoretical and phenomenological aspects. One is that PBHs could be small enough for Hawking radiation to be observationally significant \cite{Hawking:1974rv}. PBHs with masses $\lesssim 10^{15}$ g would have evaporated by now due to the emission of Hawking radiation, and the emitted particles may impact the abundances of light elements produced by the big bang nucleosynthesis (BBN) and the extragalactic gamma-ray background (EGB) \cite{Carr:2020gox}. These observables would provide indirect way to test Hawking radiation and also constrain the mass spectra of PBH. PBHs with masses greater than $10^{15}$ g could survive until the present epoch. Since the PBHs form in the radiation-dominant era as a cold non-baryonic matter, they are potential candidate for (a fraction of) cold dark matter (DM) \cite{Sasaki:2018dmp, Carr:2020xqk}. In addition, PBHs with the intermediate mass could generate the observed LIGO/Virgo coalescences of black holes with masses in the range $10 - 50 M_\odot$ \cite{Nakamura:1997sm, Sasaki:2016jop}, and massive PBHs could seed supermassive black holes and perhaps even galaxies themselves \cite{Bean:2002kx}. PBHs can be tested through their effects on a variety of cosmological and astronomical processes. For example, PBHs with masses greater than $10^{15}$ g are expected to be constrained by their gravitational effects such as gravitational lensing and dynamical effects on baryonic matter \cite{Sasaki:2018dmp}. A particular attention has also been paid to the gravitational waves (GWs) induced from the enhanced primordial density perturbations associated with PBH formation \cite{Kohri:2018awv, Bartolo:2018rku, Cai:2018dig, Cai:2019jah, Cai:2019cdl, Inomata:2019yww, Lu:2020diy}. The GW survey is thus a promising in revealing physical processes of PBH formation in the near future.

%High-density regions in the very early Universe are expected to have been engulfed in the PBH formation. One possibility is that there were large primordial inhomogeneities, and over-dense regions might collapse to form PBHs. This motivates many theoretical mechanisms of generating PBHs, which often require a power spectrum of primordial density perturbations to be suitably large on certain scales \cite{Pi:2017gih, Garcia-Bellido:2017mdw, Carr:2017edp, Cai:2018tuh, Fu:2019ttf}. Besides the collapse from the primordial density perturbation sourced during the inflationary stage, PBHs could also be produced at QCD phase transition \cite{Crawford:1982yz, Jedamzik:1998hc}, and collapse through bubble collisions \cite{Hawking:1982ga, Kodama:1982sf} and cosmic strings \cite{Hawking:1987bn}.

It has been well understood that the mass distribution of PBHs is extended (i.e., with a width $\Delta M \gg M$) in many scenarios. For example, PBHs forming from a scale-invariant fluctuations have a power-law form of mass function \cite{Carr:1975qj}. On the other hand, those from a smooth symmetric peak in the inflationary power spectrum have the lognormal mass function \cite{Dolgov:1992pu, Green:2016xgy}, which includes a large class of inflation models for PBH formation, like the axion-curvaton \cite{Kohri:2012yw, Kawasaki:2012wr} and running-mass inflation models \cite{Drees:2011yz}. However, even for PBHs forming at a single epoch, they would have an extended mass distribution when critical phenomena of gravitational collapse are taken into account \cite{Choptuik:1992jv}. Previous studies \cite{Yokoyama:1998xd, Green:1999xm} have shown that the horizon-mass approximation (i.e., the mass of a PBH is close to the horizon mass at the formation epoch) is still good enough for critical collapse mass function, and the relative fraction of PBHs located within the low-mass tail is small. It is well known that it is non-trivial to extend constraints for the monochromatic PBH distribution to extended cases. Constraints in the non-monochromatic cases are dependent on the PBH mass function, and those for a general case of mass function cannot be easily derived from those for the simplest monochromatic mass function \cite{Green:2016xgy, Carr:2020gox}. The mass function in the critical collapse model is relatively narrow and is thought of being practically indistinguishable from the monochromatic mass function \cite{Carr:2020gox}. However, we for the first time show in this paper that even small amounts of PBHs within the low-mass tail in the mass function of the critical collapse model would affect the elemental abundances. More stringent BBN constraints are then acquired than those on the monochromatic one, due to the simple fact that the Hawking radiation in the high energy region is stronger for low-mass PBHs. This leads to a caution that other sorts of constraints for the critical collapse model may be also altered compared to the monochromatic case.

In the standard evaporation model (SEM) which incorporates the standard model of particle physics, a {black hole} would directly radiate fundamental standard model particles whose de Broglie wavelengths are of the order of {black hole} size \cite{MacGibbon:1990zk}. The radiated particles could form into composite particles after radiation, and the effective number of species is determined by the mass of black hole. For a solar-mass black hole, the Hawking radiation is extremely weak and can be neglected. However, for a PBH with mass of $5 \times 10^{14}$ g or less, its radiation could turn into strong observational signals. One of the plausible constraints on PBHs derives from the primordial abundances of D, $^3$He, $^4$He and $^7$Li. Those light nuclei are produced via BBN, which is one of the robust tools to probe the physics in the early Universe. The primordial elemental abundances can be significantly altered if there are extra particle injections during or after BBN \cite{1979MNRAS.188P..15L,1985NuPhB.259..175E,1988ApJ...330..545D,Dimopoulos:1988ue,Dimopoulos:1988zz,Terasawa:1988my,1992NuPhB.373..399E,Kawasaki:1993gz,1995ApJ...452..506K,Jedamzik:1999di,Kawasaki:2000qr,2003PhRvD..67j3521C,Kawasaki:2004yh,2005PhRvD..71h3502K,Ellis:2005ii,2006PhRvD..74j3509J,2006PhRvD..74b3526K,Kawasaki:2008qe,2009PhRvD..79l3513K,Kusakabe:2012ds,2013PhRvD..87h5045K,2014PhRvD..90h3519I,Kawasaki:2020qxm}. Previous studies \cite{Carr:2009jm,Carr:2020gox,2020JCAP...06..018A} show that elemental abundances can provide the strongest constraints on the PBH mass spectrum for the {case that} $M< 10^{13}$ g via the extra particle{s} injection{s}. Their results suggest that except for photons, all species only make contributions in the epoch before (or during) the BBN via weak interactions and (or) hadronization processes. For $M >10^{12}$ g, the photodisintegration of light nuclei triggered by Hawking radiation is the main effect on the primordial elemental abundances \cite{Carr:2020gox}.

The nonthermal photodisintegration of primordial elements has been studied in various models (e.g., \cite{Kawasaki:2000qr,Kawasaki:2004yh,Kusakabe:2012ds,2014PhRvD..90h3519I,Salvati:2016jng,Goudelis:2015wpa}). In the past, although in different models, for energetic photons source{s}, the emitted photons are reckoned to experience electromagnetic (EM) cascades and finally reach a steady power-law spectrum \cite{2014PhRvD..90h3519I,2013PhRvD..87h5045K,2009PhRvD..79l3513K,2006PhRvD..74j3509J,2006PhRvD..74b3526K,2005PhRvD..71h3502K,1995ApJ...452..506K}. However, recent studies \cite{2015PhRvD..91j3007P,2015PhRvL.114i1101P,Kawasaki:2020qxm} suggest that this power-law spectrum is only valid for primary photons with energies higher than the threshold for the $e^{\pm}$ pair creation with background photons, i.e., $E_{th}\equiv10$ MeV $T^{-1}_{\rm keV}$ where $T_{\rm keV}$ is the cosmic temperature in the unit of keV (hereafter, we use the natural units with $c=k_B=\hbar =1$). For low-energy photons below {$E_{th}$} the cascade process is not fully triggered. In such a situation, an enhancement from the case with the power-law spectrum is found in the high-energy tail of the nonthermal photon{s} spectrum by solving the Boltzmann equation.  Such a sub-threshold photon emission scenario is essential for PBH in mass range $M=10^{12}-10^{13}$ g since the main radiation component for those PBHs consists of the secondary photons  (i.e., photons produced by the directly radiated particles, see Sec. \ref{sec:Hawk_rad}), and at later cosmic times with low temperatures, the power-law spectrum is not formed.

In this work, we correct the nonthermal photons spectrum properly taking into account the energy degradation processes of low-energy photons emitted from PBHs at low cosmic temperature, and we consider both monochromatic and critical collapse mass functions of PBH at a single formation epoch. For the monochromatic case, we find a discrepancy between our results and the previous $^3$He constraints. Then, we carry on careful photodisintegration calculations and also make analytical estimation by taking into account hadrodissociation for confirmation. As far as we know, there was no study which focused on $^3$He constraint on the critical collapse mass function within the same mass range. Then, we derive a $^3$He constraint on critical collapse mass function which differs from the existing constraint due to significantly distinguished nonthermal photon spectra. Unlike the power-law spectrum, the corrected photons spectrum depends on the initial emission spectrum and  because of the low-mass tail, the emission spectrum could be enhanced in the range $E_\gamma\geq{\mathcal O}(10)$ MeV, and it leads to the enhancement of the final steady-state spectrum in the same energy region.  For the critical collapse mass function, the stronger $^3$He constraint is obtained compared with the monochromatic one, which is basically due to the presence of the low-mass tail.

This article is organized as follows: in Sec. \ref{sec:Hawk_rad}, we describe the monochromatic and critical collapse mass functions, respectively, stemming from an inflationary spectrum with a narrow peak; and the corresponding photon emission spectra from Hawking radiation are calculated. In Sec. \ref{sec:photodis}, we calculate the photodisintegration of light nuclei from the photon{s} emission of PBHs. The reaction rates are calculated based on the photon{s} spectrum solved by Boltzmann equation with the up-to-date cross section data. In Sec. \ref{sec:constrn}, we provide constraints from nonthermal nucleosynthesis on initial mass spectrum of {PBH} with {the} horizon mass of $M_{H} =10^{12}-10^{13}$ g for the monochromatic and critical collapse mass functions by using an observational limit on the Galactic $^3$He abundance. Finally we conclude with discussion in Sec. \ref{sec:concl}.

\section{Primordial Black Holes and Hawking Radiation}
\label{sec:Hawk_rad}

\subsection{PBH formation and monochromatic mass function}

When the small-scale primordial inhomogeneities re-enter the Hubble horizon at the radiation-dominant epoch, they would collapse into a black hole under the influence of gravitation if their energy densit{ies} contrast $\delta$ exceeds the threshold value $\delta_c$. The simple analysis shows that the mass of  {the} formed PBH is roughly the horizon mass within the Hubble horizon at the formation epoch, which is so called horizon-mass approximation \cite{Carr:1974nx}, and we yield the PBH mass as
\be \label{horizon_mass}
M = \gamma M_H = \gamma \frac{1}{2 G} H^{-1} ,
\ee
where $M_H$ is the horizon mass, $H$ is the Hubble parameter, and $\gamma$ (somewhat below unity) is a correction factor that depends on the details of gravitational collapse \cite{Carr:2020gox}. The initial PBH mass spectrum $\beta$ is defined as the ratio of the energy density of PBH to the total energy density at the formation time \cite{Carr:2009jm}
\bl \label{beta}
\beta(M) \equiv \frac{\rho_\text{PBH}}{\rho_\text{tot}}
= & \frac{M n_\text{PBH}(t_f)}{\rho_\text{tot}(t_f)}  \nn
\\ \simeq & 7.98 \times 10^{-29} \gamma^{-1/2} \l( \frac{g_{*,\text{form}}}{106.75} \r)^{1/4} \l( \frac{M}{M_\odot} \r)^{3/2} \l( \frac{n_{\text{PBH}}(t_0)}{1 \text{Gpc}^{-3}} \r) ,
\el
where we have assumed an adiabatic cosmic expansion after PBH formation, i.e., the ratio of PBH number density to the entropy density $n_\text{PBH} / s$ is conserved. And we choose the value of the entropy density $s(t_0) = 8.55 \times 10^{85}\ \text{Gpc}^{-3}$ at present time $t_0$. $n_\text{PBH}(t_f)$ is the physical number density of PBHs at formation time $t_f$. $g_{*,\text{form}}$ is now normalised to the value of $g_*$ at around $10^{-5}$ s since it does not increase much before that in the Standard Model and most PBHs are likely to form before then. Notice that the second line of \eqref{beta} is valid for PBHs which survive today. Analogous to Refs. \cite{Carr:2009jm, Carr:2020gox}, we introduce a new parameter
\be \label{betaprime}
\beta'(M) \equiv \gamma^{1/2} \l( \frac{g_{*,\text{form}}}{106.75} \r)^{- 1/4} \l( \frac{h}{0.67} \r)^{-2} \beta(M) .
\ee
For the massive PBHs with initial mass{s} $\gtrsim 10^{15}$ g which accounts for DM, one can define the energy fraction of PBHs against the total DM component at present time \cite{Sasaki:2018dmp}
\be\label{fpbh}
f_\text{PBH}(M) \equiv \frac{\Omega_\text{PBH}}{\Omega_\text{DM}}
\simeq 1.52 \times 10^8 \l( \frac{\gamma}{0.2} \r)^{1/2} \l( \frac{g_{*,\text{form}}}{106.75} \r)^{-1/4}
\l( \frac{M}{M_{\odot}} \r)^{-1/2} \beta(M) ,
\ee
which can be bounded by various observations presented in Fig. 1 in Ref. \cite{Carr:2020xqk}.

Using the Press-Schechter formalism \cite{Press:1973iz}, once the probability distribution function of the primordial density fluctuations $P(\delta)$ is given, $\beta$ can be regarded as the probability that the energy density contrast exceeds the threshold for PBH formation. For simplicity, we assume a Gaussian initial perturbation profile
\be \label{Gauss_PDF}
P(\delta) = \frac{1}{\sqrt{2 \pi} \sigma} \exp\l( - \frac{\delta^2}{2 \sigma^2} \r) ,
\ee
which accords well with the current CMB experiments \cite{Akrami:2018odb}. The parameter $\sigma$ is the variance of the primordial density perturbation. And thus, we can evaluate the initial PBH mass spectrum as
\be \label{beta_mono}
\beta
= 2 \gamma \int_{\delta_c}^\infty d\delta P(\delta)
= \gamma \text{erfc}\l[ \frac{\delta_c}{\sqrt{2} \sigma} \r] .
\ee
The ``fudge factor $2$'' is introduced here to account for a finite probability for the collapse of underdense regions enclosed by the overdense regions \cite{Press:1973iz}. Note that we have extended the upper limit of the integral to $\sigma>1$ in contrast to some previous {literature} (e.g. \cite{Sasaki:1986hm}). It has been shown in Ref. \cite{Kopp:2010sh} that the large $\delta$ does not lead to the separate universe, it is merely a gauge chose, and the same topic is also discussed in Ref. \cite{Carr:2014pga}. However, this infinite upper limit is still sensible in practice since the integrand for large $\delta$ is exponentially suppressed, and these two choices of upper limit are nearly equivalent. The early analysis \cite{Carr:1975qj} shows that the threshold $\delta_c$ is simply related to the equation of state parameter of matter components when collapse occurs in the radiation-dominant era $\delta_c \simeq \omega = 1/3$. More precise numerical \cite{Musco:2012au} and analytic \cite{Harada:2013epa} investigations suggest $\delta_c = 0.45$, and we will adopt this value in our study. The other studies on the threshold can be found e.g., in Refs. \cite{Escriva:2020tak,Escriva:2019phb}.

Let us stress that the above energy density $\delta$ and the variance $\sigma$ always refers to the smooth ones after smoothing process over the horizon scale at PBH formation, which can be calculated from the power spectrum of primordial comoving curvature perturbation $\mathcal{P}_{\zeta}(k)$ generated during inflation in a concrete PBH formation scenario \cite{Sasaki:2018dmp}
\be \label{sigma}
\sigma(M)^2 = \int_0^\infty d \ln k W(k/k_{M})^2 \frac{16}{81} ( k/k_{M} )^4 \mathcal{P}_{\zeta}(k) ,
\ee
where $W(x) = \exp(- x^2/2)$ is a Gaussian window function and $k_{M}$ is the comoving wavenumber corresponding to the mass $M$. In order to produce PBHs efficiently, the value of $\mathcal{P}_{\zeta}(k)$ is required to be amplified on small scales and remains scale-invariant on large scales, which is confirmed by CMB observations \cite{Akrami:2018odb}.

In this paper, we focus on the simplest case that all PBHs formed at a single epoch, which is the situation for a narrow enhanced inflationary spectrum. Such a narrow spectrum as expected to be responsible for the single epoch formation could exits in several scenarios, such as some phase-transition models \cite{Jedamzik:1999am}, Starobinsky's $R^2$-gravity \cite{Pi:2017gih} and sound speed resonance (SSR) mechanism \cite{Cai:2018tuh, Chen:2019zza, Chen:2020uhe}. And it is straightforward to extend to the broad spectrum case by applying the treatment used in Ref. \cite{Kuhnel:2015vtw}, i.e., binning the inflationary spectrum and each bin corresponding to a particular horizon mass, and then calculate PBH mass function similar to the narrow spectrum case. The final result is just the total contributions from all individual bins. Since the scale-invariant part of the power spectrum is almost smaller than the critical density (extremely exponentially suppressed), no black holes would form except at scales around the amplified peak in inflationary spectrum. Analogous to Refs. \cite{Cai:2018tuh, Chen:2019zza, Chen:2020uhe}, we use the delta-function {$\Delta$} to parametrize the narrowly-enhanced primordial curvature perturbation in the following form {
\be \label{Pzeta}
\mathcal{P}_{\zeta}(k) \simeq A_s \l(\frac{k}{k_p}\r)^{n_s-1} \Big( 1 + \lambda \alpha k_* \Delta(k - k_*) \Big) ,
\ee }
where $A_s = H^2/( 8 \pi^2 \epsilon M_p^2 )$ is the amplitude of the power spectrum predicted by the conventional inflationary paradigm, here $\epsilon$ is the slow-roll parameter, $M_p$ is the reduced Planck mass, and $n_s$ is the spectral index at the pivot scale $k_p = 0.05\ {\rm Mpc}^{-1} $ \cite{Akrami:2018odb}. $k_*$ is the position of the peak, and $\lambda$ measures the amplitude of peak and satisfies $A_s \lambda \leq 1$. Note that we also introduce an additional parameter $\alpha$ to account for the effective parametrization of delta-function of inflationary spectrum. In a realistic circumstance, the shape of the narrow amplified peak is usually close to a Gauss function. The effective parameter $\alpha$ can be calculated based on the Guassian approximation so that the area enclosed by the Gauss function around peak is roughly $\lambda \alpha k_*$. For example, $\alpha = \xi/2$ in SSR mechanism \cite{Cai:2018tuh}, since we use the area of the triangle to represents the peak approximately. The parameter $\alpha$ is model-dependent and restricted to be less than unity if $\lambda$ is identical with the peak of the Gauss function. As we are working in the perturbative regime, the height of the peak in $\mathcal{P}_{\zeta}(k)$ should be no more than unity, corresponding to a maximal variance \cite{Cai:2018tuh}. And hence, we yield
\be \label{sigma_max}
\sigma(M)^2 \lesssim
\frac{16}{81} \alpha \l(\frac{k_\odot}{k_p}\r)^{n_s-1} \l( \frac{M_\odot}{M_*} \r)^{\frac{n_s + 3}{2}} \l( \frac{M}{M_\odot} \r)^2 e^{- M / M_*} ,
\ee
where we have used the relation between the PBH mass and the corresponding comoving wavenumber: $M \propto k_{M}^{-2}$ \cite{Sasaki:2018dmp}. $M_*$ is the mass scale corresponding to the characteristic scale $k_*$, and it is straightforward to see from \eqref{sigma_max} that $\sigma(M)^2$ would peak when $M \simeq 2 M_*$. $k_\odot \simeq 1.9 \times 10^6\ \text{Mpc}^{-1}$ is the comoving scale corresponding to the solar-mass PBHs at the formation epoch. Since the PBHs formed at a single epoch, they would have the same mass expressed in \eqref{horizon_mass}, i.e., the monochromatic mass function (i.e., with a width $\Delta M \sim M$). Numerical estimates of $\beta(M_H)$ at various horizon masses $M_H$ are shown by the blue solid curves in Fig. \ref{fig:beta}, for $\gamma = 1$, $g_\text{form} = 106.75$, $\delta_c = 0.45$ \cite{Carr:2020gox} and $n_s = 0.968$, $h= 0.674$ \cite{Akrami:2018odb}, and choosing the effective parameter $\alpha = 0.055$ in \eqref{Pzeta}.

\subsection{The critical collapse and extended mass function}

It is well known that PBH formation is associated with the critical phenomena of gravitational collapse, and the early studies showed that the PBH mass at the formation epoch has a universal scaling property \cite{Niemeyer:1997mt}
\be \label{crit_mass}
M = K M_H ( \delta - \delta_c )^\nu ,
\ee
near the threshold $\delta \simeq \delta_c$ for PBH formation. This scaling relation indicates that a PBH forms from the primordial density perturbation possible with an arbitrary small mass, in contrast to the usual assumption that all the PBHs have the same horizon mass at a single formation epoch. The results in Refs. \cite{Yokoyama:1998xd, Green:1999xm} show that the horizon-mass approximation is still reasonably good and that the mass function peaks around the horizon mass. This conclusion depends on, however, the assumption that all the PBHs formed at a single epoch. The dimensionless constant $K$, the universal critical exponent $\nu$ and the threshold $\delta_c$ all depend on the nature of the background fluid when the overdensity $\delta$ re-enters the Hubble horizon \cite{Musco:2012au}. It also turns out that the critical exponent $\nu$ is independent of the initial fluctuations profile \cite{Musco:2012au, Neilsen:1998qc}, though $K$ and $\delta_c$ may depend on it. In the radiation-dominant epoch, which are focused on this paper, many studies have shown that $\nu \simeq 0.35$ and $K \simeq 3.3$ \cite{Niemeyer:1997mt, Musco:2012au, Musco:2008hv, Musco:2004ak}.

\begin{figure}[h]
	\centering
	\includegraphics[width=2.9in]{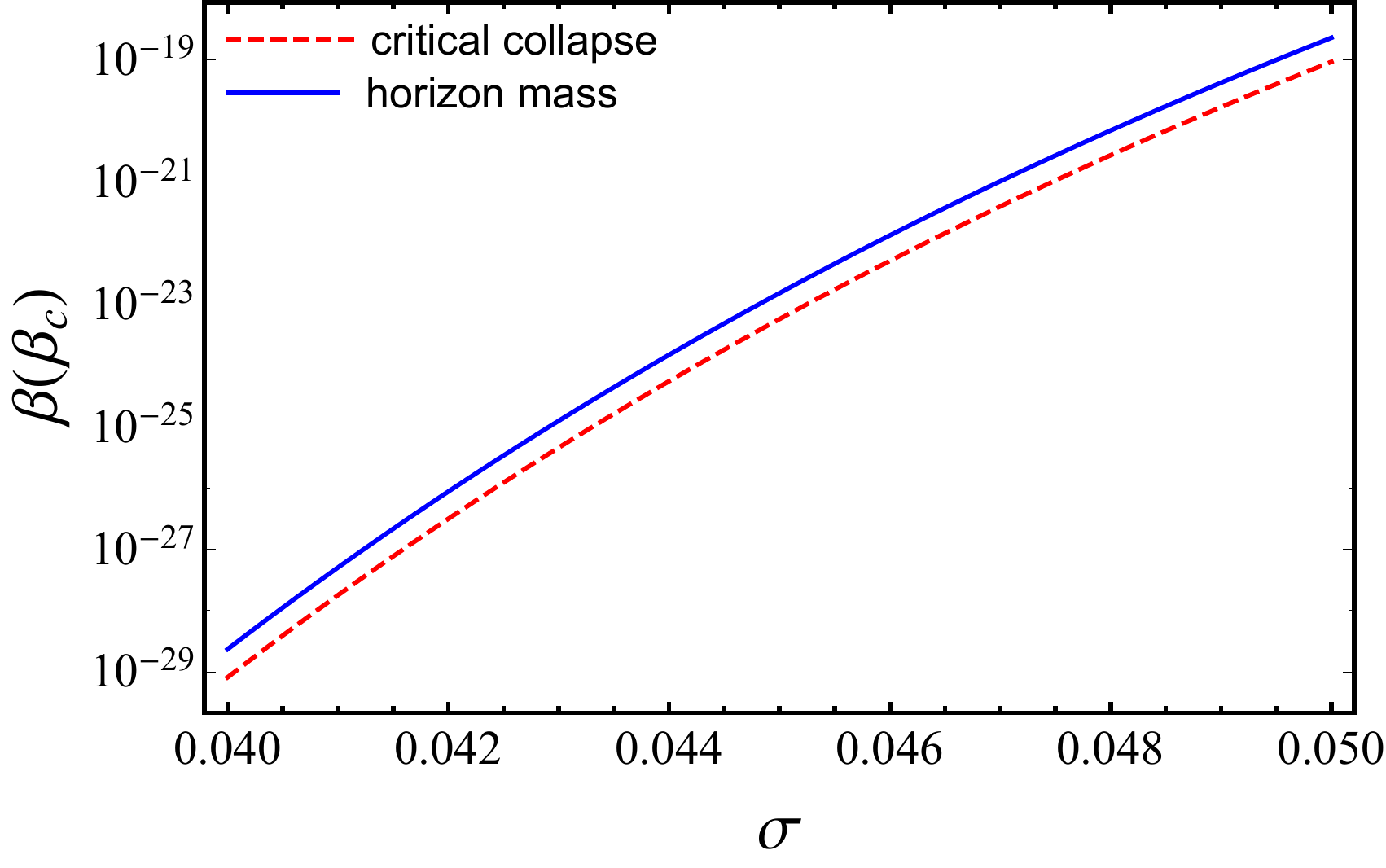}
	\includegraphics[width=2.9in]{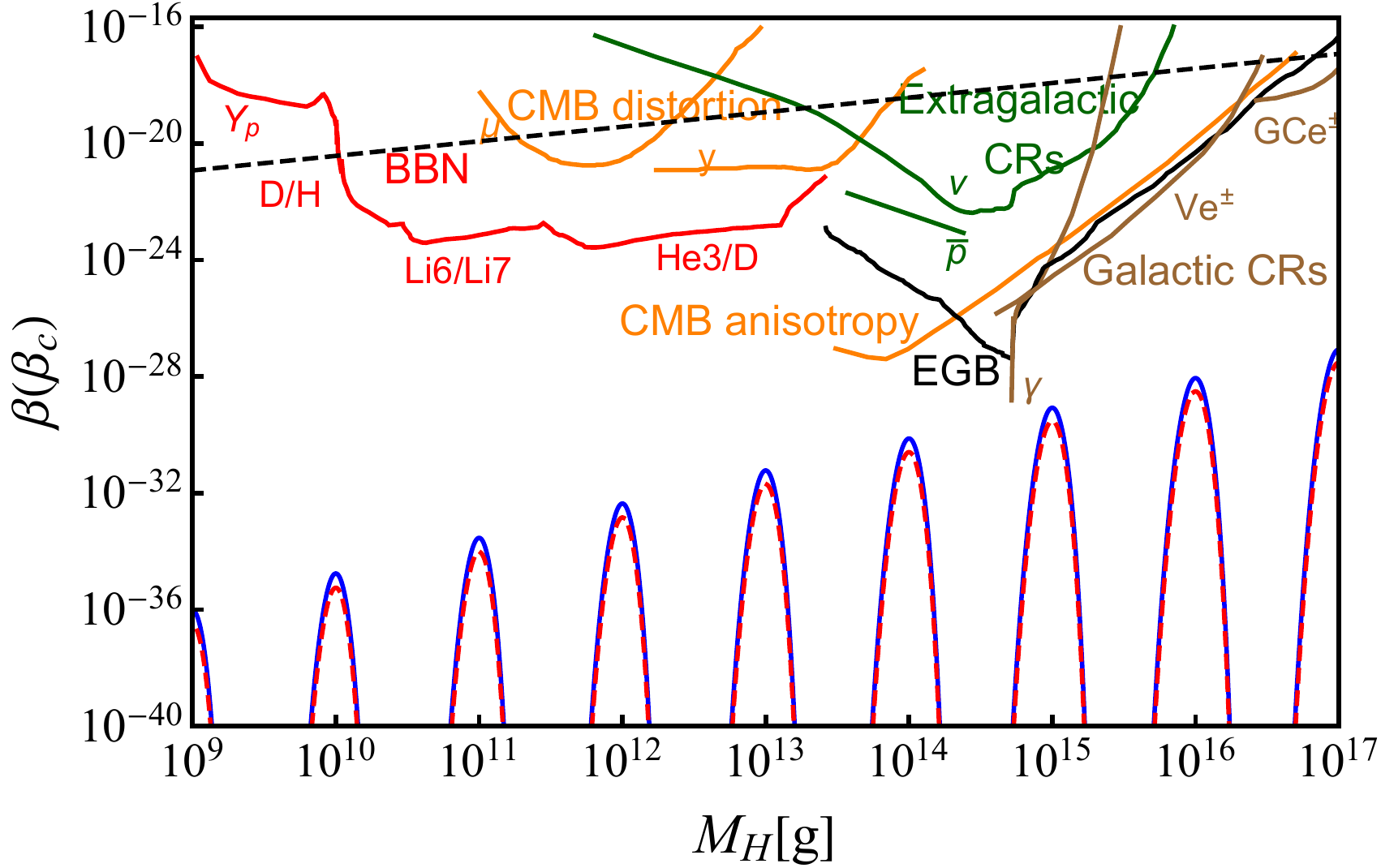}
	\caption{
	The comparison between $\beta_c$ (the red dashed curves) in \eqref{beta_crit} and $\beta$ (the blue solid curves) in \eqref{beta_mono}. The left panel shows $\beta_c$ and $\beta$ as functions of $\sigma$ lying in $(0.04, 0.05)$; the right panel represents the functions $\beta_c(M_H)$ and $\beta(M_H)$ in the mass range $(10^{9}, 10^{17})$ g, where the variance $\sigma(M_H)$ is taken by the maximum value in \eqref{sigma_max}. The various monochromatic form of constraints displayed in Fig. 4  of \cite{Carr:2020gox} are also shown: BBN (red, from $Y_p$, D/H, $^6$Li/$^7$Li and $^3$He/D observations), EGB (black) constraints, CMB distortion and anisotropy (orange), extragalactic cosmic rays (green, involving extragalactic antiprotons and neutrinos), galactic cosmic rays (brown, involving galactic $\gamma$-ray background, e$^{\pm}$ annihilations in the Galactic centre and e$^{\pm}$ observations by Voyager 1). The parameters are fixed as follows: $\gamma = 1$, $g_\text{form} = 106.75$, $\delta_c = 0.45$ and $n_s = 0.968$, $\alpha = 0.055$.
	}
	\label{fig:beta}
\end{figure}

Using the Press-Schechter formalism similar to \eqref{beta_mono}, the initial mass spectrum is calculated as \cite{Carr:2016drx}
\be \label{beta_crit}
\beta_c(M_H) = 2 \int_{\delta_c}^{\infty} d\delta K ( \delta - \delta_c )^\nu P(\delta)
\simeq K \sigma(M_H)^{2 \nu} \text{erfc}\l[ \frac{\delta_c}{\sqrt{2} \sigma(M_H)} \r] ,
\ee
here the subscript ``c'' refers to the critical collapse and we also infinitely extend the upper limit of the integral above $\delta = 1$ as we did in the monochromatic case. The standard deviation $\sigma(M_H)^2$ at the horizon mass $M_H$ is given by \eqref{sigma}. Let us stress that although the relation \eqref{crit_mass} only holds in the neighborhood of the threshold $\delta_c$, it is usually assumed still valid as $\delta$ becomes much larger (e.g. \cite{Carr:2016drx, Kuhnel:2015vtw}). Nevertheless, this is always sensible in practice since the contribution to the energy density fraction $\beta_c$ \eqref{beta_crit} is mainly from the $\delta \simeq \delta_c$ owing to the exponential form of $P(\delta)$ in \eqref{Gauss_PDF}. As shown in Ref. \cite{Kuhnel:2015vtw}, the inclusion of critical collapse for a variety of inflationary models (e.g., running mass inflation, hybrid inflation, axion-like curvaton inflation and first-order phase transitions) can generally lead to a shift, lowering and broadening of initial PBH mass spectrum. This effect is model- and parameter- dependent and cannot be contained by a constant rescaling of the spectrum, and it should be taken into account when comparing to observational constraints. As we have mentioned above, for a monochromatic mass function or a narrow inflationary spectrum, the horizon-mass approximation is still a good approximation, and the difference between $\beta_c$ in \eqref{beta_crit} and $\beta$ in \eqref{beta_mono} is relatively small. As an illustration, we plot $\beta_c$ and $\beta$ as functions of the variance $\sigma$ and the horizon mass $M_H$ in Fig. \ref{fig:beta}, respectively. For $\beta_c(M_H)$ and $\beta(M_H)$, we choose a set of horizon masses $M_H = ( 10^9, 10^{10}, 10^{11}, 10^{12}, 10^{13}, 10^{14}, 10^{15}, 10^{16}, 10^{17})$ g and $\alpha = 0.055$. The amplitude of $\beta_c$ ($\beta$) can be efficiently enhanced by setting larger values of $\alpha$.

The rescaling relation between PBH{s} mass{es} $M$ and density contrast $\delta$ implies that the PBH{s} masses are naturally extended due to the Gaussian distribution of $\delta$ in \eqref{Gauss_PDF}. Non-monochromaticity is usually described by the differential comoving number density, which is defined as $n(M) \equiv dn / dM$, where $dn$ is the comoving number density in the mass range $(M, M + dM)$. And we also assume that there is only one PBH formed within each horizon volume, i.e., the comoving number density is expressed as $n_\text{PBH}(t_f) = \beta / V_f$, $V_f$ is the comoving horizon volume at PBH formation, here $\beta = \int_{\delta_c}^{\infty} P(\delta) d \delta$ is included to account for the probability for the collapsing horizon. This assumption is sensible in the common cases since all PBHs formed at the same time would have the same horizon mass. Using the mass relation \eqref{crit_mass} and the Gaussian distribution \eqref{Gauss_PDF}, we yield the differential comoving number density as \cite{Carr:2016hva}
\bl
n(M) =& \frac{1}{\beta} n_\text{PBH}(t_f) P[\delta(M)] \frac{d \delta(M)}{d M} \nn
\\=&
\frac{1}{\sqrt{2 \pi} V_f \nu K M_H \sigma(M_H)} \l( \frac{M}{K M_H} \r)^{1/\nu - 1} \exp \l[ - \frac{ \l( \l( M / K M_H \r)^{1/\nu} + \delta_c \r)^2 }{2 \sigma(M_H)^2} \r] \label{nM_precise}
\\\simeq&
A(M_H) M^{1/\nu - 1} \exp\l[ - ( 1 - \nu ) \l( \frac{M}{ M_\text{peak} } \r)^{1/\nu} \r] ~, \label{nM_appro}
\el
for each horizon mass $M_H$, which satisfies the normalization$\int n(M) dM = n_\text{PBH}(t_f) = \beta / V_f$. Note that the second line \eqref{nM_precise} is the accurate solution also appeared in Refs. \cite{Niemeyer:1997mt, Yokoyama:1998xd}, while the third line \eqref{nM_appro} is the approximated one which is frequently used in literature, e.g., \cite{Carr:2016hva, Carr:2017jsz, Carr:2018poi}. This approximate solution is obtained when we expand the exponential up to the first order around the threshold $\delta_c$, and this approximation is applicable if $\sigma \ll \delta_c$ for Gaussian distribution of $\delta$. Since this condition is satisfied for a realistic abundance of PBHs (otherwise PBHs are overproduced), which can be easily seen from the relation \eqref{beta_mono}, this approximation is widely applicable \cite{Yokoyama:1998xd}. The detailed calculation shows that
\be \label{nM_A}
A(M_H) = \frac{\beta(M_H)}{V_f M_\text{peak}^{1/\nu} } \l( \frac{1 - \nu}{\nu} \r)
= \frac{3 a(M_H)^3 \beta(M_H)}{4 \pi ( 2 G M_H)^3 M_\text{peak}^{1/\nu} } \l( \frac{1 - \nu}{\nu} \r) ,
\ee
where the relation $V_f = \frac{4 \pi}{3} \l( \frac{2 G M_H}{a(M_H)} \r)^3$ and the approximation for \eqref{beta_mono} is used \cite{Sasaki:2018dmp}, $\beta(M_H) \simeq {\sigma(M_H) \over \sqrt{2 \pi} \delta_c} \exp\l[- { \delta_c^2 \over 2 \sigma(M_H)^2 } \r]$, and $a(M_H)$ is the scale factor at the formation epoch associated with the horizon mass $M_H$. The parameter $M_\text{peak}$ is the mass where $n(M)$ peaks for a given $M_H$, we yield
\be
M_\text{peak} = K M_H \l( \frac{1 - \nu}{q} \r)^\nu,~~
q = \frac{\delta_c}{\sigma(M_H)^2} .
\ee
Note that $\beta(M_H)$ in our result \eqref{nM_A} is slightly different from $\beta(M_f)$ in Eq. (3.30) of Ref. \cite{Carr:2016hva}, and the horizon mass $M_f$ in their notation refers to $k M_H$ here. We should emphasize that there are only one parameter involved in $n(M)$, i.e., the variance of primordial density perturbations $\sigma(M_H)$ which is model-dependent. In Sec. \ref{sec:constrn}, we will place upper limits on $\sigma(M_H)$ by using the measurements of $^3$He/H abundance ratio.

\begin{figure}[h]
	\centering
	\includegraphics[width=3.5in]{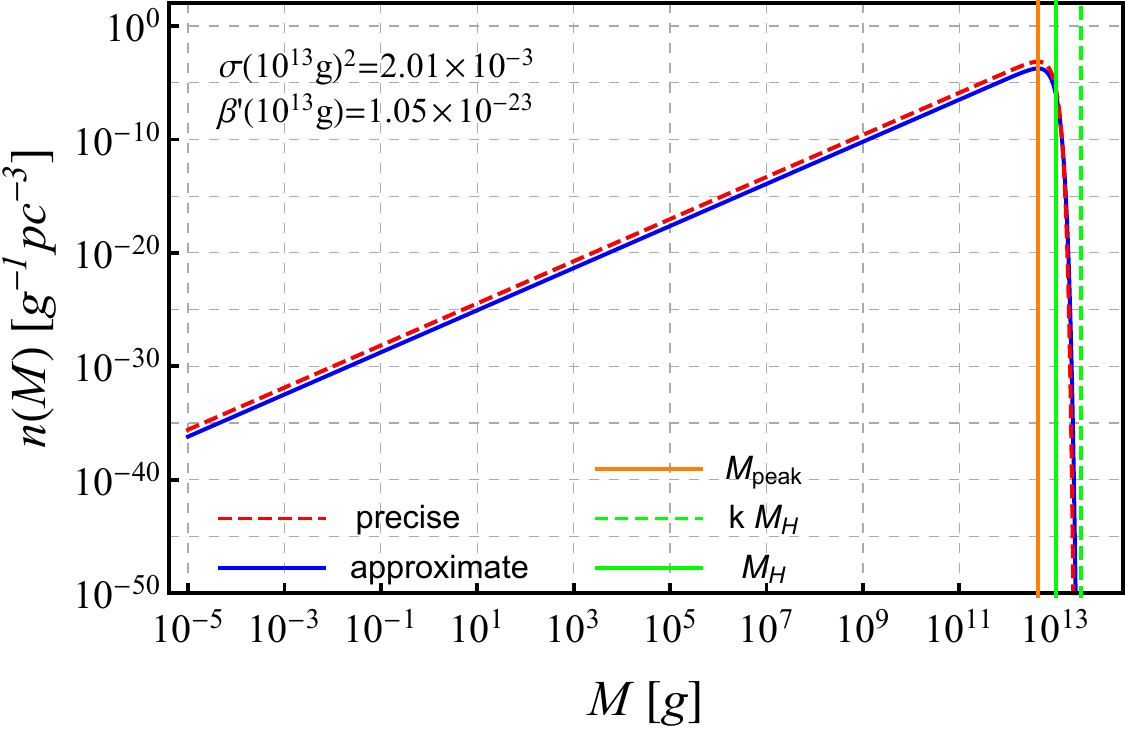}
	\caption{The comparison of the comoving number density $n(M)$ between the accurate formula (the red dashed curve) and the approximated formula (the blue solid curve) associated with the horizon mass $M_H = 10^{13}~\text{g}$. The green vertical solid line and the dotted line refer to the mass scale $M_H$ and upper cut-off scale $k M_H$, respectively; the orange solid line represents the position of peak mass $M_\text{peak}$. The value of variance is chosen as  $\sigma(10^{13}~\text{g})^2 = 2.01 \times 10^{-3}$, and the corresponding initial PBH mass spectrum is $\beta'(10^{13}~\text{g}) \simeq 1.05 \times 10^{-23}$, both of which are below the bounds from $^3$He /H abundance ratio. }
	\label{fig:nM}
\end{figure}

For the purpose of numerical computation, we normalize the present scale factor $a(t_0) = 1$, and chose the Hubble parameter $H_0 = 67.4\ \text{km}~ \text{s}^{-1}~ \text{Mpc}^{-1}$ . Taking into account the horizon mass relation $M_H = H^{-1}/(2 G)$ for $\gamma = 1$ in \eqref{horizon_mass}, we can solve $a(M_H)$ in \eqref{nM_A} from the Friedmann equation $H^2(a) = H_0^2 (\Omega_r a^{-4} + \Omega_k a^{-2} + \Omega_m a^{-3} + \Omega_\Lambda )$, here $\Omega_r \simeq 10^{-4}$, $\Omega_k \simeq 0$, $\Omega_m \simeq 0.315$ and $\Omega_\Lambda \simeq 0.6847$ are normalized radiation, curvature, baryon and dark energy density parameters, respectively. We thus yield the scale factor at the formation epoch for PBHs with the horizon mass $10^{13} \text{g}$: $a(10^{13} \text{g}) \simeq 1.4719 \times 10^{-23}$. Fig. \ref{fig:nM} displays the comparison between the accurate expression \eqref{nM_precise} and the approximated expression \eqref{nM_appro}. It is clearly seen that the approximated value is very close to the precise one at each mass scale as we expect, since the small value of $\sigma(10^{13}~\text{g})^2$ is constrained by $^3$He/H abundance ratio, see the discussions in Sec. \ref{sec:constrn}.

Using the differential comoving number density $n(M)$, the initial mass function is defined as
\be
\varphi(M, M_H) \equiv \frac{1}{\rho_\text{tot} a(M_H)^3} M \frac{d n}{d M} ,
\ee
and the total energy faction of PBHs at their formation is thus given by
\be \label{beta_c}
\beta_c(M_H) = \int_{M_\text{min}}^{M_\text{max}} \varphi(M, M_H) dM ,
\ee
which is consistent with \eqref{beta_crit}. Here $M_\text{min}$ is the minimal mass that is always identified as Planck mass $\sim 2 \times 10^{-5}$ g (e.g., assuming stable Planck relics as the final product of evaporation of PBHs \cite{Carr:2020gox, Carr:2020xqk}), and $M_\text{max}$ is the PBH mass corresponding to $\delta_\text{max}$. Since $\delta_\text{max}$ tends to be infinite in \eqref{beta_crit}, $M_\text{max}$ can also be set to infinity. Note that this does not mean arbitrary large PBHs can be produced, the probability in fact tends to zero in the such a large mass region. In practice, it is reasonable to set $M_\text{max}$ to the cutoff mass scale $k M_H$ as depicted in Fig. \ref{fig:nM}. Analogous to the definition \eqref{betaprime}, one can also define $\beta_c'$ related to $\beta_c$.

\subsection{Photons spectra from Hawking radiation}

In 1974, Hawking found that a black hole could emit particles similar to the black-body radiation, with energies in the range $(E,E + dE)$ at a rate \cite{Hawking:1974rv,Hawking:1974sw}
\be \label{emission_rate}
\frac{d^2 N}{dt dE} = \frac{1}{2 \pi} \frac{\Gamma_s(E,M)}{e^{8 \pi G M E} - (-1)^{2s}} ,
\ee
per particle degree of freedom (e.g. spin, electric charge, flavor and color). Here $M$ is the mass of the black hole, $s$ is the particle spin and the {black hole} temperature is thus defined as
\be \label{BH_Temperature}
T_\text{BH} = \frac{1}{8 \pi G M} \simeq 1.06 \times M_{10}^{-1} \text{TeV} ,
\ee
where $M_{10}$ is related to the {black hole} mass $M \equiv M_{10} \times 10^{10} ~\text{g}$. And $\Gamma_s(E,M)$ is the dimensionless absorption coefficient which accounts for the probability that the particle would be absorbed if it were incident in this state on the black hole. It appears in the emission formula on account of detailed balance between emission and absorption. In general, $\Gamma_s(E,M)$ depends on the spin, the energy of emitted particle and the {black hole} mass. The absorption coefficient is expressed as  $\Gamma_s(E,M) = E^2 \sigma_s(E,M) / \pi$, here $\sigma_s(E,M)$ is the corresponding absorption cross section. In the high-energy limit $E \gg T_\text{BH}$, $\sigma_s(E,M)$ approaches to geometric optics limit $\sigma_g = 27 \pi G^2 M^2$ which is independent of the energy of emitted particle. The functional expressions of $\Gamma_s(E,M)$ for massless and massive particles can be found in Refs. \cite{Page:1976df, Page:1976ki, Page:1977um}. Hawking temperature \eqref{BH_Temperature} tells us that a smaller {black hole} is much hotter than a larger {black hole}, naturally, the emission is also stronger. So that in this sense, PBHs can be small enough for Hawking radiation to be significant.

\begin{figure}[h]
	\centering
	\includegraphics[width=3.3in]{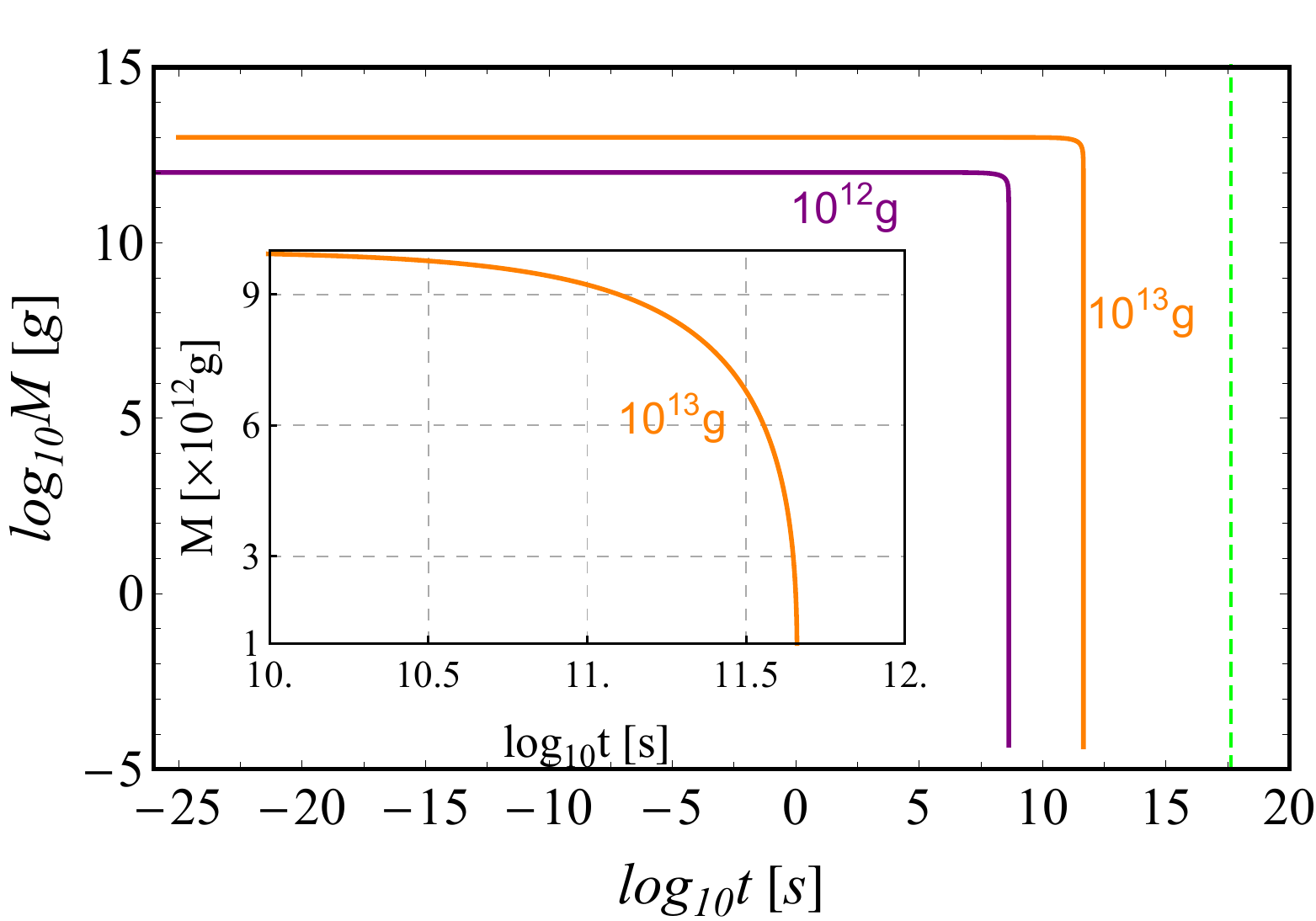}
	\caption{The mass evolutions of two types of PBHs with initial masses $10^{12}$ g (the purple curve) and $10^{13}$ g (the orange curve). The green dashed vertical line refers to the present time. i.e., the age of Universe $t_0 = 13.8$ Gyr. The numerical estimation is performed by the code BlackHawk \cite{Arbey:2019mbc}.}
	\label{fig:mass}
\end{figure}

Note that we adopt the assumption that the black hole has no charge or angular momentum, which is reasonable since charge and angular momentum will also be lost through quantum emission on a shorter time scale than the mass loss time scale; extension to the charged and rotational black holes is straightforward \cite{Page:1976df, Page:1976ki, Page:1977um}. Since the black hole continuously emits particle, its mass decreases while the temperature goes up. The approximate formula for the mass loss rate is written as \cite{MacGibbon:1990zk, Carr:2009jm}
\be \label{massloss_rate}
\frac{d M_\text{10}}{d t} \simeq - 5.34 \times 10^{-5} \phi(M) M_{10}^{-2} ~~ \text{s}^{-1} ,
\ee
where $\phi(M)$ measures the number of emitted particle species and is normalized to unity for the {black hole}s with $M \gg 10^{17}$ g, emitting only massless photons, three generations of neutrinos and graviton. The relativistic contributions to $\phi(M)$ per degree of particle freedom are \cite{MacGibbon:1990zk}
\bl
&\phi_{s=0} = 0.267,~~ \phi_{s=1} = 0.060,~~\phi_{s=3/2} = 0.020 \nn
\\&\phi_{s=2} = 0.007,~~ \phi_{s=1/2} = 0.147 ~(\text{neutral}),~~ \phi_{s=1/2} = 0.142 ~(\text{charge} \pm e) .
\el
Integrating the mass loss rate \eqref{massloss_rate} over time then gives the lifetime of a {black hole}
\be
\tau \sim 407 \l( \frac{\phi(M)}{15.35} \r)^{-1} M_{10}^3 ~~\text{s} .
\ee
If we sum up the contributions from all the particles in the Standard Model up to 1 TeV, corresponding to $M_{10} \sim 1$, this gives $\phi(M) = 15.35$. The mass of a PBH evaporating at $\tau$ after Big Bang is given by \cite{Carr:2009jm}
\be
M \simeq 1.35 \times 10^9 \l( \frac{\phi(M)}{15.35} \r)^{1/3} \l( \frac{\tau}{1 \text{s}} \r)^{1/3} ~\text{g} .
\ee
Thus, the mass of a PBH evaporating at present is roughly $M_* \simeq 5.1 \times 10^{14} ~\text{g}$ (corresponding to $T_\text{BH} = 21 \text{MeV}$). Fig. \ref{fig:mass} shows the mass evolutions of two types of PBHs with initial masses $10^{13}$ g and $10^{16}$ g. It is clear that for the most of lifetime, PBH mass remains nearly unchanged and it would drop rapidly at the end stage of evaporation. Also, it usually assumes that PBH evaporation leave stable Planck-mass relics \cite{MacGibbon:1987my, Carr:2020xqk}. For the heavy PBHs with masses greater than $M_*$, the instantaneous emission rate \eqref{emission_rate} is almost time-independent.

We adopt a standard emission picture that a black hole emits only those particles which appear elementary on the scale of the radiated energy (or equivalently the black hole size) \cite{MacGibbon:1990zk}. The emitted particles could form into composite particles after emission. A {black} hole should emit all elementary particles whose rest masses are less than or of the order of $T_\text{BH}$. The spectra of the particles emitted through the life time of PBHs is calculated from BlackHawk code \cite{Arbey:2019mbc}. In order of increasing $T_\text{BH}$, the black hole initially directly emits only photons (and gravitons), then neutrinos, electrons, muons and eventually direct pions join in the emission as $T_\text{BH}$ surpasses successive particle rest mass thresholds. Once the black hole temperature exceeds QCD energy scale $\Lambda_{QCD} = 250 - 300 \text{MeV}$, the particles radiated can be regarded as asymptotically free, leading to the emission of quarks and gluons. After their emission, quarks and gluons fragment into further quarks and gluons until they cluster into the observable hadrons including protons and antiprotons, electrons, and positrons. Since there are 12 quark degrees of freedom per flavor and 16 gluon degrees of freedom, one would expect the emission rate (i.e., the value of $\phi$) to increase suddenly once the QCD temperature is reached. Thus, Hawking radiation is dominated by the decay of QCD particles when the PBH{s} mass{es} falls below $M_q \simeq 0.4 M_* \simeq 2 \times 10^{14} ~\text{g}$ \cite{MacGibbon:1990zk, MacGibbon:1991tj, Carr:2009jm}.

As discussed above, particles injected from a PBH have two components: the primary component, which is the direct Hawking emission; the secondary component, which comes from the decay of gauge bosons or heavy leptons and the hadrons produced by fragmentation of primary quarks and gluons \cite{Carr:2009jm}. For photons, we have
\be
\frac{d \dot{N}_\gamma}{d E_\gamma} (E_\gamma, M)
= \frac{d \dot{N}_\gamma^\text{pri}}{d E_\gamma}(E_\gamma, M)
+ \frac{d \dot{N}_\gamma^\text{sec}}{d E_\gamma}(E_\gamma, M) ,
\ee
with similar expressions to other particles. The average energy of the emitted particles are $4.22 T_\text{BH}$ for $s=1/2$, neutral, $4.18 T_\text{BH}$ for $s=1/2$, charged, and $5.71 T_\text{BH}$ for $s=1$, respectively. The peak energies of the flux and power are within $7\%$ of these values. For example, the energy peak for primary photons is $5.8 T_\text{BH}$. For secondary photons, the average and peak energy are both $m_{\pi^0}/2 \simeq 68 \text{MeV}$ independent of {black hole} temperature, because the secondary photons is dominated by 2 $\gamma$-decay of soft neutral pions which are practically at rest. The emission rate for primary photons at the peak energy is given by
\be
\frac{d \dot{N}_\gamma^\text{pri}}{d E_\gamma} (E_\gamma = E^\text{peak})
= 1.4 \times 10^{21} ~ \mathrm{s}^{-1} \text{GeV}^{-1} ,
\ee
and peak flux for the second photons is expressed as
\be
\frac{d \dot{N}_\gamma^\text{sec}}{d E_\gamma} (E_\gamma = m_{\pi^0}/2)
= 2 \sum_{i = q,g} B_{i \rightarrow  \pi^0}(\bar{E}, E_{\pi^0}) \frac{\bar{E}}{m_{\pi^0}} \frac{d \dot{N}_i^\text{pri}}{d E_i} (E_i \simeq \bar{E}) ~,
\ee
where $B_{i \rightarrow  \pi^0}(E_\text{jet}, E_{\pi^0})$ is the fraction of the jet energy $E_\text{jet}$ going into the neutral pions of energy $E_{\pi^0}$. This is of order 0.1 and fairly independent of jet energy. If we assume that most of primary particles have average energy $\bar{E} \simeq 4.4 T_\text{BH}$, last factor becomes $d \dot{N}_i^\text{pri} / d E_i \simeq 1.6 \times 10^{-3} \hbar^{-1}$. Fig. 1 in Ref. \cite{Carr:2009jm} shows the instantaneous emission rate of photons for four typical black hole temperatures, and Fig. 2 in Ref. \cite{Carr:2009jm}  shows the ratio of the secondary peak energy (flux) to the primary peak energy (flux), we can see that the secondary emission becomes significant once the {black hole} temperature reaches $\Lambda_{QCD}$, or equivalently the {black hole} mass falls to $M_q$.

The above analysis based on a single black hole. If we consider PBHs with an extended mass function, the total emission should be contributed by all individual mass scales. The time-dependent comoving number density of elementary particle, emitted by a distribution of PBHs per unit time and per unit energy is computed through the integral
\be \label{Hawk_spectra}
\frac{d \dot{N}_\text{tot}}{d E}(E_\gamma) = \int_{M_\text{min}}^{M_\text{max}} \frac{d^2 N}{dt dE}(E_\gamma,M) {dn \over dM} dM ~,
\ee
for a given $M_H$.
\begin{figure}[h]
	\centering
	\includegraphics[width=3.2in]{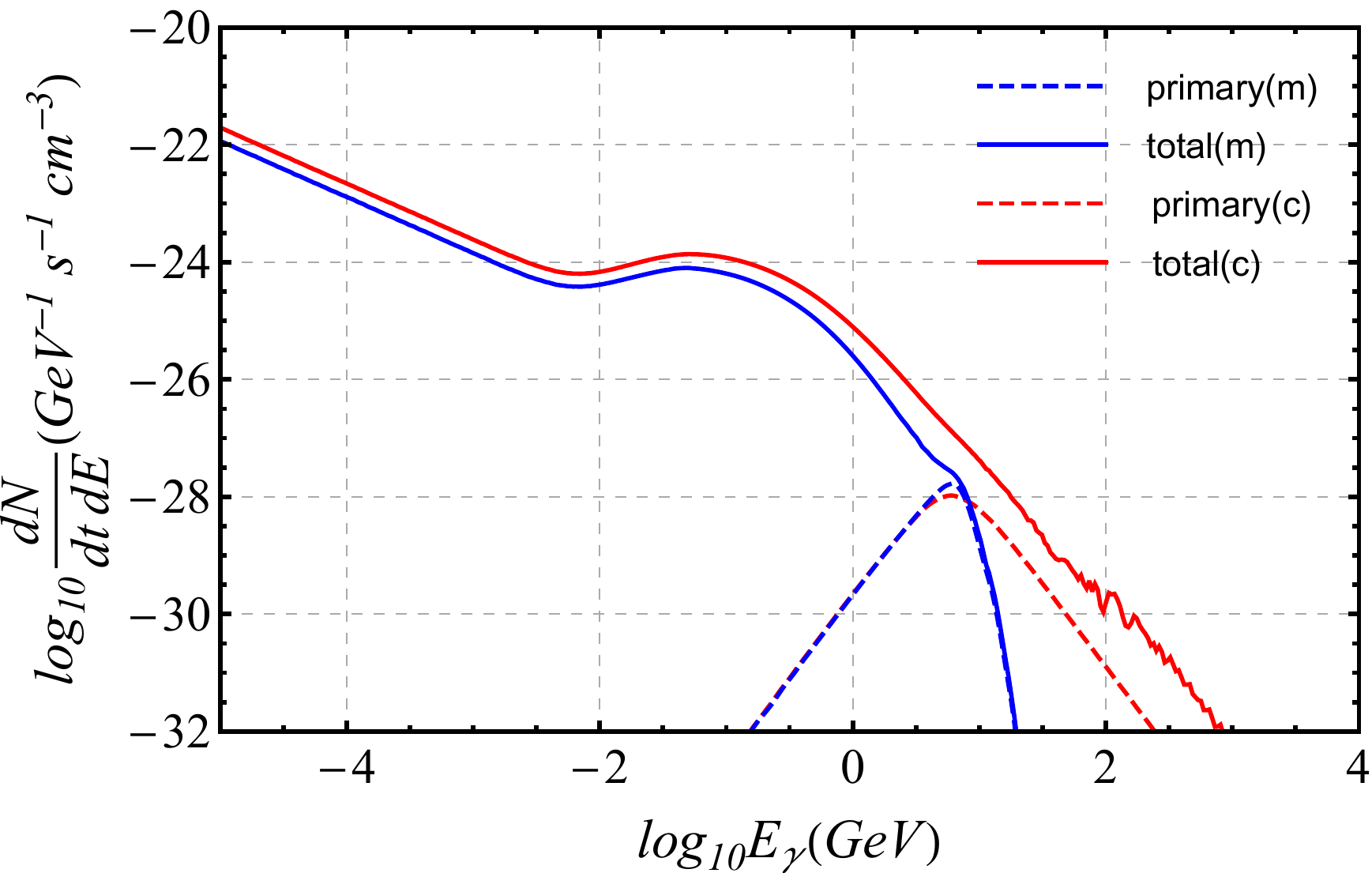}
	\caption{Instantaneous emission rate of photons per comoving $\text{cm}^3$ for PBHs with horizon mass $10^{13}$ g, for the same energy fraction of the monochromatic (denoted by the script ``m'') and critical collapse mass functions (denoted by the script ``c''): $\beta_m = \beta_c = 10^{-25}$. The blue curves and the red curves refer to the monochromatic and critical collapse mass function, respectively, while the dashed and the solid ones represent the instantaneous primary and total (primary + secondary) emission rates, respectively. We used the open source code BlackHawk to calculate the above photons radiated spectra by setting the total number of PBHs $N = 100$.}
	\label{fig:hwaking}
\end{figure}
For the case of critical collapse, $dn / dM$ at PBH formation epoch is given by $n(M)$ shown in \eqref{nM_precise} and \eqref{nM_appro}. Fig. \ref{fig:hwaking} plots the instantaneous emission rate of photons for PBHs with the horizon mass $10^{13}$ g for monochromatic (blue) and critical collapse mass functions (red) at the formation epoch, in which we set the total number density of PBHs with these two types of mass functions to be equal. As we expect, the primary photons instantaneous spectrum for the critical collapse spreads more wider than that of monochromatic {mass function}, and the total primary spectrum is the superposition of the emission from many (actually infinite) individual PBHs of distinct masses, with different peak strengths determined by differential mass function $n(M)$ in \eqref{nM_precise}. With the evaporation of PBHs with small masses, the low-mass tail of $n(M)$ would be deformed and the associated relative fraction of PBHs decreases. This deformation is significant for small horizon masses and for the long time evolution of PBHs (for more discussions, see Refs \cite{Carr:2016hva}).

\section{Nonthermal nucleosynthesis via Hawking radiation}
\label{sec:photodis}

\subsection{Electromagnetic cascade}
The PBHs would emit all the elementary particles (stable and unstable) and the final components which consist of neutrinos, photons, electrons, positrons, (anti-) protons, (anti-) neutrons and tiny fraction of gravitons\cite{MacGibbon:1991tj}. In this study, we are mainly interested in the spectrum of non-thermal photons at low temperature, which cannot be described as the power-law spectrum. Therefore, we focus on PBHs with horizon masses between $10^{12}$ g and $10^{13}$ g lifetimes of which are between $10^{8}$ s and $10^{12}$ s. In such a late epoch, neutrinos have already decoupled and cannot contribute to the nonthermal nucleosynthesis. For neutrons (anti-neutrons), they quickly decay to protons (anti-protons). Therefore only the hadronic shower from protons (anti-protons) can trigger the hadrodissociation process to destroy nucleus. \textbf{In the case of large branching ratio} (i.e., the main decay channel is hadron production), previous studies \cite{Kawasaki:2000qr,Kawasaki:2004yh,2005PhRvD..71h3502K,2006PhRvD..74j3509J} have concluded that the hadrodissociation can dramatically destroy the $^4$He nuclei via $^4$He($p$,$X$)$^3$A, so that the $^3$He would be overproduced. Recently, Ref. \cite{Kawasaki:2017bqm} also pointed out that if processes of anti-protons and anti-neutrons are taken into account, the constraints on DM abundance from $^3$He becomes stronger due to the hadrodissociation from these anti-particles, while the constraints become weaker if one includes the inelastic scattering between protons and neutrons. This is  because energetic neutrons change into protons and stop without causing hadrodissociation. For PBHs within horizon mass between $10^{12}$ g and $10^{13}$ g, proton and neutron emissions contribute to $<20\%$ of the final production \cite{MacGibbon:1991tj}. \textbf{Therefore we mainly focus on the} EM cascade process and the following photodisintegration in this work. The impact of hadronic emission from PBHs is discussed briefly in Sec. \ref{sec:constrn} and Appendix \ref{apendB}.

Once the photons are emitted from PBHs via Hawking radiation, they could interact with the cosmic background radiation (including photons, electrons and positrons). The injections of energetic photons would initiate the EM cascade showers which are mainly made of energetic photons and pairs of electron and positron \cite{1985NuPhB.259..175E,1995ApJ...452..506K,1995PhRvD..51.4134P}. {At different cosmic temperature $T$, the photons would form a quasi-static equilibrium spectrum $f_\gamma(E_\gamma;T)$ very quickly, and $f_\gamma(E_\gamma;T)$ is given by the solution of the steady Boltzmann equation}:
\be
\frac{\partial f_\gamma(E_\gamma;T)}{\partial t}=\frac{\partial}{\partial E}\Big[b_{exp}(E;T)f_\gamma(E_\gamma;T)\Big]-\Gamma_\gamma(E_\gamma;T)f_\gamma(E_\gamma;T)+S(E_\gamma;T)\equiv0;
\ee
Here, $S(E_\gamma;T)$ is the source term that depends on the emitted photons energies. We include the cosmic expansion effect in the Boltzmann equation with $b_{exp}(E;T)$ the energy loss rate via the cosmic expansion. It is given by \cite{Montmerle:1977vc}
\be
b_{exp}(E;T)=-H(T)E,
\ee
where $H(T)$ is the Hubble expansion rate. Excepting a special case that the energy distribution is a non-smooth function of energy, the energy derivative term roughly scales as
\be
\frac{\partial}{\partial E}\Big[b_{exp}(E,T)f_\gamma(E_\gamma;T)\Big]\sim -H(T)f_\gamma(E_\gamma;T) ,
\ee
where $\Gamma_\gamma(E_\gamma;T)$ describes the degradation rate of zeroth generation nonthermal photons given by the summation of 3 slow energy-loss processes of photons \cite{Berezinskii,Maximon,1990ApJ...349..415S}:
\be\label{Gamma}
\l\{
\begin{aligned}
    & \Gamma_{\rm CS}=n_e\sigma_{\rm CS}, \ \ \ \ &&(\gamma + e^\pm_{bg} \to \gamma + e^\pm),
   \\& \Gamma_{\rm NP}=n_N\sigma_{\rm PC}, \ \ \ \ &&(\gamma + N_{bg} \to e^{\pm} + N),
    \\& \Gamma_{\rm PP}=-\frac{1946}{50625}\alpha^2r_e^2m_e^{-6}E^3_\gamma\int_0^\infty \bar{\epsilon}^3\bar{f}_\gamma(\bar{\epsilon}) d\bar{\epsilon}, \ \ \ \ &&(\gamma + \gamma_{bg} \to \gamma + \gamma) .
\end{aligned}  \r.
\ee
Here, $\sigma_{\rm CS}$ is the cross section of Compton scattering given by
\be
\sigma_{\rm CS}
=
2\pi r_e^2 \frac{1}{x} \Big[\Big(1-\frac{4}{x}-\frac{8}{x^2}\Big) \ln(1+x)+\frac{1}{2} + \frac{8}{x} - \frac{1}{2(1+x)^2 }\Big] ,
\ee
where $r_e$ is the classical radius of electron and $x\equiv 2E_\gamma/m_e$ where $m_e$ is the electron mass. The Bethe–Heitler pair creation cross section $\sigma_{\rm PC}$ for low-energy photons is
\be
\sigma_{\rm PC}(E_\gamma)|_{k<4}=\alpha r_e^2\frac{2\pi}{3}\Big(\frac{k-2}{k}\Big)^3\Big[1+\frac{1}{2}\rho+\frac{23}{40}\rho^2+\frac{11}{60}\rho^3+\frac{29}{960}\rho^4\Big],
\ee
where
\be
k \equiv \frac{E_\gamma}{m_e}, \ \ \rho \equiv \frac{2k-4}{k+2+2\sqrt{2k}} ,
\ee
and the $\alpha$ is the fine structure constant. For high-energy photons, the cross section is
\ba
\sigma_{\rm PC}(E_\gamma)|_{k>4}&=&
\alpha r_e^2\Big\{\frac{28}{9} \ln2k - \frac{218}{27} \nonumber\\
&+& \Big(\frac{2}{k}\Big)^2 \Big[\frac{2}{3}(\ln 2k)^3 -(\ln 2k)^2 +\Big(6 - \frac{\pi^2}{3}\Big) \ln2k + 2\zeta(3) +\frac{\pi^2}{6} - \frac{7}{2}\Big]\nonumber\\
&-&\Big(\frac{2}{k}\Big)^4 \Big(\frac{3}{16} \ln2k+\frac{1}{2}\Big)
-\Big(\frac{2}{k}\Big)^6 \Big(\frac{29}{2304} \ln2k-\frac{77}{13824}\Big)\Big\} .
\ea
For the Hawking-radiated photons at a cosmic temperature $T$, we define the initial spectrum as
\be
f_\text{Hawk}(E_\gamma;T)\equiv \frac{d\dot{N}_\text{tot}(E_\gamma;T)}{dE_\gamma} .
\ee
The RHS of the above definition is given by \eqref{Hawk_spectra} involving the time evolution of Hawking spectrum that is carried by $T$. Such {an} initial photons spectrum covers energy range from keV to GeV, for energetic emitted photons, the two fast scattering processes: pair production ($\gamma+\gamma_{bg}\to e^{\pm}$, where $\gamma_{bg}$ represents the background photons) and inverse Compton scattering ($e^{\pm}+\gamma_{bg} \to e^{\pm}+\gamma$) lead to a power-law source term $p_{\gamma,EC}(E_{\gamma0};E_\gamma;T)$ on time scales much shorter than the thermodynamical equilibration \cite{Berezinskii,1995ApJ...452..506K,1995PhRvD..51.4134P}:
\be\label{steadyspect}
p_{\gamma,EC}(E_{\gamma0};E_\gamma;T)
=
\l\{
\begin{aligned}
    &K \l( {  E_X \over E_\gamma  } \r)^{3/2}  \ \ &&\text{for}  \ \ E_\gamma<E_X,
   \\& K \l( {  E_X \over E_\gamma  } \r)^{2}  \ \ &&\text{for}  \ \ E_X<E_\gamma<E_C,
    \\& 0                                      \ \ &&\text{for}    \ \ E_\gamma > E_C,
\end{aligned}
\r.
\ee
 where $E_X \sim m_e^2/(80T)$ refers to the threshold energy of inverse Compton scattering, $E_C \sim m_e^2/(22T)$ is the threshold energy of pair production, and $T$ is the cosmic temperature. $K=E_{\gamma0}/E_X^2[2+ \ln(E_C/E_X)]$ is the normalization constant. Note that the spectrum has a cutoff because for photons with energy larger than $E_{C}$, they are quickly destroyed via electron-positron pair production. The above power-law spectrum is only valid for the energetic photon emission since these photons quickly constitutes the power-law spectrum $p_{\gamma,EC}$ in \eqref{steadyspect}, and the high-energy photons contribute to the final spectrum as
\be\label{pl_spec}
f_{pl}(E_\gamma;T)=\frac{1}{\Gamma_\gamma(E_\gamma;T)+H(T)}\cdot \int_{E_{th}}^\infty dE_\gamma'f_\text{Hawk}(E_\gamma';T)  p_{\gamma,EC}(E_\gamma';E_\gamma;T).
\ee
For low-energy emitted photons, Ref. \cite{2015PhRvL.114i1101P} showed that in the condition that $E_\gamma \leq 10 T_\text{keV}^{-1}$ MeV, the {sub-threshold} photons injection can not trigger the pair production and the final spectrum in this case is significantly altered with respect to \eqref{steadyspect}. The spectrum of low-energy photons is
\be\label{low_e_f}
f_{\rm low}(E_\gamma;T)=\frac{S_{\rm low}(E_\gamma;T)}{\Gamma_\gamma(E_\gamma;T)+H(T)} .
\ee
For these photons, $S_{\rm low}(E_\gamma;T)$ is initially given by the Hawking radiation spectrum after normalization:
\be\label{s_pou_ini}
S_{ini}(E_\gamma;T)
=
f_\text{Hawk}(E_\gamma;T)\cdot \theta(E_{th} - E_\gamma) .
\ee
where $\theta$ is the step function. However, in reality, not all scattered photons will be ``lost'' for low-energy photons injection. Low-energy photons can remain in the final state even after Compton scattering and the $\gamma \gamma$ scattering. Therefore, $S(E_\gamma;T)$ becomes
\be\label{s_pou}
S_{\rm low}(E_\gamma;T)
=
S_{ini}(E_\gamma;T) + \int^{E_{th}}_{E_\gamma}dx K_\gamma(E_\gamma,x;T)f_{\rm low}(x;T) ,
\ee
where $K_\gamma(E_\gamma,x;T)$ is the summation of the differential rate of Compton scattering and the $\gamma \gamma$ scattering \cite{1990ApJ...349..415S,Berezinskii}:
\be
\begin{aligned}
&K_\gamma(E_\gamma,E'_\gamma;T)=\frac{1112}{10125}\alpha^2 r_e^2 m_e^{-6}\cdot\frac{8\pi^4 T^6}{63}E'^2_\gamma\Big[1-\frac{E_\gamma}{E'_\gamma}+\Big(\frac{E_\gamma}{E'_\gamma}\Big)^2\Big]^2 \\
&\quad\quad\quad\quad\quad\quad\quad
\quad
+\pi r_e^2n_e\frac{m_e}{E'^2_\gamma}\Big[\frac{E'_\gamma}{E_\gamma}+\frac{E_\gamma}{E'_\gamma}+\Big(\frac{m_e}{E'_\gamma}-\frac{m_e}{E_\gamma}-1\Big)^2-1\Big] .
\end{aligned}
\ee
Then \eqref{low_e_f} can be rewritten as
\be\label{f_pou}
f_{\rm low}(E_\gamma;T)
=
\frac{S_{ini}(E_\gamma;T)+ \int^{E_{th}}_{E_\gamma}dx K_\gamma(E_\gamma,x;T)f_{\rm low}(x;T)}{\Gamma_\gamma(E_\gamma;T)+H(T)} ,
\ee
and the RHS of this equation also contains $f_{\rm low}(E_\gamma;T)$. Therefore, we apply the same numerical method to solve \eqref{f_pou} as Ref. \cite{2015PhRvD..91j3007P} by using an iterative method: the initial spectrum $f^{ini}_{\rm low}(E_\gamma;T)$ is first calculated directly by $S_{ini}(E_\gamma;T)/\Gamma_\gamma(E_\gamma;T)$; (i) this $f^{ini}_{\rm low}(E_\gamma;T)$ is plugged into \eqref{s_pou} to obtain the new ``effective'' source term $S_{\rm low}(E_\gamma;T)$; (ii) this new $S_{\rm low}(E_\gamma;T)$ is put into \eqref{f_pou} to obtain new ``effective'' spectrum, and step (i) and (ii) are repeated. The iteration stops after the error reach below few percent. Finally, the steady spectrum after the photons injection from PBHs are given by:
\be\label{f_pou2}
f_{fin}(E_\gamma;T)=f_{pl}(E_\gamma;T)+f_{\rm low}(E_\gamma;T) .
\ee
\begin{figure}[h]
\centering
\includegraphics[scale=0.35]{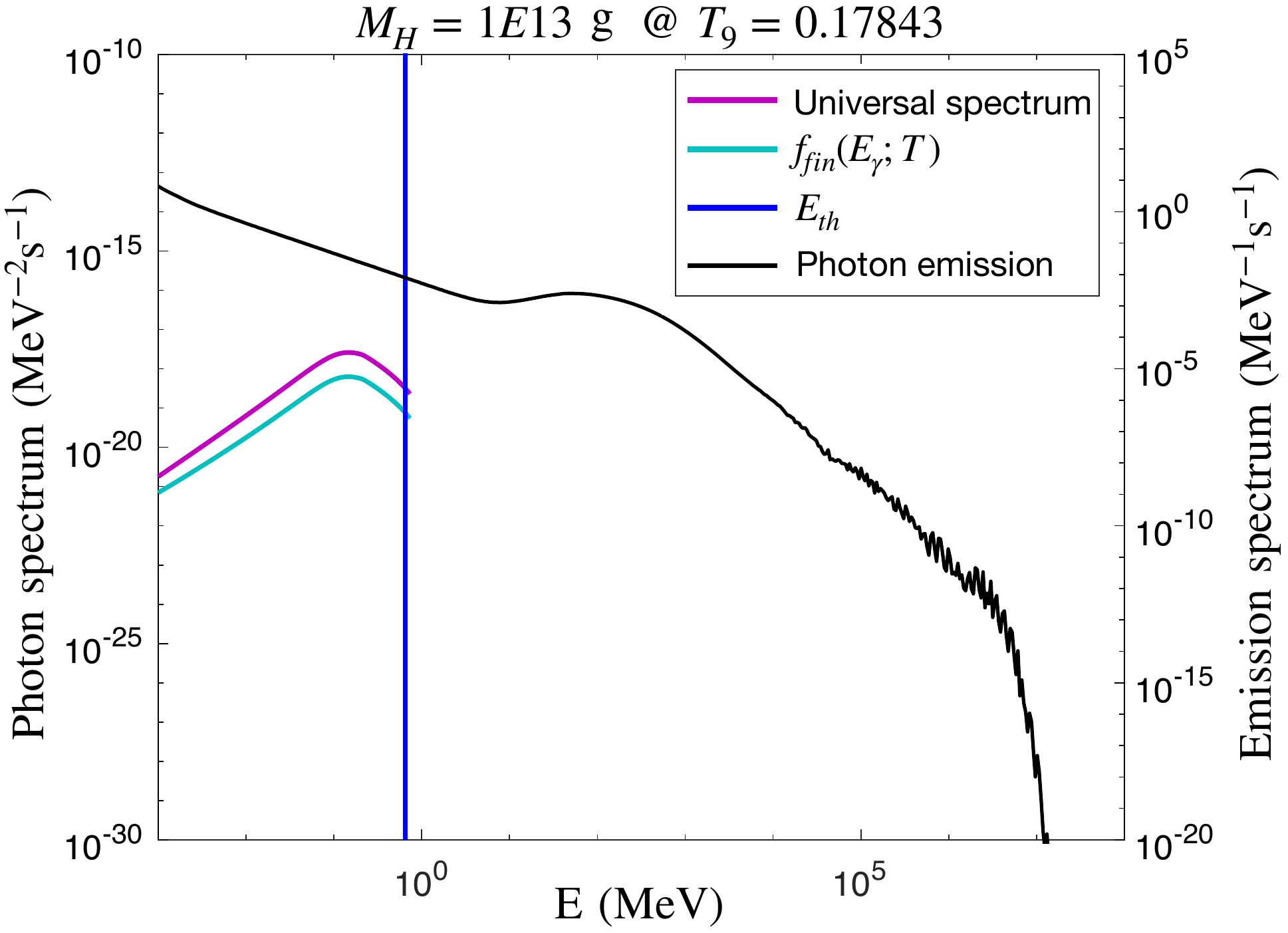}
\includegraphics[scale=0.35]{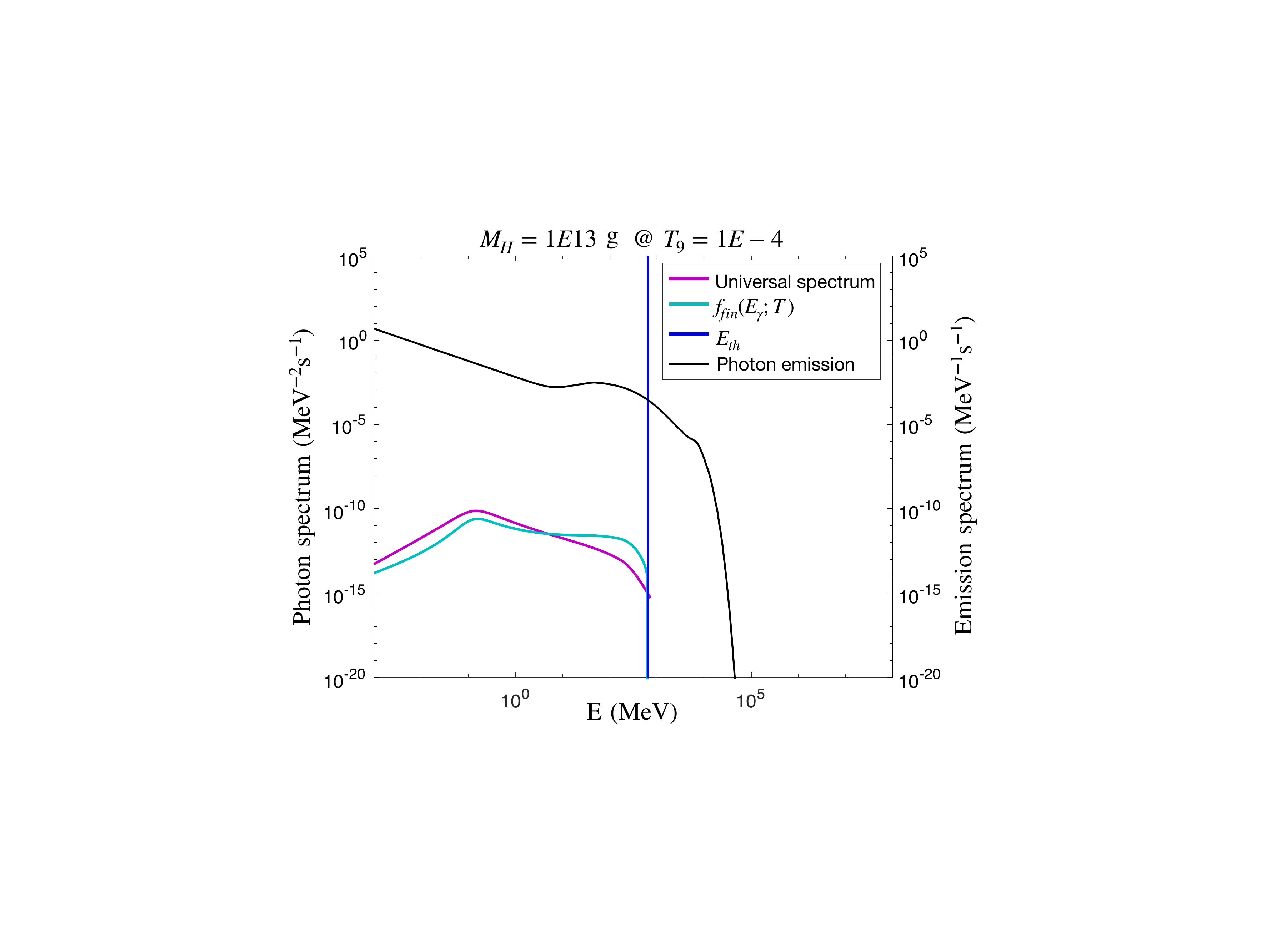}
\caption{The power-law spectrum and $f_{fin}(E_\gamma;T)$ as a function of $E$ for PBHs with a monochromatic mass function at various cosmic times. All the spectra are normalized against total number of emitted photons. Left panel corresponds to $10^{13}$ g PBH{s} at time when $T_9=0.1$ and  right panel corresponds to $T_9=10^{-4}$, respectively. Black curves are the total emission spectrum of nonthermal photons via Hawking radiation, purple curves are the power law spectrum and cyan curves refer to $f_{fin}(E_\gamma;T)$ derived in this work.}
\label{fig:spect_dirac}
\end{figure}
The high energy electrons and positrons emitted from PBH also participate in the EM cascade starting from the Compton scattering, and their energy finally converts to the nonthermal photons. We add such contribution from electrons and positrons to the amplitude of the final nonthermal photon spectrum. Fig. \ref{fig:spect_dirac} shows the comparison between the power-law spectrum and the spectrum $f_{fin}(E_\gamma;T)$ we derived for PBHs with a monochromatic mass function. Left panel is the spectrum at $T_9=0.1$ and right panel is at $T_9=10^{-4}$.
%For this figure, a monochromatic mass function has been adopted for PBHs.
On both panels, we show the total photon emission spectrum (primary photons + secondary photons + $e^\pm$) from PBHs with black solid curve (the amplitude read by right vertical axis). The blue vertical line is located at $E_{th}$, beyond which all the emitted photons lose their energies quickly due to the pair production and inverse Compton scattering and the spectrum is given by $f_{pl}(E_\gamma;T)$. As shown in this figure, even for the $10^{13}$ g PBH{s}, the low-energy photons account for a relatively large portion of the nonthermal spectra at low temperature. Therefore, at low temperature, e.g., $T_9=10^{-4}$ (right panel), the spectrum shows the slight enhancement in the energy range just below $E_{th}$ from the {power-law} spectrum.

\begin{figure}[h]
\centering
\includegraphics[scale=0.35]{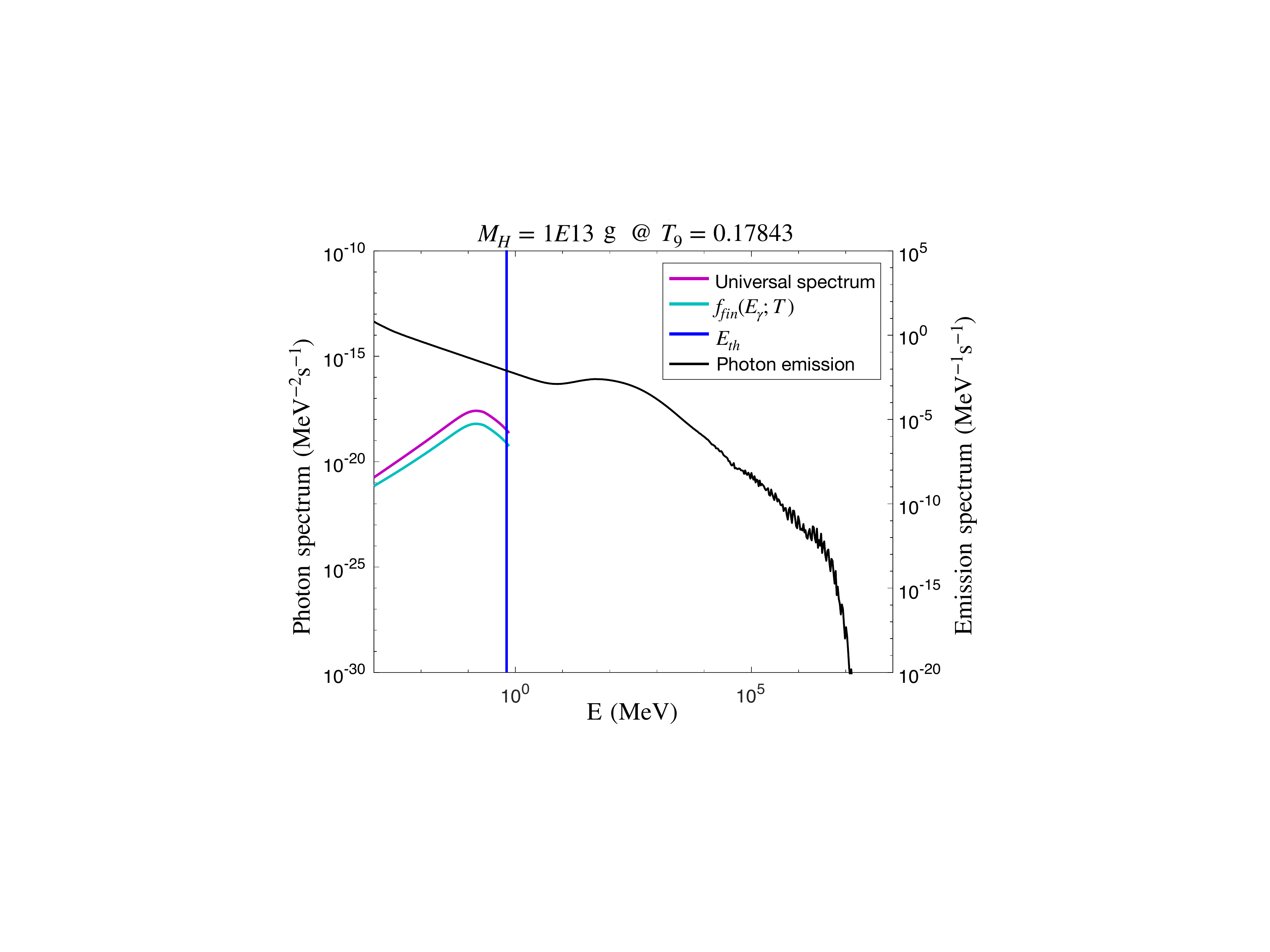}
\includegraphics[scale=0.35]{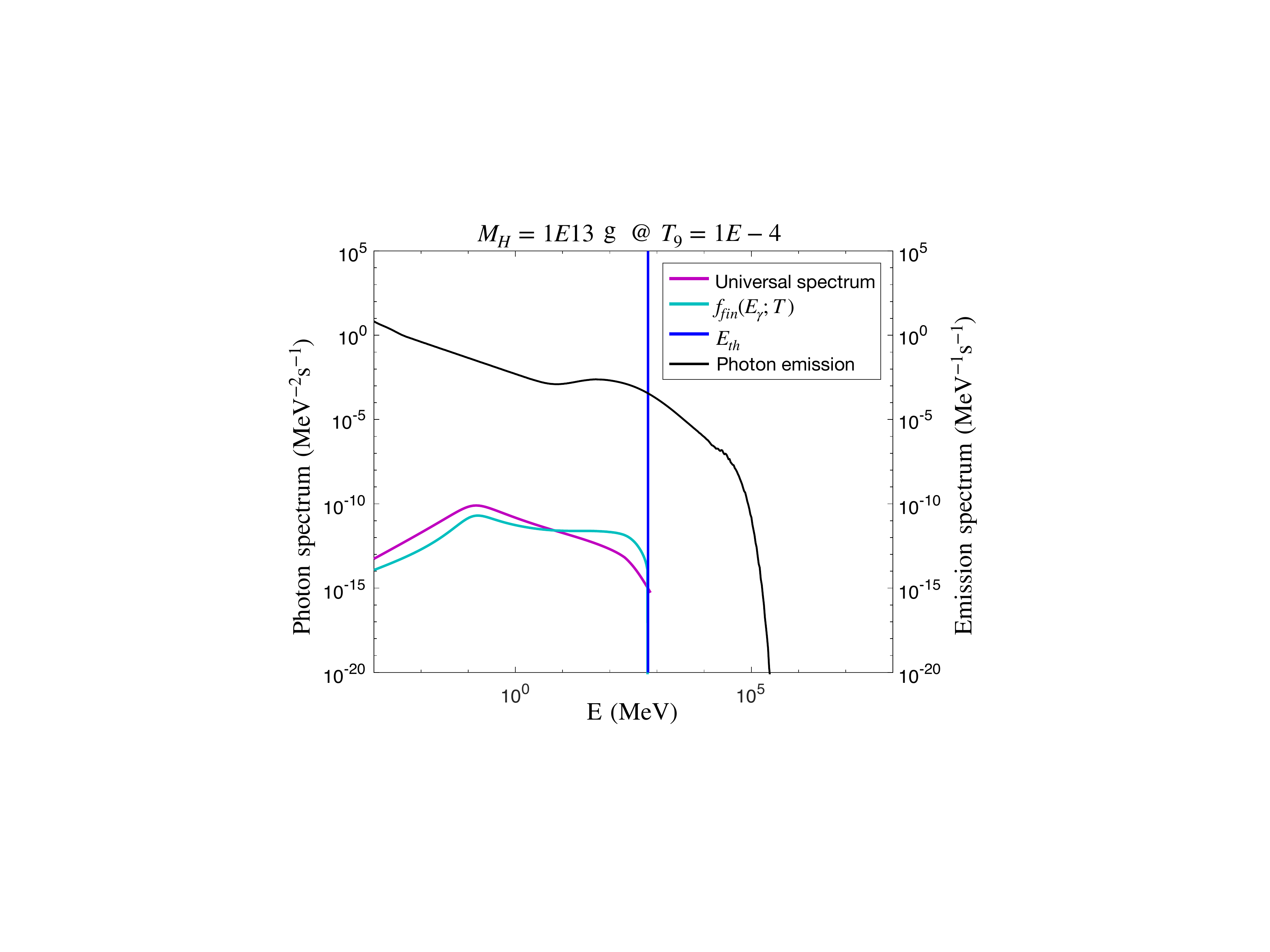}
\caption{The same plot as in Fig. \ref{fig:spect_dirac} but for the critical collapse model of PBHs.}
\label{fig:spect_cr}
\end{figure}
Fig. \ref{fig:spect_cr} shows the the same plot as Fig. \ref{fig:spect_dirac} but for the critical collapse model of PBH formation. The photon emission from PBHs with horizon mass $10^{13}$ g shows that almost the same Hawking radiation spectrum as the monochromatic mass function, this is because the secondary photon emission is dominant in such a low-mass range (see Sec. \ref{sec:Hawk_rad}), the final spectra $f_{fin}(E_\gamma;T)$ are the same for the critical collapse and the monochromatic mass function cases \footnote{However, one should notice that here the comparison is between normalized spectra, the critical collapse provides more high-energy photons, so the actual amplitude of spectra in critical collapse mass function  is still larger than monochromatic case.}

\subsection{Nonthermal nucleosynthesis}

The photodisintegration of nuclei can be triggered by photons with energies larger than the photodisintegration energy threshold. Several previous investigations \cite{1979MNRAS.188P..15L,1985NuPhB.259..175E,1988ApJ...330..545D,1992NuPhB.373..399E,1995ApJ...452..506K,Jedamzik:1999di,Ellis:2005ii,2006PhRvD..74j3509J,2006PhRvD..74b3526K,2009PhRvD..79l3513K,2013PhRvD..87h5045K,2014PhRvD..90h3519I} already studied the photons injection after $t>10^4$ s and the primordial nuclear abundances could be changed by such process{es}. We focus on nonthermal nucleosynthesis induced by PBH{s'} Hawking radiations in this work.

The time evolution of the nuclear abundances is governed by
\be\label{eq:time_evol}
\frac{dY_A}{dt}=\sum_P N_{AC}[P\gamma]_AY_T(T)-\sum_P [A\gamma]_PY_A(T) ,
\ee
where the first term on RHS represents production of nuclei $A$ via the reaction $\gamma+T\to A$, the second term is for destruction of nuclide $A$ via $\gamma + A \to P$ for any product nuclide $P$. $Y_i \equiv n_i/n_B$ is the mole fraction of a nuclear species $i$ with $n_i$ and $n_B$ number densities of nuclide $i$ and total baryon, respectively. $N_{AC}$ is the number of identical nuclear species in the final state: $N_{AC}=2$ when particles $A$ and $C$ are identical and $N_{AC}=1$ when they are not, for the case of two-body final state $A+C$. $[A\gamma]_P$ is the reaction rate per unit time for destroying the nuclei $A$ given by:
\be\label{eq:rate}
[A\gamma]_P(T)=\int^\infty_0 dE_\gamma f_{fin}(E_\gamma;T) \sigma_{\gamma+A\to P} ,
\ee
where $n_\text{PBH}(z)$ is the physical number density of PBHs as a function of the redshift $z$.
Secondary nonthermal reactions also occur if products of the primary photodisintegration reactions are energetic enough. The secondary nonthermal nuclear reactions affect the nuclear mole fractions \cite{2006PhRvD..74b3526K} as
\be\label{eq:time_evol_sec}
\frac{dY_S}{dt}=\sum Y_P Y_{P'}\frac{N_{AX_1}N_{SX_2}}{N_{AP'}}[P(A)P']_s- \rm (destruction~term) ,
\ee
 where the first terms on RHS describes the secondary productions of nuclei via the reaction sequence $P(\gamma,X_1)A(P',$
$X_2)S$, and the second term is for destruction. The reaction rate per unit time is given by
\be
\begin{aligned}\label{eq:secondary}
&[P(A)P']_s=\int^\infty_0 dE_A \frac{\sigma_{A+P'\to S}(E_A)\beta_A}{b_A(E_A)/n_b(z)} \\
&\quad\quad\quad\quad\quad \quad \times
\int^\infty_{\epsilon_A^{-1}}dE_\gamma f_{fin}(E_\gamma;T)\sigma_{\gamma+P\to A}\exp{\Big[-\int^{\epsilon_A(E_\gamma)}_{E_A}}dE''_A\frac{\Gamma_A(E''_A)}{b_A(E''_A)}\Big] .
\end{aligned}
\ee
The integration against $E_A$ represents the  total rate for the secondary process of primary product $A$ for production of nuclei $S$. The integration against $E_\gamma$ stands for the production of nuclei $A$  via the photodisintegration and its subsequent decay described with the exponential term. The quantity $\beta_A$ is the velocity of the primary product $A$, and $b_A = -dE/dt$ is the rate of energy loss of  $A$ during propagation through the background. The $\Gamma_A$ is the destruction rate of $A$ until it is thermalized. The energy-loss rate is usually much faster than any destruction rate (cf. Ref. \cite{2006PhRvD..74b3526K}), and the destruction can be ignored from the abundance evolution of the primary products for stable nuclei. Here, only for unstable nuclei, finite values of $\beta$-decay rates are inputs in $\Gamma_A$. The quantity $\epsilon_A(E_\gamma)$ is the energy of the primary product $A$ produced by the photodisintegration process $\gamma+P \to A$. Correspondingly, $\epsilon_A^{-1}(E_A)$ is the energy of the nonthermal photons which  produces primary product $A$ with energy $E_A$.
For secondary nonthermal reactions, we only consider the $^6$Li produced via secondary processes $^4$He($t$,$n$)$^6$Li and $^4$He($^3$He,$p$)$^6$Li \cite{Jedamzik:1999di,2003PhRvD..67j3521C,2005PhRvD..71h3502K,2006PhRvD..74b3526K}. $^6$Li nuclei can be produced by the secondary reactions at levels much higher than that of the Standard BBN.

The nonthermally produced $^6$Li can be destroyed via the nuclear reaction $^6$Li($p$,$^3$He)$^4$He before it is thermalized. The secondary $^6$Li production rate taking into account its tertiary destruction is given by
\be\label{eq:secondary_li}
\begin{aligned}
  &[P(A)P']_s=
      \sum_{^3A=t,^3\mathrm{He}}
        \int^{(E_C-E_{\gamma,th})/4}_{E_{p,th}} dE_3 \frac{\sigma_{^3 A+\alpha}(E_3)\beta_3}{b_3(E_3)/n_b(z)}P_{^6\text{Li}\to ^6\text{Li}}(E^{in}_6) \\
&\quad\quad\quad\quad\quad \quad\times
\int^{E_C}_{4E_3+E_{\gamma,th}}dE_\gamma f_{fin}(E_\gamma;T)\sigma_{\gamma+\alpha\to ^3A}\exp{\Big[-\int^{\epsilon_3(E_\gamma)}_{E_3}}dE''_3\frac{\Gamma_3(E''_3)}{b_3(E''_3)}\Big] ,
\end{aligned}
\ee
where the survival probability of nonthermal $^6$Li produced via the secondary process at initial kinetic energy $E^{in}_6$ is given by
\be\label{eq:Li6_decay}
P_{^6\text{Li}\to ^6\text{Li}}(E^{in}_6)= \exp\Big[-\int_{E^{th}_{^6\text{Li}+\text{p}}}^{E_6^{in}}dE_6 b_6^{-1} n_p\sigma_{^6\text{Li}+\text{p}}\beta_6\Big] ,
\ee
where the subscript 6 refers to $^6$Li, {$E^{th}$} is the threshold energy for the $^6$Li(p,$^3$He)$^4$He reaction and $\sigma_{^6\rm Li+p}$ is the  cross section \cite{Abramovich}. Nuclei with $A=3$ are mainly produced from the two primary reactions, i.e., $^4$He($\gamma$,$p$)$^3$H and $^4$He($\gamma$,$n$)$^3$He, in which the energy of product $^6$Li is given by
\be
\epsilon_{^6\text{Li}}(E_3)=\frac{1}{7}(m_\alpha+E_3-\gamma E^{th}_{^6\text{Li}+\text{p}}) ,
\ee
where $\gamma=\sqrt{1-\beta^2}$, and $\beta=m_3\beta_3\gamma_3/(m_\alpha+E_3)$, $E_3$, $m_3$, $\beta_3$ and $\gamma_3$ are the kinetic energy, mass, velocity and the Lorentz factor of the $A=3$ nuclide, $m_\alpha$ is the mass of $^4$He.

\begin{table}%[H] add [H] placement to break table across pages
	\centering
  \caption{Included primary photodisintegration reactions}
  \label{tab1}
    \begin{tabular}{c c c}
    \hline\hline
 No. & Reaction & Threshold  (MeV)\\

     \hline
  1&  $^2$H($\gamma$,$n$)$^1$H   &2.2246  \\
      \hline
 2       &      $^3$H($\gamma$,$n$)$^2$H  &6.2572  \\
 3       &      $^3$H($\gamma$,$2n$)$^1$H  &8.4818  \\
      \hline
 4       &     $^3$He($\gamma$,$p$)$^2$H  &5.4934  \\
 5       &     $^3$He($\gamma$,$np$)$^1$H  &7.7180  \\
 6       &      $^4$He($\gamma$,$p$)$^3$H  &19.8138  \\
      \hline
 7       &      $^4$He($\gamma$,$n$)$^3$He  &20.5776  \\
 8       &      $^4$He($\gamma$,$d$)$^2$H  &23.8465  \\
 9       &     $^4$He($\gamma$,$np$)$^2$H  &26.0710  \\
      \hline
 10       &     $^6$Li($\gamma$,$np$)$^4$He &3.6982  \\
 11      &     $^6$Li($\gamma$,$X$)$^3$A  &15.7940  \\
      \hline
 12       &      $^7$Li($\gamma$,$t$)$^4$He  &2.4675  \\
 13       &      $^7$Li($\gamma$,$n$)$^6$Li  &7.2511  \\
 14       &      $^7$Li($\gamma$,$2np$)$^4$He  &10.9493  \\
      \hline
 15       &     $^7$Be($\gamma$,$^3$He)$^4$He  &1.5869  \\
 16       &     $^7$Be($\gamma$,$p$)$^6$Li  &5.6067  \\
 17       &      $^7$Be($\gamma$,$2pn$)$^4$He &9.3049 \\
	\hline\hline
    \end{tabular}
\end{table}

In this study, we take into account the primary photodisintegration reactions listed in Table \ref{tab1}. The photodisintegration threshold energies are based on atomic mass data \cite{Wang2017} and electron binding energies \cite{Huang:1976jyl}. The cross section for these reactions are taken from Ref. \cite{2003PhRvD..67j3521C} and updated for the reactions $^4$He($\gamma$,$p$)$^3$H and $^4$He($\gamma$,$n$)$^3$He \cite{2009PhRvD..79l3513K}, $^7$Be($\gamma$,$p$)$^6$Li \cite{He:2013ica}, and $^7$Li($\gamma$,$t$)$^4$He \cite{2014PhRvD..90h3519I}. Among those reactions, $^7$Be and $^7$Li photodisintegration cross sections were corrected in Ref. \cite{2014PhRvD..90h3519I}. However, the cross section of $^7$Be photodisintegration \cite{2014PhRvD..90h3519I} includes an error, i.e., the contribution of the first excited state of $^7$Be has been included in using the detailed balance relation between the forward and reverse reactions. We should take into account only the ground state to derive the rate of the reverse reaction, i.e., $^7$Be($\gamma$,$t$)$^4$He.
The cutoff energy of nonthermal photons spectrum, i.e., $E_C$ in \eqref{steadyspect}, becomes high enough that photodisintegrations can be operative only long after the BBN. In such a low temperatures, there are very low abundance of the first excited states of $^7$Be, and the effect from any excited state is always negligible due to an exponential Bolzmann suppression factor.
Instead, excited states in the final state need to be considered in general. However, since the first excited state of $^4$He has $\sim 20$ MeV excitation energy, branching ratios for excited states are small as considering the softness of nonthermal photons spectrum. Therefore, only the cross section for transition to the ground state of $^4$He is used safely.
The correct cross section is given by
\be
\l\{
\begin{aligned}
    & \sigma_{^7\text{Be}+\gamma}=\frac{801 {\rm mb}}{E_\gamma^2}\exp{\Big(-\frac{5.19}{E_{CM}^{1/2}}\Big)}\frac{Q}{E_{CM}+Q}\Big[s_{00}\Big(1+a_0E_{CM}\Big)^2
    \\ &\quad\quad\quad\quad \quad
    +s_{20}\Big(1+4\pi^2\frac{E_{CM}}{E_G}\Big)\Big(1+16\pi^2\frac{E_{CM}}{E_G}\Big)\Big]\ \  &&\text{for} \ E_\gamma>Q+1.2\ \rm MeV,
    \\
	\\& \sigma_{^7\text{Be}+\gamma}=\frac{128 {\rm mb}}{E_\gamma^2}\exp{\Big(-\frac{5.19}{E_{CM}^{1/2}}\Big)}\ \  &&\text{for} \ E_\gamma<Q+1.2 \rm MeV.
\end{aligned}
\r. ~
\ee
Here, $Q=1.5866$ MeV is the binding energy of $^7$Be with respect to the separation channel of $^3$He+$^4$He, $E_{CM}=E_\gamma-Q$ is the center of mass (CM) energy.  $E_G=2\mu(
\pi Z_AZ_B\alpha)^2$ is the Gamow energy with $Z_i$ the proton number of species $i$. $s_{00}=0.406$, $s_{20}=0.007$ and $a_0=-0.207$.
%The cross section is shown in Fig. \ref{fig:cross_sec}.

%\begin{figure}[h!]

%\centering
%\includegraphics[scale=0.5]{cross_section.eps}
%\caption{Cross sections of reactions $^7$Be($\gamma,\alpha$)$^3$He as a function of the photon energy. Dashed line is calculated from \cite{2003PhRvD..67j3521C}, however, for the low temperature, only the ground state should be taken into account as shown in the solid line.}
%\label{fig:cross_sec}
%\end{figure}

\section{PBHs impact on primordial abundances}
\label{sec:constrn}

\subsection{An updated constraints on monochromatic mass function}
In this study, we use a nonthermal BBN nuclear reaction network code based on Refs. \cite{Kawano1992,Smith:1992yy} and have updated the reaction rates of nuclei with mass numbers $A\leq10$ using the JINA REACLIB Database \cite{Descov2004,Cyburt2010,Coc2015}. The neutron lifetime is $880.2 \pm 1.0$ s, corresponds to the central value of the Particle Data Group \cite{Patrignani}. The baryon-to-photon ratio $\eta$ is taken to be {$\eta_{10}\equiv\eta/10^{-10}=(6.16 \pm 0.02)$ corresponding to the baryon density $\Omega_b h^2=0.0224\pm0.0001$ in the standard $\rm \Lambda CDM$ model determined from Planck analysis of Ref. \cite{Aghanim:2018eyx}}. For the nonthermal photodisintegration reaction rates, we take into account the primary photodisintegration reactions listed in Table. \ref{tab1} and also the secondary reactions of $^4$He($\alpha$,$N$)$^3A$($\alpha$,$N$)$^6$Li (Sec. \ref{sec:photodis}). The cross sections of those nonthermal photodisintegration reaction rates are taken from Refs. \cite{2003PhRvD..67j3521C,2009PhRvD..79l3513K,He:2013ica,2014PhRvD..90h3519I}. {The effect of the conversion of primordial $^7$Be to $^7$Li via the electron capture decay is also taken into account in the current calculation. Once the cosmological recombination of $^7$Be$^{4+}$ ions occurs at $z\sim 3 \times 10^4$, the $^7$Be nucleus instantaneously captures the orbital electron with the half-life $T_{1/2} =106$ d, and is converted to $^7$Li \cite{Khatri:2010ed}.}
At the end of PBH’s life, the evaporation rate increases dramatically. Then for this time period, we check the convergence of our calculation. Namely, we change the time step $dt$ down to smaller values and we obtain the same numerical result with less than $0.1\%$ difference.
The observational constraints on $^4$He are taken from measurements for metal-poor extragalactic H II regions, $Y_p = 0.2449 \pm 0.0040$ \cite{Aver2015} where $Y_p$ is the mass fraction of $^4$He. The deuterium can only be destroyed during stellar evolution long after the production at BBN.  The D/H abundance ratio is constrained with observations of metal-poor Lyman-$\alpha$ absorption systems towards quasi-stellar objects. We use the weighted mean value of D/H= ($2.527 \pm 0.030) \times 10^{-5}$ \cite{Cooke2018}. Contrary to the case of $^4$He and D, the time evolution of the $^3$He abundance to the present epoch is not simple \cite{VangioniFlam:2002sa}. The $^3$He abundance can change via the Galactic chemical evolution although the net effect of Galactic chemical evolution is not constrained sufficiently since stars can both destroy and synthesize $^3$He. However, it is not expected that the $^3$He abundance has decreased significantly over galactic history as this would require that a large fraction of Galactic baryonic material have participated in star formations and experienced $^3$He destruction, while the present interstellar deuterium abundance limits the amount of astration to not more than about a factor of two. Refs. \cite{Carr:2009jm, Carr:2020gox} adopted the upper limit on the $^3$He abundance ratio as $^3$He/D<1.37 \cite{2003SSRv..106....3G}. However, in this study, we use a more stringent $^3$He constraint as $^3$He/H $<1.5 \times 10^{-5}$, which is the $2\sigma$ upper limit among several Galactic H II regions using the $8.665$ GHz hyperfine transition of $^3$He$^+$ ion \cite{Bania:2002yj}.
\begin{figure}[h]
\centering
\includegraphics[width=3.5in]{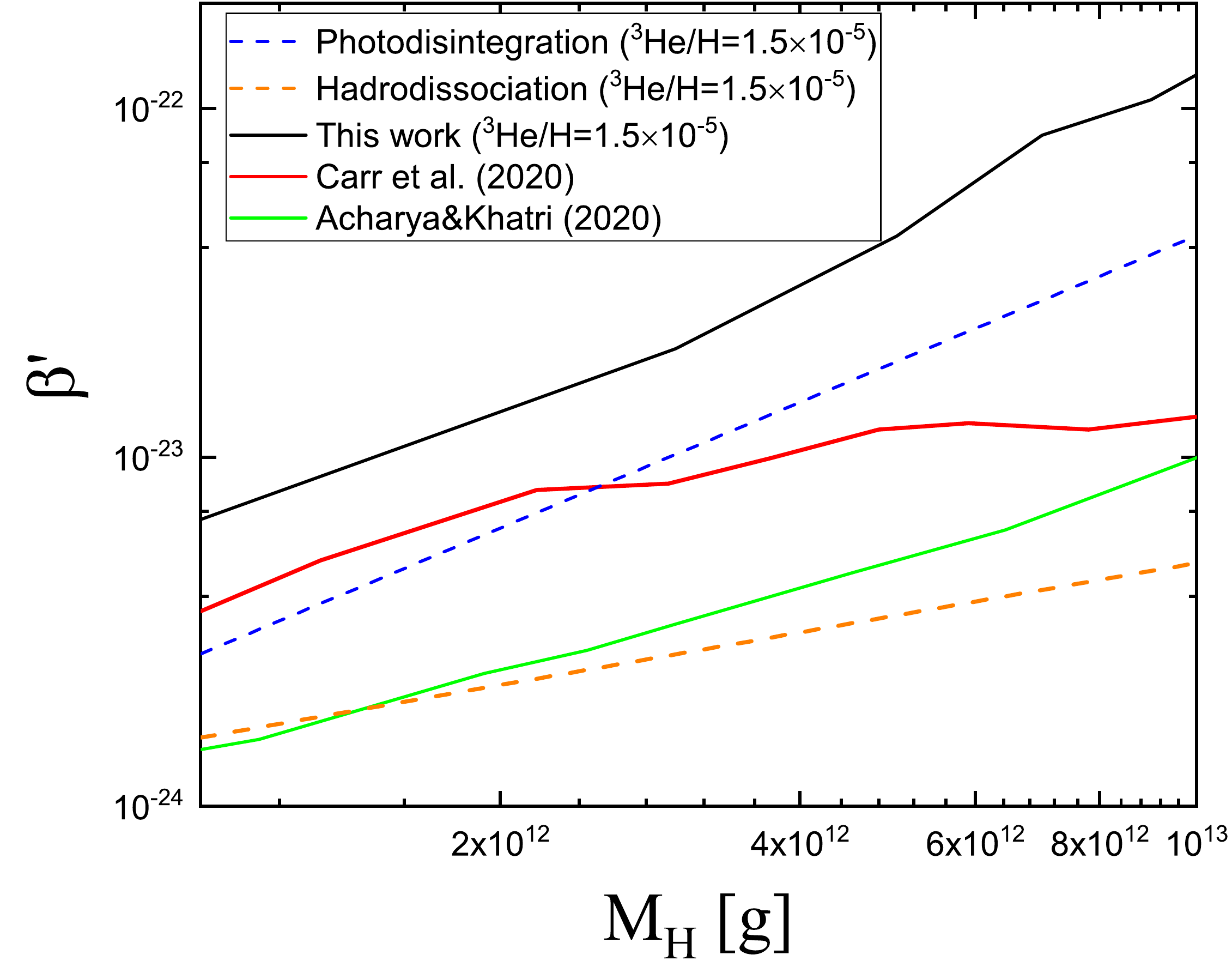}
\caption{The upper bounds on $\beta'(M_H)$ for the monochromatic mass function in the horizon mass range $10^{12}-10^{13}$ g. The dashed curves are the analytical bound based on the constraint $^3$He/H$<1.5\times10^{-5}$ measured in the present Galaxy. The blue dashed curve is the strongest possible bound obtained for 100 \% EM energy injection (all PBH mass turns into photons, see Appendix \ref{apendA} for details), and the orange dashed curve is the strongest possible bound for 100 \% hadronic energy injection (all PBH mass turns into protons, see Appendix \ref{apendB} for details).}
\label{dirac_cons}
\end{figure}
Fig. \ref{dirac_cons} shows our updated constraints on the monochromatic mass function of PBHs within mass range $10^{12}-10^{13}$ g. The black solid curve shows the constraint on $\beta'$ by using the observational limit $^3$He/H $<1.5 \times 10^{-5}$. Compared with the constraint in Ref. \cite{Carr:2009jm} (the red solid curve), we found one order of magnitude difference between our constraints and theirs. Considering that they use the more conservative constraint $^3$He/D<1.37, the constraint derived in this study is much weaker. One might think that since the hadrodissociation processes are not included, it is natural that our constraint is weaker. However, we derived an analytical expression of the $^3$He constraint on $\beta'$, including both photodisintegration and hadrodissociation, and found that the hadrodissociation effect is not so large that it accounts for this difference (see following discussion and Appendix \ref{apendB} for details). In addition, we note that although our treatment of PBHs radiation is basically similar to that of Ref. \cite{2020JCAP...06..018A} and uses the same observational $^3$He abundance ($^3$He/H $<1.5 \times 10^{-5}$), they also obtain a stronger constraint (green solid line) than ours. Thus, both of Refs. \cite{Carr:2009jm,2020JCAP...06..018A} derived severer constraints on PBHs than our constraint.

For photodisintegration, we assume: 1). All the PBH mass turns into nonthermal photons during its evaporation; 2). All the nonthermal photons carry the same energy with typical value $E_\gamma^{\rm typ}=30\ \rm MeV$, which corresponds to the peak of the photon disintegration reaction cross section for $^3$A production, i.e., $\sigma_{\rm dis} (E_\gamma^{\rm typ})=1 \rm mb$. Thus, this estimation should give the strongest $^3$He constraint on PBH abundances from photodisintegration. The detailed derivation is showed in Appendix. \ref{apendA}, the analytical estimation is given by
\be
\begin{aligned}
\label{estimate_beta}
\Delta \frac{^3{\rm He}}{\rm H}
=&
c_{\rm eff}(T) \frac{\Delta n_\gamma^{\rm nt}} {n_{\rm H}}(E_{\rm typ}) P(E_{\rm typ})
\\\simeq&
2.7 \times 10^8 c_{\rm eff}(T)
\l( \frac{E_\gamma^{\rm typ}}{30~{\rm MeV}} \r)^{-1} \l( { \eta \over 6 \times 10^{-10} } \r)^{-1} \l( { X_p \over 0.75 } \r)^{-1}  \l( {M \over M_\odot} \r)^{-1/2} \beta' ,
\end{aligned}
\ee
where $c_{\rm eff}(T)$ is the efficiency factor of photodisintegration reactions (see Fig. \ref{fig1} in Appendix), $\Delta n_\gamma^{\rm nt}/n_{\rm H}$ is the number density of nonthermal photons emitted from PBH evaporation, normalized to the Hydrogen number density, $X_p$ is the proton fraction and $P$ is the probability of nonthermal photons to react with background $^4$He via $^4$He($\gamma$,$N$)$^3A$ reactions. For the PBHs with mass $M=10^{13}$ g, it evaporated at $\tau=4.587 \times 10^{11}$ s, and this cosmic time corresponds to $T=1.70$ eV, $E_X=1.92$ GeV, and $E_C=7.00$ GeV. At this temperature, the efficiency factor is $c_{\rm eff} \sim 3 \times 10^{-2}$. Therefore, if we set $\Delta( ^3$He/H)$=2.42\times10^{-5}$, which corresponds to $^3$He/D $<1.37$, the upper bound on $\beta'(10^{13} \text{g})$ can be calculated by using \eqref{betaprime} and \eqref{fpbh}:
$\beta'(10^{13} \text{g})
\simeq
2.09 \times 10^{-22}$.

For hadrodissociation process, we make a similar estimation (see details in Appendix \ref{apendB}), and the result is given by
\ba
\Delta \frac{^3{\rm He}}{\rm H} &=& \frac{\Delta n_p^{\rm nt}}{n_{\rm H}}P_p  \nonumber \\
&\simeq& 7.1 \times 10^7
  \left( \frac{\langle E_p \rangle} {2.24~{\rm GeV}} \right)^{-1}
  \left( \frac{P_p}{0.07} \right)\nonumber\\
&\times&  \left( \frac{\eta}{6 \times 10^{-10}} \right)^{-1}
  \left( \frac{X_p}{0.75} \right)^{-1} \l( {M \over M_\odot} \r)^{-1/2} \beta',
\ea
where $P_p$ is the probability of nonthermal protons to destroy background $^4$He and produce $^3$He, and $E_p$ is the averaged energy of proton. The blue and orange dashed curves on Fig. \ref{dirac_cons} are the estimated constraints by photodisintegration and hadrodissociation processes, respectively. Both lines are based on the presumption that the mass of PBH turns into energies of the corresponding particles by 100 \%. Therefore, these estimated bounds should be the conceivably strongest $^3$He constraint on PBHs. Notice that in the realistic PBH evaporation for this mass range, the branching ratio of proton energy normalized to the total available mass of PBH is less than $20\%$. Therefore even considering the hadrodissociation process in our calculation, our result (black solid line) could only move down utmost by half an order of magnitude, which still cannot explain the present inconsistency. Moreover, although Ref. \cite{Carr:2009jm} used a relatively loose observational constraint on $^3$He, if the same value as $^3$He/H$<1.5\times10^{-5}$ is applied to their result, the red solid line would become comparable with the orange dashed curve, which implies nearly a $100\%$ hadronic energy injection from the PBH in our estimate. Also, the result from Ref. \cite{2020JCAP...06..018A} exceeds the blue dashed line, which should be the stringent bound for the case that only photodisintegration is included.
% The possible reasons of such discrepancy are: 1) in Ref. \cite{Carr:2020gox}, for energies lower than the GeV-scale, the secondary emission and decay rates are obtained by using the approximation in Ref. \cite{MacGibbon:1990zk}, which overestimate the magnitude of the emission spectrum for 2-3 times; 2) The time evolution of a single PBH we calculated did not assume a fixed value of $\phi(M)$ in \eqref{massloss_rate}, if this value is fixed for some approximation, then the mass loss rate of the PBH will be enhanced for more than one order of magnitude when time approaches to the end of its lifetime.

\subsection{The $^3$He constraints for critical collapse mass function}

It is well understood that it is non-trivial to extend the constraints for the monochromatic mass function to the extended case, i.e., one cannot just simply compare an extended mass function with the monochromatic form using the same constraints as shown in Fig. \ref{fig:beta}, since the form of constraints in the extended case itself is dependent on the PBH mass function \cite{Carr:2017jsz}. Several approaches to calculate the non-monochromatic constraints are suggested. One is to break each constraint up into narrow mass bins \cite{Carr:2016drx} which is a complicated procedure and has been criticized by Ref. \cite{Green:2016xgy}; Ref. \cite{Bellomo:2017zsr} introduced an equivalent mass for each specific extended mass function, in order to convert constraints on the monochromatic mass case into those on non-monochromatic cases, while Ref. \cite{Carr:2017jsz} proposed a general approach to place bounds on the parameters of mass functions, and also modify the constraints themselves. However, the previous studies mainly focus on the lognormal and power-law types of mass functions, and argue that the critical collapse mass function is equivalent to the lognormal one with small mass variance $\sigma = 0.26$, which is relatively narrow and even the monochromatic form provides a good fit \cite{Carr:2017jsz}. This argument is also consistent with the previous conclusion that the horizon-mass approximation is reasonably good for the critical collapse mass function, i.e., only a small fraction of PBHs is associated with the low-mass tail. For an illustration, one can also simply check that the equivalent mass of the critical collapse mass function (defined in Ref. \cite{Bellomo:2017zsr}) for EGB constraints is close to the horizon mass. We also notice that several studies reported the potential observational differences between the critical collapse and the monochromatic mass functions in the corresponding DM density \cite{Yokoyama:1998xd}, the spectral index of primordial power spectrum and the spectra of diffuse $\gamma$-ray \cite{Kribs:1999bs,Bugaev:2008gw} and neutrino \cite{Bugaev:2008gw} backgrounds, and gravitational wave \cite{Bugaev:2009kq}. However, few studies give the explicit observational constraints on PBH initial mass spectrum $\beta(M_H)$ for the critical collapse mass function.

Nevertheless, our following results will show that even though the relative fraction of PBHs within the low-mass tail of the critical collapse mass function is small, this tail \eqref{nM_appro} would certainly have an impact on light elemental abundances produced during BBN. This is because the stronger Hawking radiation than in the monochromatic mass function case emerges in the presence of the low-mass tail of critical collapse mass function. Hence, the usual BBN constraints on PBHs should be considered more carefully for the critical collapse model. In principle, one can apply the unified method presented in Ref. \cite{Carr:2017jsz} to calculate the constraints on mass function and alter the constraints themselves for PBHs which already evaporated by now, however, the constraints on monochromatic PBHs from the light elements are not simple functions of PBH mass. We therefore adopt a simple and direct approach to obtain the constraints on the critical collapse model. For each initial horizon mass $M_H$, we derive the Hawking radiation spectra \eqref{Hawk_spectra} from the critical collapse mass function \eqref{nM_appro} by the BlackHawk code, and calculate the nonthermal reaction rates of photodisintegration triggered by the Hawking radiation of PBHs as \eqref{eq:rate} and \eqref{eq:secondary_li} in Sec. \ref{sec:photodis}. Finally, we include those photodisintegration reaction rates into BBN calculation. By comparing the theoretical prediction and observations of light elemental abundances, we obtain the constraints on the normalization constant $A(M_H)$ in the mass function \eqref{nM_appro} of PBHs and those on $\beta_c(M_H)$ using \eqref{beta_c}. Since the difference between $\beta_c(M_H)$ and $\beta(M_H)$ is quite small as shown in Fig. \ref{fig:beta}, the constraints on $\beta_c(M_H)$ is nearly equivalent to that of $\beta(M_H)$. Additionally, one can extract constraints on the variance of primordial density perturbations $\sigma(M_H)$ from the relation \eqref{beta_mono}. If one considers an inflationary spectrum with a narrow peak, the upper bounds for $\sigma(M_H)$ also limit the maximum value of the effective parameter $\alpha(M_H)$ via \eqref{sigma_max}. In the following subsections, we investigate the constraints for PBHs in the horizon mass ranges $10^{12} - 10^{13}$ g in detail.

\begin{figure}[h!]
	\centering
	\includegraphics[width=3.5in]{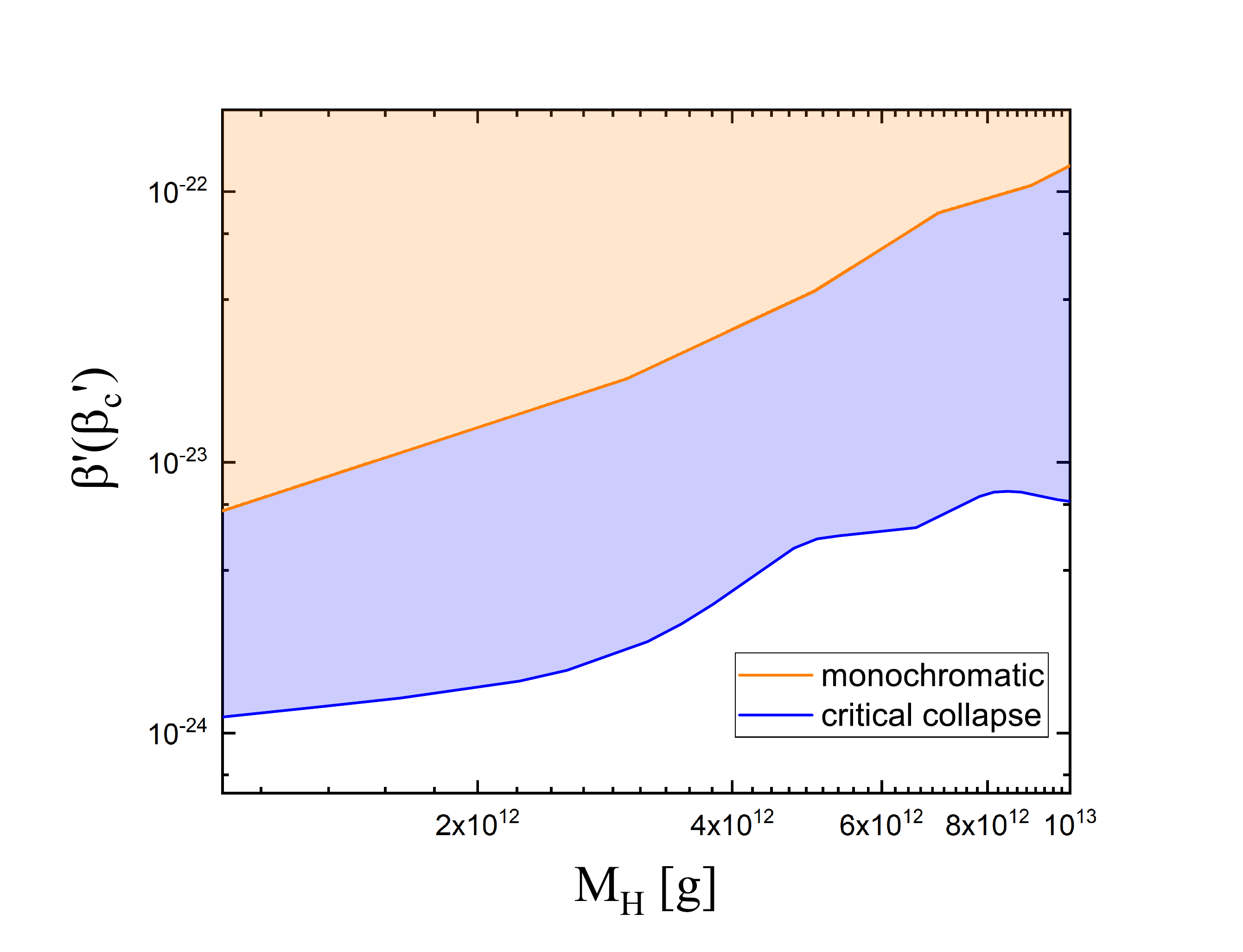}
	\caption{The upper bounds on $\beta'(M_H)$ (the orange curve) and $\beta_c'(M_H)$ (the blue curve) for the monochromatic and critical collapse mass functions, respectively, from the $^3$He/H ratio in the present Galaxy in the horizon mass range $10^{12}-10^{13}$ g. The colored regions refer to the ruled-out PBH initial mass spectrum.}
	\label{fig:1213beta}
\end{figure}

For PBHs within the horizon mass range $10^{12}-10^{13}$ g, we run the nonthermal BBN code to put constraints on the initial PBH mass spectrum. By using the $^3$He abundance, we plot the upper limits of $\beta'(M_H)$ and $\beta_c'(M_H)$ for the monochromatic and critical collapse mass functions in Fig. \ref{fig:1213beta}, respectively. For both cases, instead of the power-law spectrum, we apply \eqref{f_pou} for the {sub-threshold photons (i.e., emitted photons with energies $E_\gamma<E_{th}$)}. However, for the PBHs in this mass range, those {sub-threshold} photons are not the leading component in forming the final photons spectrum $f_{fin}(E_\gamma;T)$, and there is only a slight enhancement of the final photons spectra below $E_{th}$ (see {comparison between the purple and cyan curve in} left panel of Fig. \ref{fig:spect_dirac}). Although the normalized {final} photons spectra are the same for both monochromatic and critical collapse cases (see the comparison of right panels of Fig. \ref{fig:spect_dirac} and Fig. \ref{fig:spect_cr}), the low-mass tail of the critical collapse mass function could provide an enhancement of high energy photons number, which provides higher nonthermal photodissociation rates. The constraint on $\beta_c'(M_H)$ (blue curve) is therefore much lower than that of the monochromatic mass function.

\begin{figure}[h!]
	\centering
	\includegraphics[height=2.2in]{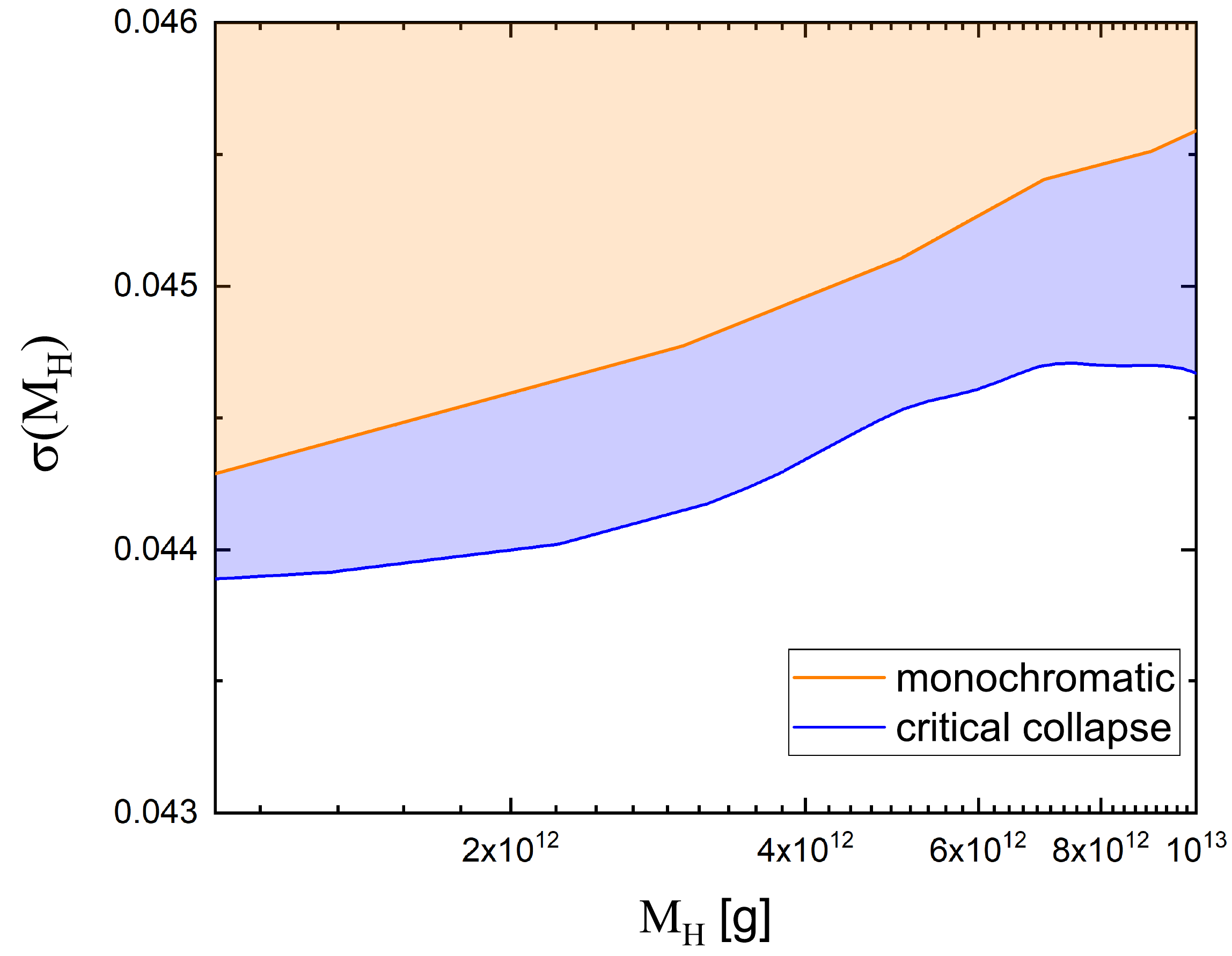}
	\includegraphics[height=2.2in]{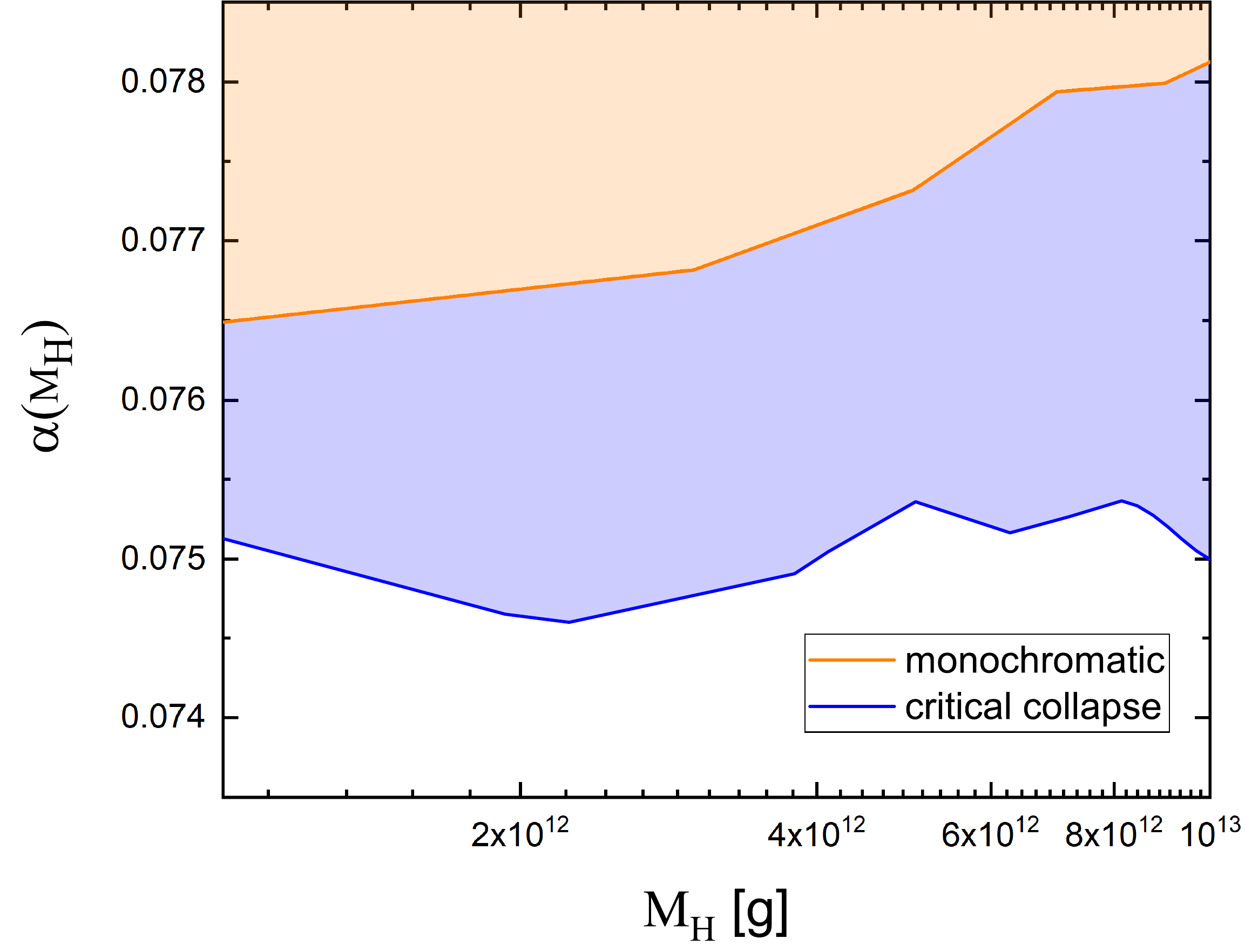}
	\caption{The upper bounds for $\sigma(M_H)$ and $\alpha(M_H)$ for the monochromatic (the orange curve) and critical collapse mass function (the blue curve), respectively, in the horizon mass range $10^{12}-10^{13}$ g. The colored regions refer to the forbidden values. }
	\label{fig:1213sigma}
\end{figure}

Additionally, the constraints on the variance of primordial density perturbations $\sigma(M_H)$ are derived in Fig. \ref{fig:1213sigma} (left panel) from the relation \eqref{beta_mono} and the constraints on $\beta'(M_H)$ and $\beta_c'(M_H)$. Assuming that the PBHs formed from a narrow inflationary spectrum, the maximum of the variance $\sigma(M_H)$ is determined by the effective parameter $\alpha$ in \eqref{sigma_max}. The upper bounds on $\alpha$ corresponding to the limit on $\sigma(M_H)$ is then displayed in the right panel of Fig. \ref{fig:1213sigma}. The upper bounds on $\sigma(M_H)$ and $\alpha(M_H)$ are more stringent for the critical collapse than that of the monochromatic case. The larger nonthermal photons spectra originating from the low-mass tail of mass function in the critical collapse model influence light elemental abundances more significantly than those in the case of single-horizon-mass PBHs.

\subsection{Impact on Li abundance}
The standard BBN model predicts $^6$Li/H $\sim 10^{-14}$ \cite{Coc:2011az,Coc:2014oia,Pitrou:2018cgg}. However, it has been suggested that the $^6$Li abundance is sensitive to processes operating in non-standard BBN models, such as hadronic and (or) radiative particle decays \cite{1988ApJ...330..545D,Dimopoulos:1988zz,Dimopoulos:1988ue,Jedamzik:1999di,Kawasaki:2000qr,2006PhRvD..74b3526K}. The $^6$Li can be also produced via $\alpha$+$\alpha$ fusion of cosmic rays accelerated by structure formation shocks \cite{Suzuki:2002qa} and supernova shocks \cite{Rollinde:2004kz} as well as via $^{3,4}$He+$\alpha$ fusion by flare-accelerated nuclei \cite{Tatischeff:2006tw} up to a level of $^6$Li/$^7$Li$\lesssim 10$ \% at metallicity [Fe/H] $< -2$ \cite{Prantzos:2012wt}. In the past, a spectroscopic determination of $^6$Li abundances in the metal-poor stellar atmosphere suggested that $^6$Li abundances in nine stars are at a plateau with abundance ratio $^6$Li/H$\sim 6 \times 10^{-12}$ \cite{Asplund:2005yt}, which is $\sim 3$ orders of magnitude higher than the standard BBN prediction. Although it seemed like an indication of the ``cosmic $^6$Li problem'', an asymmetry in the Li I absorption line caused by photospheric convective motions can mimic the existence of $^6$Li \cite{Cayrel:2007te}. The latest investigation based upon 3D non-local thermal equilibrium model \cite{Lind:2013iza} concludes no detection and sets upper limits derived on the isotopic ratios as $^6\rm Li/^7 Li=0.051$.

\begin{figure}[h]
\centering
\includegraphics[width=3.5in]{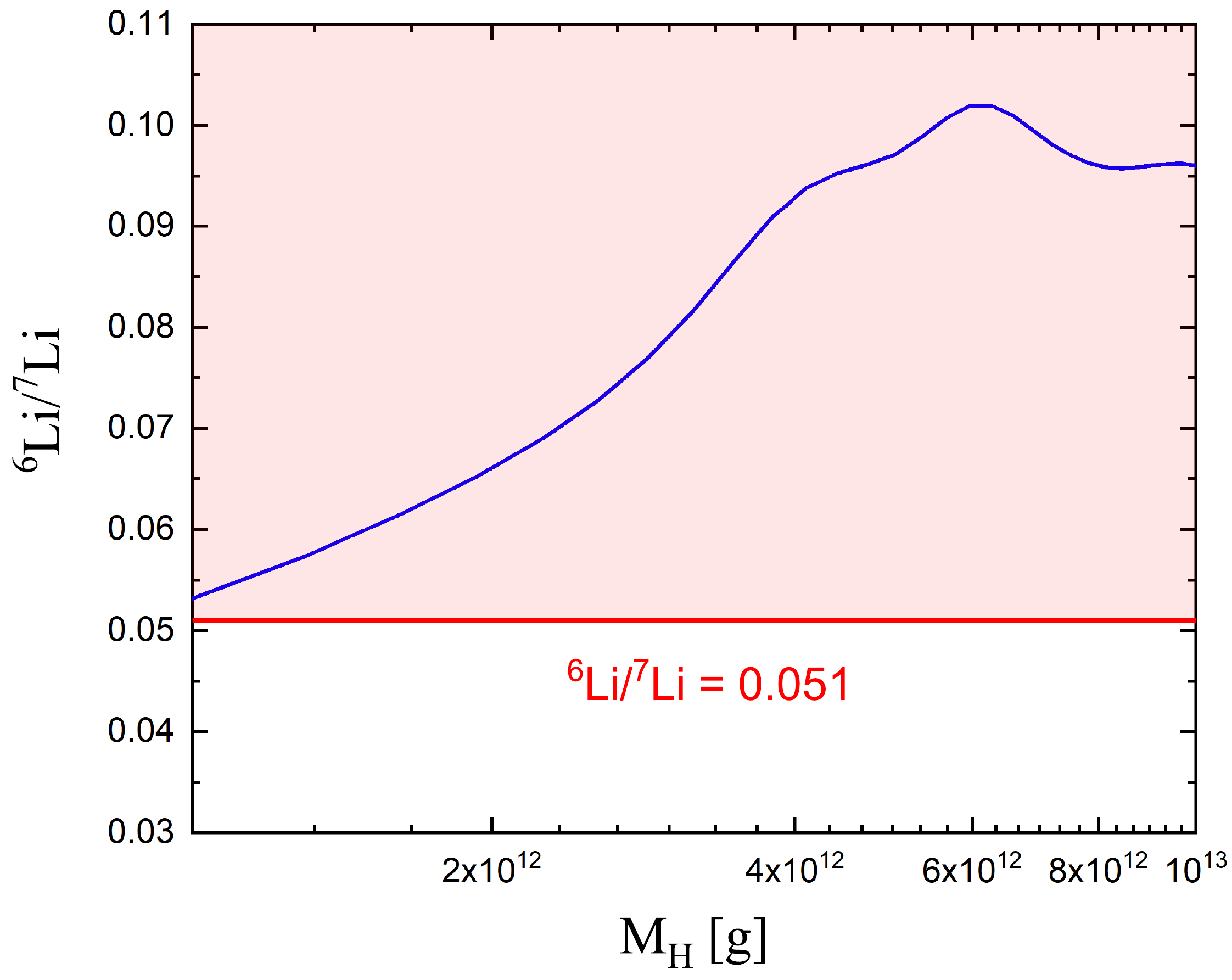}
\caption{Li isotopic ratio $^6$Li/$^7$Li as a function of the horizon mass of PBHs. The amplitude of PBH initial mass function is set at the upper limit from the the $^3$He abundance $^3$He/H$=1.5\times10^{-5}$.}
\label{fig:li6}
\end{figure}

Fig. \ref{fig:li6} shows the Li isotopic ratio $^6$Li/$^7$Li as a function of $M_{\rm H}$ for critical collapse model (blue curve). The amplitude of PBH initial mass function is set at the upper limit from the the $^3$He abundance as $^3$He/H$=1.5\times10^{-5}$. The $^6$Li is produced significantly in this mass range to the level of $^6$Li/H$\sim 10^{-11}$. Although the theoretical calculation is above the observed upper limits on the isotopic ratio, many scenarios have been proposed for $^7$Li reduction. If a stellar depletion involving an atomic diffusion \cite{Richard:2001qp,Richard:2004pj} or rotational mixing \cite{Korn2006} are responsible for the Li problem, i.e., the present contradiction of cosmic $^7$Li abundance between theory and the observations from metal-poor halo stars, then, the primordial $^6$Li abundance should also be larger than the value on the surfaces of these stars \cite{Pinsonneault2002,Korn2006}. This is because $^6$Li destruction is more effective than $^7$Li destruction during the pre-main-sequence stage, and the $^6$Li isotopic ratio is predicted to be lower than the initial value. Theoretical calculations indicate a reduction of $^6$Li/$^7$Li by a factor of $\gtrsim 2$ in the pre-main-sequence epoch with high temperatures of $T_{\rm eff}>6000$ K \cite{Richard:2004pj}. In addition, if rotational mixing occurred in metal poor stars \cite{Pinsonneault2002}, the  $2\sigma$ upper bound of $^6$Li/$^7$Li should be increased by taking into account the stellar depletion factors as in Ref. \cite{Carr:2009jm} which adopted $^6$Li/$^7$Li$<0.302$. Therefore, observed isotopic ratios are more than a factor of two smaller than the initial value.

For $^7$Li, it has been suggested that $^7$Li abundance can be reduced via photodisintegration of $^7$Be if the nonthermal photons spectrum is very soft as in the monoenergetic photons generation via a two-body decay \cite{2013PhRvD..87h5045K,2015PhRvL.114i1101P,2015PhRvD..91j3007P} or MeV-scale sterile neutrino decay \cite{2014PhRvD..90h3519I}. However, finite-width distributions of PBH mass function in the current model result in hard spectra in which low mass PBHs contribute to the high energy part. Therefore, the $^4$He photodisintegration is unavoidable. The $^7$Li abundance in metal-poor stars still remains an unsolved problem that is a factor of $\sim 3-4$ lower than the standard BBN prediction. In this work, the $^7$Li abundance, i.e., Spite plateau, could not be explained by the nonthermal photodisintegrations since the constraint from a $^3$He overproduction caused by the photodisintegration of $^4$He is stronger.

\section{Conclusion}
\label{sec:concl}
In this work, we study the {nuclear photodisintegration reactions triggered by nonthermal electromagnetic Hawking radiation from PBHs in the critical collapse model. We consider the simplest case that all PBHs formed at a single epoch, referring to a single horizon mass. This situation can realize in several PBH formation scenarios, in which a relatively narrow enhancement in the small-scale inflationary spectrum is required. As a consequence of Hawking radiation, the low-mass tail of the critical collapse mass function would lead to an enhancement of high-energy photon emission compared with the monochromatic mass function. It has been well understood that the constraints on PBH mass spectrum depend on the shape of PBH mass function, so that one cannot simply compare the extended mass function with the monochromatic one in considering the constraints. Previous studies have already shown that the horizon-mass approximation is good enough for the critical collapse model if PBHs formed at a single cosmic time. Then, the critical collapse model was thought to be practically indistinguishable from the monochromatic model. However, we for the first time find in this work that the high-energy photons radiated from the low-mass tail of the critical collapse mass function could significantly affect the primordial light elemental abundances via nonthermal BBN processes. This result indicates that tail contribution of the critical collapse model should be referred with caution when considering the constraints on PBH evaporation.

 For the low-energy nonthermal photons, we solve the Boltzmann equation because those photons are sensitive to the energy of photons emitted from PBHs, and the power-law spectrum is invalid any more. We carry out a sophisticated calculation of primary photodisintegration reactions. For the $^7$Be($\gamma, \alpha$)$^3$He reaction, we provide a correct analytical function for the cross section. The secondary nuclear fusion reactions related to $^6$Li production, i.e., reactions of energetic products from the primary reactions and the background nuclei, have been taken into account.

For PBHs with the initial horizon masses of $10^{12}-10^{13}$ g, we make an analytical estimation of $\beta'$ constraint and carry out the detailed numerical calculations including accurate treatment of EM cascade spectrum and updating nuclear reaction cross section data. We also report that the discrepancy between our updated $^3$He constraints and the previous ones, deriving easy analytical bounds for photodisintegration process and hadrodissociation process triggered by PBH Hawking radiation. We also update a constraint on the initial mass spectrum $\beta'$ of PBHs with a monochromatic mass function by using the $2\sigma$ observational upper limit on the Galactic $^3$He abundance, i.e., $^3$He/H$=1.5\times10^{-5}$. We also provide new constraints on initial mass spectrum $\beta_c'$ for the critical collapse mass function and the variance of primordial density perturbation $\sigma$ as well as the model-dependent effective parameter $\alpha$ for a narrow inflationary spectrum. In principle, constraints on the extended mass function are different from the monochromatic one. Our results show that the constraints on the critical collapse model from $^3$He abundance is more stringent than that on the monochromatic model.

The secondary processes could produce $^6$Li dramatically abundantly as high as $^6$Li/H$\sim10^{-11}$ for the critical collapse mass function. This high abundance level can be comparable to the observational upper limit to the isotopic ratio $^6$Li/$^7$Li$<0.302$ if the stellar depletion processes are taken into account. For $^7$Li, the Spite plateau could not be explained by the nonthermal photodisintegrations triggered by PBH Hawking radiation because the constraint from a $^3$He overproduction caused by the photodisintegration of $^4$He is stronger.
}

\appendix
\section{The analytical estimation: photodisintegration}
\label{apendA}
Here, we derive a simple analytic estimation, based on the assumption: All the PBH mass turns into nonthermal photons during its evaporation.

The $^3$He abundance change can be written as
\be\label{a1}
\Delta \frac{^3{\rm He}}{\rm H} = \frac{\Delta n_\gamma^{\rm nt}}{n_{\rm H}} P ,
\ee
where $\Delta n_\gamma^{\rm nt}/n_{\rm H}$ is the number density of nonthermal photons emitted from PBH evaporation,  normalized to the {\color{blue}h}ydrogen number density, and $P$ is the average probability of nonthermal photons to react with background $^4$He via $^4$He($\gamma$,$N$)$^3A$ reactions. The PBHs within the mass range $10^{12}-10^{13}$ g have already evaporated by now. Then, the total number of nonthermal photons emitted from a single PBH with an initial mass $M$ is given by
\be
N_{\gamma}(M) =\frac{M}{\langle E_\gamma \rangle},
\ee
where $\langle E_\gamma \rangle$ is the average energy of nonthermal photons.
At the end of PBHs' evaporation at the redshift $z_{\rm end}$, we evaluate
\ba\label{a3}
{ \Delta n_\gamma^{\rm nt} \over n_{\rm H} }
 \left( \langle E_\gamma \rangle \right)
&=&
{ n_{\rm PBH}(z_{\rm end}) N_{\gamma}(M) \over n_{\gamma}(z_{\rm end}) \eta X_p } \nonumber \\
&=&
 \Omega_{\rm PBH} \rho_{\rm crit,0} \over \langle E_\gamma \rangle n_{\gamma,0} \eta X_p
\nonumber \\ &\simeq&
2.28 \times 10^2 \l( {\langle E_\gamma \rangle}\over{30 {\rm MeV}} \r)^{-1}
\l( { \eta \over 6 \times 10^{-10} } \r)^{-1} \l( { X_p \over 0.75 } \r)^{-1} f_{\rm PBH} ,
\ea
where $n_{\rm PBH}(z_{\rm end})$ and $n_{\gamma}(z_{\rm end})$ are the physical number densities of PBHs and the background photons at $z_{\rm end}$, respectively. The quantity $n_{\gamma,0} =  [2 \zeta(3) /\pi^2] T_0^3$ is the current number density of the background photons with the temperature $T_0 = 2.73$ K. In the above equation, we have used the relation $n_H = n_{\gamma,0} (1 + z_{\rm end})^3 \eta X_p$. For the convenience of calculation, we define $\Omega_{\rm PBH} \equiv n_{\rm PBH,0} M {\color{blue}/} \rho_{\rm crit,0}$ and $f_\text{PBH} \equiv \Omega_{\rm PBH} / \Omega_{\rm DM}$, where $\rho_{\rm crit,0}$ is the current critical density, and $\Omega_{\rm DM}$ is the current normalized energy fraction of dark matter and $n_{\rm PBH,0}$ is related to $n_{\rm PBH}(z_{\rm end})$ as $n_{\rm PBH,0} = n_{\rm PBH}(z_{\rm end}) (1 + z_{\rm end})^{-3}$, which is the comoving number density of PBHs before evaporation. Then, the parameters $\Omega_{\rm PBH}$ and $f_\text{PBH}$ defined here are related by Eq. \eqref{fpbh}, although the PBHs within the mass range $10^{12} - 10^{13}$ g have already evaporated by now. Furthermore, $f_\text{PBH}$ is also related with $\beta$ by Eq. \eqref{fpbh}.

In the radiation-dominated epoch, the dominant energy-loss process for the nonthermal photons is the Compton scattering off of the background electrons, and the rate is
\ba
\Gamma_{\rm Com}(E)
&=& n_e \sigma_{\rm loss}\nonumber \\
&=& 8.5 \times 10^{-47} \left( \frac{E/m_e}{60} \right)^{-1}
\left( \frac{\ln (2E/m_e)}{\ln (120)} \right) (1+z)^3~{\rm GeV}.
\label{eq_comp1}
\ea
The photodisintegration rate of $^4$He is given by
\ba
\Gamma_{\rm dis}(E) &=& n_{\rm b} \frac{Y}{4} \sigma_{\rm dis}\nonumber \\
&=& 1.5 \times 10^{-8}~{\rm Gyr}^{-1} \left( \frac{\eta}{6 \times 10^{-10}} \right)
\left( \frac{Y}{0.25} \right) \left( \frac{\sigma_{\rm dis}(E)}{1~{\rm mb}} \right)
\left(1+z \right)^3.
\ea
Then the probability of $^3$He production by a nonthermal photon with energy $E$ can be estimated as
\ba
P(E; z \gtrsim 400) &\simeq& \frac{\Gamma_{\rm dis}}{\Gamma_{\rm Com}}\nonumber \\
&=& 3.59 \times 10^{-3} \left( \frac{E/m_e}{60} \right)
\left( \frac{\ln (2E/m_e)}{\ln (120)} \right)^{-1}.
\label{prob_early}
\ea
The average probability $P$ is then given by
\be
P(z) = \int p_\gamma(E_\gamma; z) P(E_\gamma; z) dE_\gamma,
\ee
where $p_\gamma(E_\gamma)$ is the zeroth-generation quasi-equilibrium photon spectrum attained after a quick EM cascade.

\begin{figure}[h]
\centering
\includegraphics[width=3.5in]{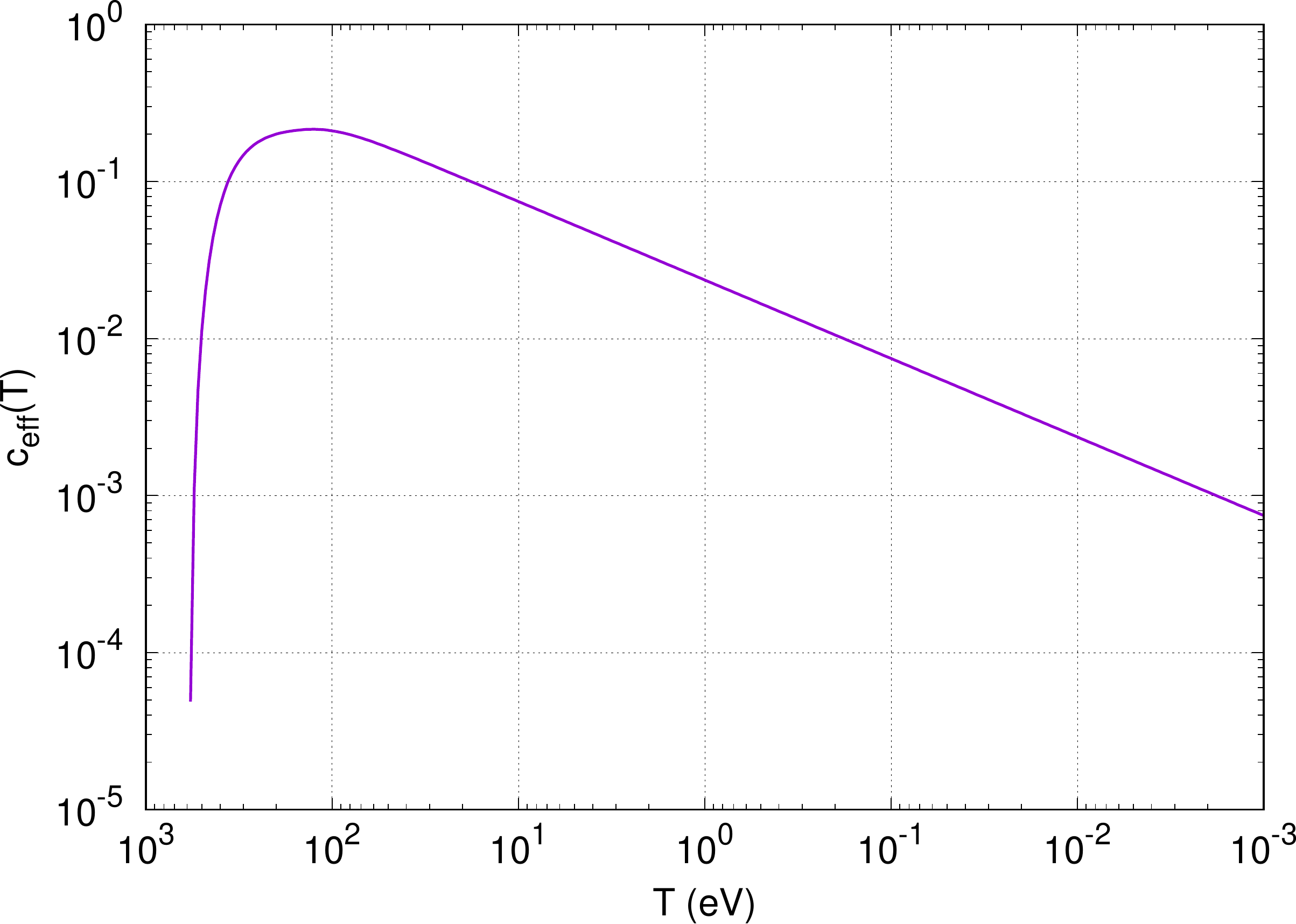}
\caption{The efficiency factor $c_{\rm eff}(T)$ as a function of cosmic temperature.}
\label{fig1}
\end{figure}

The efficiency factor $c_{\rm eff}$ is defined by
\ba
c_{\rm eff}(T) &=& \frac{\int p_\gamma(E_\gamma) \sigma_{\rm dis}(E_\gamma) /\sigma_{\rm loss}(E_\gamma) dE_\gamma}
{\sigma_{\rm dis}(E_\gamma^{\rm typ}) /\sigma_{\rm loss}(E_\gamma^{\rm typ})},
\ea
where the zeroth-generation spectrum $p_\gamma(E_\gamma)$ is approximated with a broken power-law [Eq. \eqref{steadyspect}], and its total energy is normalized to $E_{\gamma 0} =E_\gamma^{\rm typ}$.
This is the relative efficiency of $^3$He production normalized to the case of monoenergetic nonthermal photon injection at a typical energy $E_\gamma^{\rm typ}=30\ \rm MeV$, which is taken here to be the peak of the photodisintegration cross section for $^3$A production, i.e., $\sigma_{\rm dis} (E_\gamma^{\rm typ})= \mathcal{O}(1)$ mb.
The photodisintegration cross section is the sum of those for $^3$H and $^3$He production from Ref. \cite{2009PhRvD..79l3513K}. The denominator has the value of $0.11539$ according to (A.4) and (A.5). Fig. \ref{fig1} shows the efficiency $c_{\rm eff}(T)$ as a function of temperature. At a high temperature, the cutoff energy is below the threshold energy of $^4$He photodisintegration, i.e., $E_{\rm th} \sim 20$ MeV, and the efficiency is zero. At $T\sim 10^2$ eV, the efficiency reaches maximum at $c_{\rm eff} \sim 0.2$, when the cutoff energy goes somewhat over the threshold. As $T$ deceases, the cutoff energy increases, and the EM cascade results in production of abundant high-energy photons. Since such high-energy photons are less efficient in inducing photodisintegration, the efficiency decreases again.

Finally, we have
\be
\begin{aligned}
\label{app:estimate_beta}
\Delta \frac{^3{\rm He}}{\rm H}
=&
c_{\rm eff}(T) \frac{\Delta n_\gamma^{\rm nt}} {n_{\rm H}}(E_{\rm typ}) P(E_{\rm typ})
\\\simeq&
2.7 \times 10^8 c_{\rm eff}(T)
\l( \frac{E_\gamma^{\rm typ}}{30~{\rm MeV}} \r)^{-1} \l( { \eta \over 6 \times 10^{-10} } \r)^{-1} \l( { X_p \over 0.75 } \r)^{-1}  \l( {M \over M_\odot} \r)^{-1/2} \beta' .
\end{aligned}
\ee

\section{The analytical estimation: hadrodissociation}
\label{apendB}

The number density of nonthermal protons generated by PBH evaporation normalized to the hydrogen number density is given similarly to Eqs. (\ref{a1}) and (\ref{a3}) by
\ba\label{b1}
\frac{\Delta n_p^{\rm nt}}{n_{\rm H}} &=&
 \Omega_{\rm PBH} f_{{\rm PBH},p} \rho_{\rm crit,0} \over \langle E_p \rangle n_{\gamma,0} \eta X_p
 \nonumber \\
  &=&
3.055
f_{{\rm PBH},p}
  \left( \frac{\langle E_p \rangle} {2.24~{\rm GeV}} \right)^{-1}
  \left( \frac{\eta}{6 \times 10^{-10}} \right)^{-1}
  \left( \frac{X_p}{0.75} \right)^{-1},
\ea
where
$f_{{\rm PBH},p}$ is the energy fraction of PBH given to protons, and $\langle E_p \rangle$ is the average proton energy.

\begin{figure}[h]
\begin{center}
\includegraphics[width=3.5in]{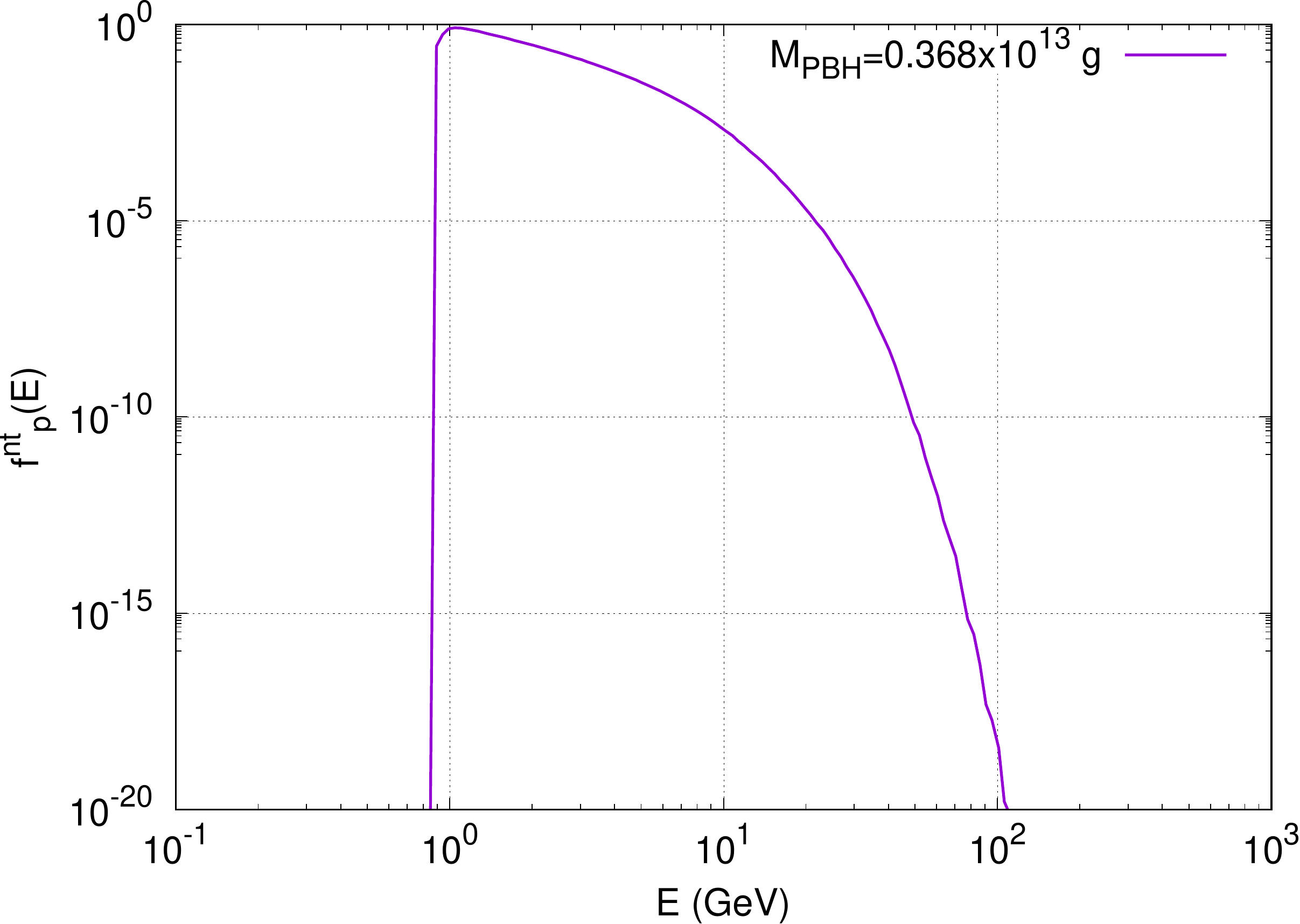}
\caption{The normalized distribution function of nonthermal proton $f^{\rm nt}_p(E)$ at typical evaporation time of $M_{{\rm PBH}}=M_{\rm BH}(0) /e$ with the initial mass of PBH $M_{\rm BH}(0)=10^{13}$ g.}
\label{dist1}
\end{center}
\end{figure}

\begin{figure}[h]
\begin{center}
\includegraphics[width=3.5in]{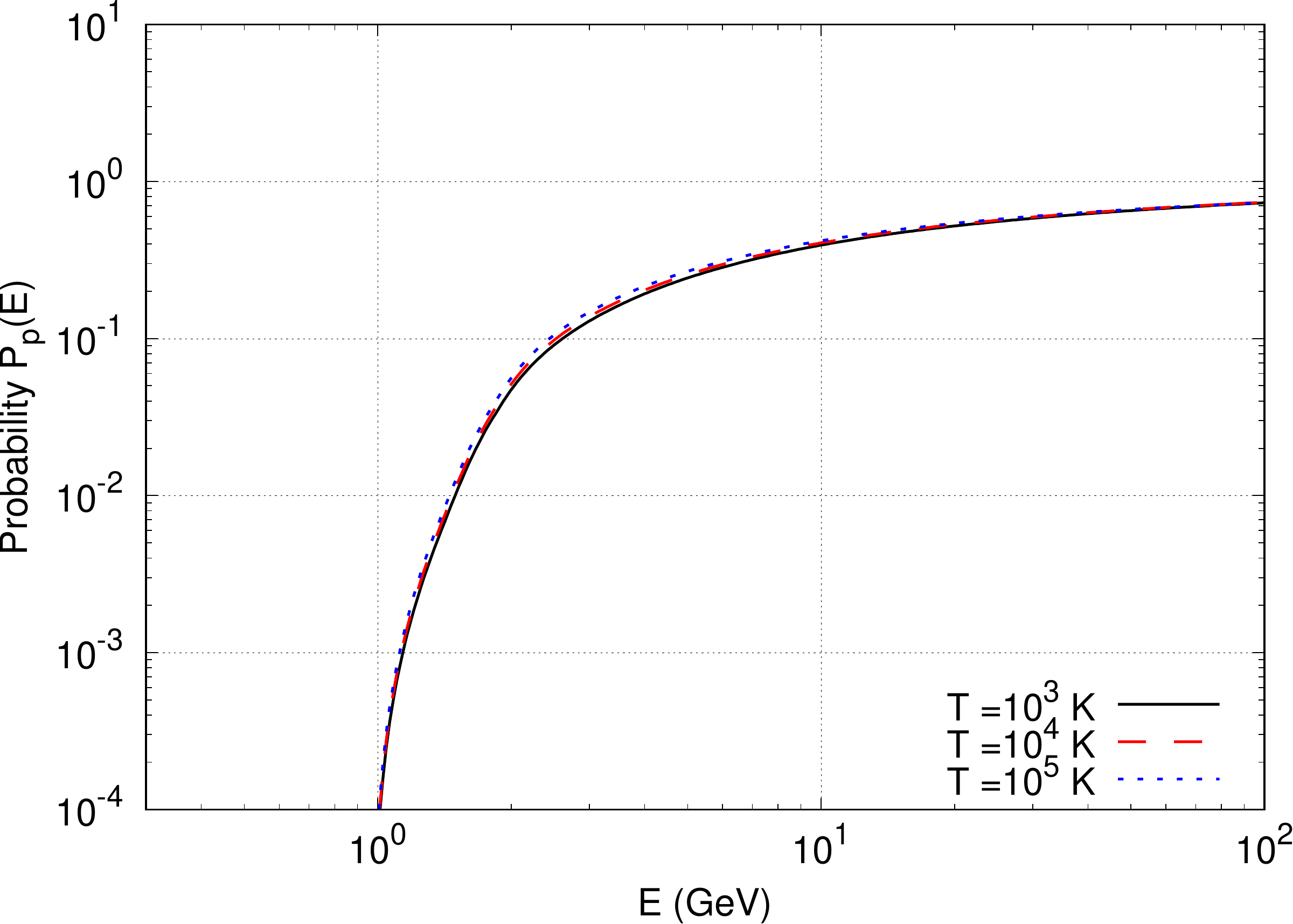}
\caption{The probability of $^3$He production for nonthermal protons $P_p(E)$ as a function of proton energy at $T =10^{3,4,5}$ K.}
\label{pro3}
\end{center}
\end{figure}

The probability of nonthermal protons to react with background $^4$He via $^4$He($p$ ,X)$^3$A reactions is given by
  \ba\label{b3}
P_p &=& \int_{m_p}^\infty dE f^{\rm nt}_p(E) P_p(E) \label{eqb2}\\
  P_p(E) &=& 1 -\exp \left[ -
    \int_0^\infty \Gamma_3(E(t)) dt \right] \nonumber \\
  &=& 1 -\exp \left[ -
    \int_{E^{\rm th}_{^3{\rm He}}}^{E} dE' \frac{1}{E'}\frac{\Gamma_3(E')}{\Gamma_{\rm Coul}(E')+\Gamma_{\rm inel}(E')}
    \right],
\ea
where
$\Gamma_i$ are reaction rates (in GeV) and $R_i =\Gamma_i/n_b$ (in GeV$^{-2}$) for three processes, i.e.,
$i =3$ for the $^4$He($p$ ,X)$^3$A reactions ($X=pp,nn$ and $d$), $i= \rm Coul$ for the Coulomb loss process, and $i= \rm inel$ for the energy loss via the $p+p \rightarrow p+p { (n+\pi^+)} +m\pi$ reaction for $m=1, 2, ...$ \cite{1969NCimA..63..529B,Mcgill:1984jq,Nagae:1986wu,1991PhRvL..67.1982C,Engel:1996ic,AbdelBary:2004qb,Kurbatov:2007zv,Maeda:2008gv}. The reaction rates are given by
\ba
R_3(E) &=& \frac{Y}{4} \left[\sigma_3 v \right](E)  \\
R_{\rm Coul}(E;T) &=& \frac{1}{n_b E}  \left. \frac{dE}{dt} \right|_{\rm Coul}  \\
R_{\rm inel}(E) & { \approx} & { \frac{1}{3}} X \left[\sigma_{\rm inel} v \right](E),
\ea
where $dE/dt|_{\rm Coul}$ is the energy loss rate via Coulomb scatterings off of background electrons \cite{Reno:1987qw}, and we assumed that about a third of the incident proton energy is lost at one inelastic scattering. At a relatively low energy $p+p\rightarrow p+n+\pi^+$ scattering at $E_p=1.05$ GeV \cite{1991PhRvL..67.1982C} shows that forward scattered neutrons have a peak energy of $\sim 1.05$ GeV which corresponds to a loss of $\sim 1/3$ of initial energy. At higher energies of $E$/GeV =[2.85, 7.88] most relevant to the current PBH mass \cite{1969NCimA..63..529B}, various nuclear isobaric resonances work and show their angular dependence in partial cross section. However, the energy of scattered protons generally distribute widely, and it is seen that about a half or a third of energy is lost at one scattering. Fig. \ref{dist1} shows the normalized distribution function of nonthermal proton $f^{\rm nt}_p(E)$ at typical evaporation time corresponding to $M_{\rm BH}(0) /e =0.368 \times 10^{13}$ g. Fig. \ref{pro3} shows the probability of $^3$He production for nonthermal protons $P_p(E)$ as a function of proton energy at $T =10^3,10^4$ and $10^5$ K. In this late epoch, the average probability of proton-dissociation [Eq. (\ref{eqb2})] is almost independent from the temperature. Then, in this estimation we take the averaged value of $P_p\simeq0.07$.

\begin{figure}[h]
\begin{center}
\includegraphics[width=3.5in]{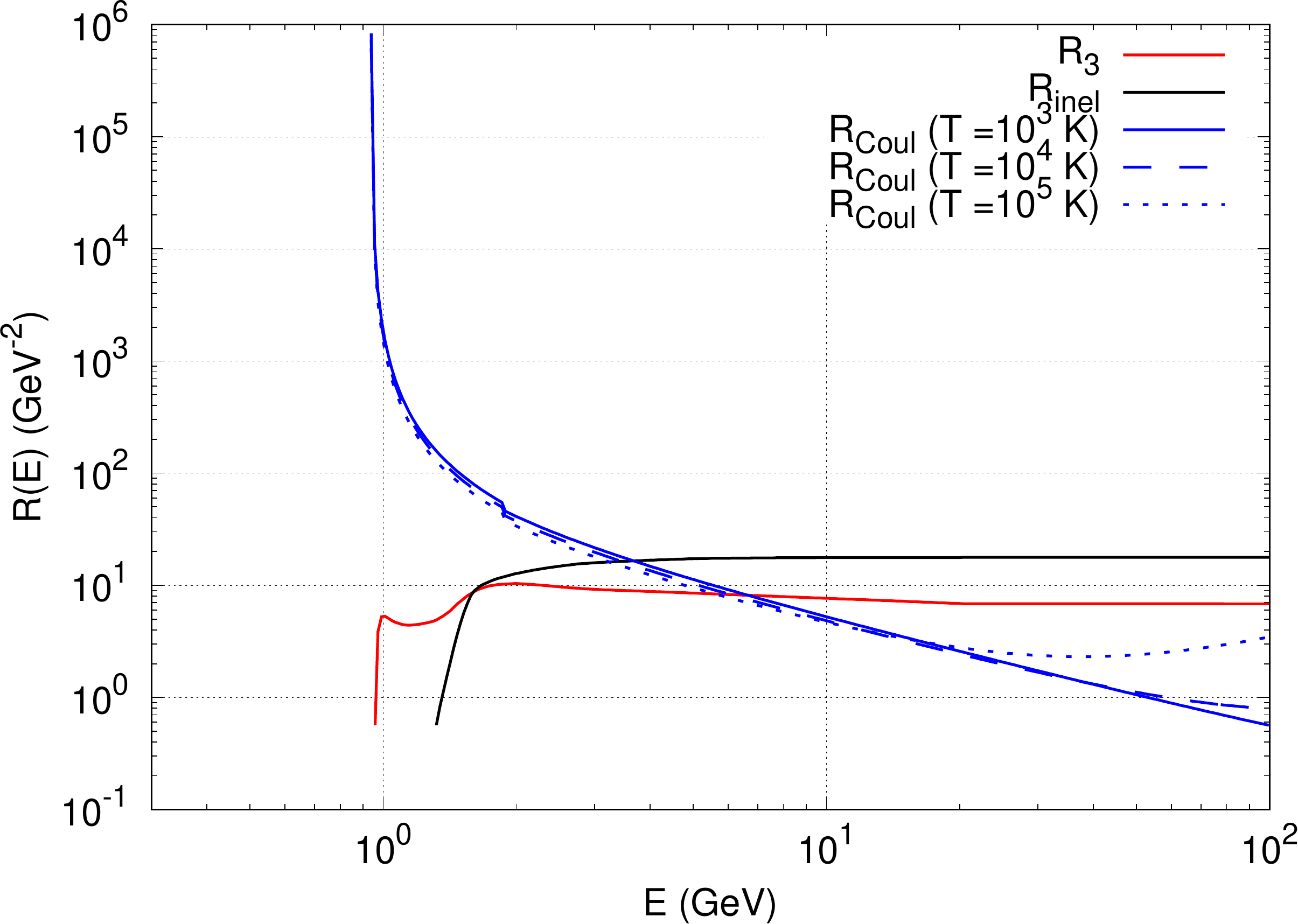}
\caption{The reaction rates as a function of proton energy at $T =10^{3,4,5}$ K.}
\label{rate2eff}
\end{center}
\end{figure}

We note that secondary process of nonthermal $^3$He can also operate. However, this effect is not very large and neglected here. As shown below (Fig. \ref{rate2eff}), the inelastic reaction rate is predominant at high energy $E$ while the Coulomb loss rate is predominant at low energy $E$. Therefore, the secondary process starting from $^4$He($p$ ,$X$)$^3$A is insignificant. In addition, the cross section of elastic scattering of $p+p$ is smaller than that of inelastic scattering at $E \gtrsim 1$ GeV. Since the elastic scatterings at the high energy are approximately the forward and backward scatterings (Fig. 11 in Ref. \cite{1972A&AS....7..417M}), after the scattering, the number of energetic protons does not double. Quite the contrary, the elastic scatterings work as energy loss since slightly energetic proton is produced at the forward or backward scattering, and it is quickly thermalized by the Coulomb loss.

Finally, combining Eqs. (\ref{b1}) and (\ref{b3}), the change of $^3$He via proton-dissociation process is
 \ba
\Delta \frac{^3{\rm He}}{\rm H} &=& \frac{\Delta n_p^{\rm nt}}{n_{\rm H}}P_p  \nonumber \\
&\simeq& 7.1 \times 10^7
  \left( \frac{\langle E_p \rangle} {2.24~{\rm GeV}} \right)^{-1}
  \left( \frac{P_p}{0.07} \right)
  \left( \frac{\eta}{6 \times 10^{-10}} \right)^{-1}
  \left( \frac{X_p}{0.75} \right)^{-1} \l( {M \over M_\odot} \r)^{-1/2} \beta'. \nonumber \\
\ea

\acknowledgments

We are grateful to Jérémy Auffinger, Kazunori Kohri, Jiewen Chen, Yi-Fu Cai and Dong-Gang Wang for stimulating discussion.
Y.L. is supported by JSPS KAKENHI Grant No. 19J22167. C.C. is supported in part by NSFC Research Fund (11722327 and 1181101398).  M. K. is supported by NSFC Research Fund for International Young Scientists (11850410441). T. K. is supported in part by Grants-in-Aid for Scientific Research of JSPS (17K05457 and 20K03958). C.C. is grateful to Weixia Chen\&Xueying Tian for their hospitality and support.


\begin{thebibliography}{999}




	\bibitem{Zeldovich:1966}
	Y.~B.~Zel'dovich and I.~D.~Novikov,
	\emph{The Hypothesis of Cores Retarded during Expansion and the Hot Cosmological Model},
	Sov.\ Astron. \ {\bf 10}, 602 (1967).

	\bibitem{Hawking:1971ei}
	S.~Hawking,
	\emph{Gravitationally collapsed objects of very low mass},
	Mon.\ Not.\ Roy.\ Astron.\ Soc.\  {\bf 152}, 75 (1971).

	\bibitem{Carr:1974nx}
	B.~J.~Carr and S.~W.~Hawking,
	\emph{Black holes in the early Universe},
	Mon.\ Not.\ Roy.\ Astron.\ Soc.\  {\bf 168}, 399 (1974).

	\bibitem{Carr:1975qj}
	B.~J.~Carr,
	\emph{The Primordial black hole mass spectrum},
	Astrophys.\ J.\  {\bf 201}, 1 (1975).
	%doi:10.1086/153853

	\bibitem{Hawking:1974rv}
	S.~W.~Hawking,
	\emph{Black hole explosions},
	Nature {\bf 248}, 30 (1974).
	%doi:10.1038/248030a0


	\bibitem{Carr:2020gox}
	B.~Carr, K.~Kohri, Y.~Sendouda and J.~Yokoyama,
	\emph{Constraints on Primordial Black Holes},
	[arXiv:2002.12778 [astro-ph.CO]].
	%52 citations counted in INSPIRE as of 10 Jul 2020





	\bibitem{Sasaki:2018dmp}
	M.~Sasaki, T.~Suyama, T.~Tanaka and S.~Yokoyama,
	\emph{Primordial black holes -- perspectives in gravitational wave astronomy},
	Class.\ Quant.\ Grav.\  {\bf 35}, no. 6, 063001 (2018)
	%doi:10.1088/1361-6382/aaa7b4
	[arXiv:1801.05235 [astro-ph.CO]].

	\bibitem{Carr:2020xqk}
	B.~Carr and F.~Kuhnel,
	\emph{Primordial Black Holes as Dark Matter: Recent Developments},
	[arXiv:2006.02838 [astro-ph.CO]].
	%5 citations counted in INSPIRE as of 09 Jul 2020





	\bibitem{Nakamura:1997sm}
	T.~Nakamura, M.~Sasaki, T.~Tanaka and K.~S.~Thorne,
	\emph{Gravitational waves from coalescing black hole MACHO binaries},
	Astrophys. J. Lett. \textbf{487}, L139-L142 (1997)
	%doi:10.1086/310886
	[arXiv:astro-ph/9708060 [astro-ph]].
	%198 citations counted in INSPIRE as of 09 Jul 2020


	\bibitem{Sasaki:2016jop}
	M.~Sasaki, T.~Suyama, T.~Tanaka and S.~Yokoyama,
	\emph{Primordial Black Hole Scenario for the Gravitational-Wave Event GW150914},
	Phys.\ Rev.\ Lett.\  {\bf 117}, no. 6, 061101 (2016)
	Erratum: [Phys.\ Rev.\ Lett.\  {\bf 121}, no. 5, 059901 (2018)]
	%doi:10.1103/PhysRevLett.121.059901, 10.1103/PhysRevLett.117.061101
	[arXiv:1603.08338 [astro-ph.CO]].



	\bibitem{Bean:2002kx}
	R.~Bean and J.~Magueijo,
	\emph{Could supermassive black holes be quintessential primordial black holes?},
	Phys. Rev. D \textbf{66}, 063505 (2002)
	%doi:10.1103/PhysRevD.66.063505
	[arXiv:astro-ph/0204486 [astro-ph]].
	%116 citations counted in INSPIRE as of 09 Jul 2020


	\bibitem{Inomata:2019yww}
	K.~Inomata and T.~Terada,
	\emph{Gauge Independence of Induced Gravitational Waves},
	Phys. Rev. D \textbf{101}, no.2, 023523 (2020)
	%doi:10.1103/PhysRevD.101.023523
	[arXiv:1912.00785 [gr-qc]].
	%12 citations counted in INSPIRE as of 09 Jul 2020

	\bibitem{Cai:2019cdl}
	R.~G.~Cai, S.~Pi and M.~Sasaki,
	\emph{Universal infrared scaling of gravitational wave background spectra},
	[arXiv:1909.13728 [astro-ph.CO]].
	%18 citations counted in INSPIRE as of 09 Jul 2020

	%\cite{Cai:2018dig}

\bibitem{Cai:2018dig}
R.~g.~Cai, S.~Pi and M.~Sasaki,
%``Gravitational Waves Induced by non-Gaussian Scalar Perturbations,''
Phys. Rev. Lett. \textbf{122}, no.20, 201101 (2019)
%doi:10.1103/PhysRevLett.122.201101
[arXiv:1810.11000 [astro-ph.CO]].
%101 citations counted in INSPIRE as of 20 Jan 2021

	\bibitem{Lu:2020diy}
	Y.~Lu, A.~Ali, Y.~Gong, J.~Lin and F.~Zhang,
	\emph{On the gauge transformation of scalar induced gravitational waves},
	[arXiv:2006.03450 [gr-qc]].
	%0 citations counted in INSPIRE as of 09 Jul 2020



	\bibitem{Kohri:2018awv}
	K.~Kohri and T.~Terada,
	\emph{Semianalytic calculation of gravitational wave spectrum nonlinearly induced from primordial curvature perturbations},
	Phys.\ Rev.\ D {\bf 97}, no. 12, 123532 (2018)
	%doi:10.1103/PhysRevD.97.123532
	[arXiv:1804.08577 [gr-qc]].

	\bibitem{Bartolo:2018rku}
	N.~Bartolo, V.~De Luca, G.~Franciolini, M.~Peloso, D.~Racco and A.~Riotto,
	\emph{Testing primordial black holes as dark matter with LISA},
	Phys.\ Rev.\ D {\bf 99}, no. 10, 103521 (2019)
	%doi:10.1103/PhysRevD.99.103521
	[arXiv:1810.12224 [astro-ph.CO]].





	\bibitem{Cai:2019jah}
	Y.~F.~Cai, C.~Chen, X.~Tong, D.~G.~Wang and S.~F.~Yan,
	\emph{When Primordial Black Holes from Sound Speed Resonance Meet a Stochastic Background of Gravitational Waves},
	Phys.\ Rev.\ D {\bf 100}, no. 4, 043518 (2019)
	%doi:10.1103/PhysRevD.100.043518
	[arXiv:1902.08187 [astro-ph.CO]].




	\bibitem{Dolgov:1992pu}
	A.~Dolgov and J.~Silk,
	\emph{Baryon isocurvature fluctuations at small scales and baryonic dark matter},
	Phys. Rev. D \textbf{47}, 4244-4255 (1993)
	%doi:10.1103/PhysRevD.47.4244
	%185 citations counted in INSPIRE as of 10 Jul 2020

	\bibitem{Green:2016xgy}
	A.~M.~Green,
	\emph{Microlensing and dynamical constraints on primordial black hole dark matter with an extended mass function},
	Phys. Rev. D \textbf{94}, no.6, 063530 (2016)
	%doi:10.1103/PhysRevD.94.063530
	[arXiv:1609.01143 [astro-ph.CO]].
	%102 citations counted in INSPIRE as of 10 Jul 2020

	\bibitem{Kohri:2012yw}
	K.~Kohri, C.~M.~Lin and T.~Matsuda,
	\emph{Primordial black holes from the inflating curvaton},
	Phys.\ Rev.\ D {\bf 87}, no. 10, 103527 (2013)
	%doi:10.1103/PhysRevD.87.103527
	[arXiv:1211.2371 [hep-ph]].

	\bibitem{Kawasaki:2012wr}
	M.~Kawasaki, N.~Kitajima and T.~T.~Yanagida,
	\emph{Primordial black hole formation from an axionlike curvaton model},
	Phys.\ Rev.\ D {\bf 87}, no. 6, 063519 (2013)
	%doi:10.1103/PhysRevD.87.063519
	[arXiv:1207.2550 [hep-ph]].



	%\cite{Drees:2011yz}
	\bibitem{Drees:2011yz}
	M.~Drees and E.~Erfani,
	\emph{Running Spectral Index and Formation of Primordial Black Hole in Single Field Inflation Models},
	JCAP \textbf{01}, 035 (2012)
	%doi:10.1088/1475-7516/2012/01/035
	[arXiv:1110.6052 [astro-ph.CO]].
	%58 citations counted in INSPIRE as of 04 Aug 2020

	%\cite{Choptuik:1992jv}
	\bibitem{Choptuik:1992jv}
	M.~W.~Choptuik,
	\emph{Universality and scaling in gravitational collapse of a massless scalar field},
	Phys. Rev. Lett. \textbf{70}, 9-12 (1993)
	%doi:10.1103/PhysRevLett.70.9
	%916 citations counted in INSPIRE as of 04 Aug 2020

	%\cite{Yokoyama:1998xd}
	\bibitem{Yokoyama:1998xd}
	J.~Yokoyama,
	\emph{Cosmological constraints on primordial black holes produced in the near critical gravitational collapse},
	Phys. Rev. D \textbf{58}, 107502 (1998)
	%doi:10.1103/PhysRevD.58.107502
	[arXiv:gr-qc/9804041 [gr-qc]].
	%75 citations counted in INSPIRE as of 04 Aug 2020


	%\cite{Green:1999xm}
	\bibitem{Green:1999xm}
	A.~M.~Green and A.~R.~Liddle,
	\emph{Critical collapse and the primordial black hole initial mass function},
	Phys. Rev. D \textbf{60}, 063509 (1999)
	%doi:10.1103/PhysRevD.60.063509
	[arXiv:astro-ph/9901268 [astro-ph]].
	%58 citations counted in INSPIRE as of 04 Aug 2020




	%\cite{MacGibbon:1990zk}
	\bibitem{MacGibbon:1990zk}
	J.~H.~MacGibbon and B.~R.~Webber,
	\emph{Quark and gluon jet emission from primordial black holes: The instantaneous spectra},
	Phys. Rev. D \textbf{41}, 3052-3079 (1990)
	%doi:10.1103/PhysRevD.41.3052
	%188 citations counted in INSPIRE as of 04 Aug 2020



	\bibitem{1979MNRAS.188P..15L}
	D.~Lindley,
	\emph{Radiative decay of massive neutrinos and cosmic element abundances},
	Monthly Notices of the Royal Astronomical Society 188, 15P. (1979)



	%\cite{Ellis:1984er}
	\bibitem{1985NuPhB.259..175E}
	J.~R.~Ellis, D.~V.~Nanopoulos and S.~Sarkar,
	\emph{The Cosmology of Decaying Gravitinos},
	Nucl. Phys. B \textbf{259}, 175-188 (1985)
	%doi:10.1016/0550-3213(85)90306-2
	%439 citations counted in INSPIRE as of 12 Jul 2020



	%\cite{Dimopoulos:1987fz}
	\bibitem{1988ApJ...330..545D}
	S.~Dimopoulos, R.~Esmailzadeh, L.~J.~Hall and G.~D.~Starkman,
	\emph{Is the Universe Closed by Baryons? Nucleosynthesis With a Late Decaying Massive Particle},
	Astrophys. J. \textbf{330}, 545 (1988)
	%doi:10.1086/166493
	%178 citations counted in INSPIRE as of 12 Jul 2020



	%\cite{Dimopoulos:1988zz}
    \bibitem{Dimopoulos:1988zz}
    S.~Dimopoulos, R.~Esmailzadeh, L.~J.~Hall and G.~D.~Starkman,
    \emph{Kiloelectronvolt-Era Nucleosynthesis and Its Implications},
    Phys.\ Rev.\ Lett.\  {\bf 60}, 7 (1988).
    %  doi:10.1103/PhysRevLett.60.7
    %%CITATION = doi:10.1103/PhysRevLett.60.7;%%
    %46 citations counted in INSPIRE as of 22 Aug 2020

    %\cite{Dimopoulos:1988ue}
    \bibitem{Dimopoulos:1988ue}
    S.~Dimopoulos, R.~Esmailzadeh, L.~J.~Hall and G.~D.~Starkman,
    \emph{Limits on Late Decaying Particles From Nucleosynthesis},
    Nucl.\ Phys.\ B {\bf 311}, 699 (1989).
    %  doi:10.1016/0550-3213(89)90173-9
    %%CITATION = doi:10.1016/0550-3213(89)90173-9;%%
    %134 citations counted in INSPIRE as of 22 Aug 2020

    %\cite{Terasawa:1988my}
    \bibitem{Terasawa:1988my}
    N.~Terasawa, M.~Kawasaki and K.~Sato,
    \emph{Radiative Decay of Neutrino and Primordial Nucleosynthesis},
    Nucl.\ Phys.\ B {\bf 302}, 697 (1988).
    doi:10.1016/0550-3213(88)90194-0
    %%CITATION = doi:10.1016/0550-3213(88)90194-0;%%
    %29 citations counted in INSPIRE as of 01 Sep 2020

	%\cite{Ellis:1990nb}
	\bibitem{1992NuPhB.373..399E}
	J.~R.~Ellis, G.~B.~Gelmini, J.~L.~Lopez, D.~V.~Nanopoulos and S.~Sarkar,
	\emph{Astrophysical constraints on massive unstable neutral relic particles},
	Nucl. Phys. B \textbf{373}, 399-437 (1992)
	%doi:10.1016/0550-3213(92)90438-H
	%397 citations counted in INSPIRE as of 12 Jul 2020

	%\cite{Kawasaki:1993gz}
    \bibitem{Kawasaki:1993gz}
     M.~Kawasaki, P.~Kernan, H.~S.~Kang, R.~J.~Scherrer, G.~Steigman and T.~P.~Walker,
    \emph{Big bang nucleosynthesis constraints on the tau-neutrino mass},
    Nucl.\ Phys.\ B {\bf 419}, 105 (1994).
    doi:10.1016/0550-3213(94)90359-X
    %%CITATION = doi:10.1016/0550-3213(94)90359-X;%%
    %101 citations counted in INSPIRE as of 01 Sep 2020




	%\cite{Kawasaki:1994sc}
	\bibitem{1995ApJ...452..506K}
	M.~Kawasaki and T.~Moroi,
	\emph{Electromagnetic cascade in the early universe and its application to the big bang nucleosynthesis},
	Astrophys. J. \textbf{452}, 506 (1995)
	%doi:10.1086/176324
	[arXiv:astro-ph/9412055 [astro-ph]].
	%126 citations counted in INSPIRE as of 12 Jul 2020



	%\cite{Jedamzik:1999di}
	\bibitem{Jedamzik:1999di}
	K.~Jedamzik,
	\emph{Lithium 6: A Probe of the early universe},
	Phys. Rev. Lett. \textbf{84}, 3248 (2000)
	%doi:10.1103/PhysRevLett.84.3248
	[arXiv:astro-ph/9909445 [astro-ph]].
	%121 citations counted in INSPIRE as of 12 Jul 2020

	%\cite{Kawasaki:2000qr}
	\bibitem{Kawasaki:2000qr}
	M.~Kawasaki, K.~Kohri and T.~Moroi,
	\emph{Radiative decay of a massive particle and the nonthermal process in primordial nucleosynthesis},
	Phys. Rev. D \textbf{63}, 103502 (2001)
	%doi:10.1103/PhysRevD.63.103502
	[arXiv:hep-ph/0012279 [hep-ph]].
	%133 citations counted in INSPIRE as of 12 Jul 2020

	%\cite{Cyburt:2002uv}
	\bibitem{2003PhRvD..67j3521C}
	R.~H.~Cyburt, J.~R.~Ellis, B.~D.~Fields and K.~A.~Olive,
	\emph{Updated nucleosynthesis constraints on unstable relic particles},
	Phys. Rev. D \textbf{67}, 103521 (2003)
	%doi:10.1103/PhysRevD.67.103521
	[arXiv:astro-ph/0211258 [astro-ph]].
	%366 citations counted in INSPIRE as of 12 Jul 2020


	%\cite{Kawasaki:2004yh}
	\bibitem{Kawasaki:2004yh}
	M.~Kawasaki, K.~Kohri and T.~Moroi,
	\emph{Hadronic decay of late - decaying particles and Big-Bang Nucleosynthesis},
	Phys. Lett. B \textbf{625}, 7-12 (2005)
	%doi:10.1016/j.physletb.2005.08.045
	[arXiv:astro-ph/0402490 [astro-ph]].
	%463 citations counted in INSPIRE as of 12 Jul 2020


	%\cite{Kawasaki:2004qu}
	\bibitem{2005PhRvD..71h3502K}
	M.~Kawasaki, K.~Kohri and T.~Moroi,
	\emph{Big-Bang nucleosynthesis and hadronic decay of long-lived massive particles},
	Phys. Rev. D \textbf{71}, 083502 (2005)
	%doi:10.1103/PhysRevD.71.083502
	[arXiv:astro-ph/0408426 [astro-ph]].
	%783 citations counted in INSPIRE as of 12 Jul 2020



	%\cite{Jedamzik:2006xz}
	\bibitem{2006PhRvD..74j3509J}
	K.~Jedamzik,
	\emph{Big bang nucleosynthesis constraints on hadronically and electromagnetically decaying relic neutral particles},
	Phys. Rev. D \textbf{74}, 103509 (2006)
	%doi:10.1103/PhysRevD.74.103509
	[arXiv:hep-ph/0604251 [hep-ph]].
	%384 citations counted in INSPIRE as of 12 Jul 2020



	%\cite{Kusakabe:2006hc}
	\bibitem{2006PhRvD..74b3526K}
	M.~Kusakabe, T.~Kajino and G.~J.~Mathews,
	\emph{Li-6 Production by the Radiative Decay of Long-Lived Particles},
	Phys. Rev. D \textbf{74}, 023526 (2006)
	%doi:10.1103/PhysRevD.74.023526
	[arXiv:astro-ph/0605255 [astro-ph]].
	%52 citations counted in INSPIRE as of 12 Jul 2020


%\cite{Kawasaki:2008qe}
\bibitem{Kawasaki:2008qe}
  M.~Kawasaki, K.~Kohri, T.~Moroi and A.~Yotsuyanagi,
  \emph{Big-Bang Nucleosynthesis and Gravitino},
  Phys.\ Rev.\  D {\bf 78}, 065011 (2008).
%  arXiv:0804.3745 [hep-ph].
  %%CITATION = ARXIV:0804.3745;%%

	%\cite{Kusakabe:2008kf}
	\bibitem{2009PhRvD..79l3513K}
	M.~Kusakabe, T.~Kajino, T.~Yoshida, T.~Shima, Y.~Nagai and T.~Kii,
	\emph{New Constraints on Radiative Decay of Long-Lived X Particles in Big Bang Nucleosynthesis with New Rates of Photodisintegration Reactions of $^4$He},
	Phys. Rev. D \textbf{79}, 123513 (2009)
	%doi:10.1103/PhysRevD.79.123513
	[arXiv:0806.4040 [astro-ph]].
	%31 citations counted in INSPIRE as of 12 Jul 2020

	%\cite{Kusakabe:2012ds}
	\bibitem{Kusakabe:2012ds}
	M.~Kusakabe, A.~B.~Balantekin, T.~Kajino and Y.~Pehlivan,
	\emph{Solution to Big-Bang Nucleosynthesis in Hybrid Axion Dark Matter Model},
	Phys. Lett. B \textbf{718}, 704-708 (2013)
	%doi:10.1016/j.physletb.2012.11.007
	[arXiv:1202.5603 [astro-ph.CO]].
	%21 citations counted in INSPIRE as of 12 Jul 2020

	%\cite{Kusakabe:2013sna}
	\bibitem{2013PhRvD..87h5045K}
	M.~Kusakabe, A.~B.~Balantekin, T.~Kajino and Y.~Pehlivan,
	\emph{Big-bang nucleosynthesis limit on the neutral fermion decays into neutrinos},
	Phys. Rev. D \textbf{87}, no.8, 085045 (2013)
	%doi:10.1103/PhysRevD.87.085045
	[arXiv:1303.2291 [astro-ph.CO]].
	%19 citations counted in INSPIRE as of 12 Jul 2020

	%\cite{Ishida:2014wqa}
	\bibitem{2014PhRvD..90h3519I}
	H.~Ishida, M.~Kusakabe and H.~Okada,
	\emph{Effects of long-lived 10 MeV-scale sterile neutrinos on primordial elemental abundances and the effective neutrino number},
	Phys. Rev. D \textbf{90}, no.8, 083519 (2014)
	%doi:10.1103/PhysRevD.90.083519
	[arXiv:1403.5995 [astro-ph.CO]].
	%31 citations counted in INSPIRE as of 12 Jul 2020

	%\cite{Kawasaki:2020qxm}
	\bibitem{Kawasaki:2020qxm}
	M.~Kawasaki, K.~Kohri, T.~Moroi, K.~Murai and H.~Murayama,
	\emph{Big-bang nucleosynthesis with sub-GeV massive decaying particles},
	[arXiv:2006.14803 [hep-ph]].
	%1 citations counted in INSPIRE as of 12 Jul 2020


    %\cite{Ellis:2005ii}
	\bibitem{Ellis:2005ii}
	J.~R.~Ellis, K.~A.~Olive and E.~Vangioni,
	\emph{Effects of unstable particles on light-element abundances: Lithium versus deuterium and He-3},
	Phys. Lett. B \textbf{619}, 30-42 (2005)
	%doi:10.1016/j.physletb.2005.05.066
	[arXiv:astro-ph/0503023 [astro-ph]].
	%121 citations counted in INSPIRE as of 12 Jul 2020

    \bibitem{Carr:2009jm}
	B.~J.~Carr, K.~Kohri, Y.~Sendouda and J.~Yokoyama,
	\emph{New cosmological constraints on primordial black holes},
	Phys.\ Rev.\ D {\bf 81}, 104019 (2010)
	%doi:10.1103/PhysRevD.81.104019
	[arXiv:0912.5297 [astro-ph.CO]].


	%\cite{Acharya:2020jbv}
	\bibitem{2020JCAP...06..018A}
	S.~K.~Acharya and R.~Khatri,
	\emph{CMB and BBN constraints on evaporating primordial black holes revisited},
	JCAP \textbf{06}, 018 (2020)
	%doi:10.1088/1475-7516/2020/06/018
	[arXiv:2002.00898 [astro-ph.CO]].
	%5 citations counted in INSPIRE as of 12 Jul 2020



    %\cite{Salvati:2016jng}
	\bibitem{Salvati:2016jng}
	L.~Salvati, L.~Pagano, M.~Lattanzi, M.~Gerbino and A.~Melchiorri,
	\emph{Breaking Be: a sterile neutrino solution to the cosmological lithium problem},
	JCAP \textbf{08}, 022 (2016)
	%doi:10.1088/1475-7516/2016/08/022
	[arXiv:1606.06968 [astro-ph.CO]].
	%8 citations counted in INSPIRE as of 12 Jul 2020

	%\cite{Goudelis:2015wpa}
	\bibitem{Goudelis:2015wpa}
	A.~Goudelis, M.~Pospelov and J.~Pradler,
	\emph{Light Particle Solution to the Cosmic Lithium Problem},
	Phys. Rev. Lett. \textbf{116}, no.21, 211303 (2016)
	%doi:10.1103/PhysRevLett.116.211303
	[arXiv:1510.08858 [hep-ph]].
	%26 citations counted in INSPIRE as of 12 Jul 2020


		%\cite{Poulin:2015opa}
	\bibitem{2015PhRvD..91j3007P}
	V.~Poulin and P.~D.~Serpico,
	\emph{Nonuniversal BBN bounds on electromagnetically decaying particles},
	Phys. Rev. D \textbf{91}, no.10, 103007 (2015)
	%doi:10.1103/PhysRevD.91.103007
	[arXiv:1503.04852 [astro-ph.CO]].
	%40 citations counted in INSPIRE as of 12 Jul 2020

	%\cite{Poulin:2015woa}
	\bibitem{2015PhRvL.114i1101P}
	V.~Poulin and P.~D.~Serpico,
	\emph{Loophole to the Universal Photon Spectrum in Electromagnetic Cascades and Application to the Cosmological Lithium Problem},
	Phys. Rev. Lett. \textbf{114}, no.9, 091101 (2015)
	%doi:10.1103/PhysRevLett.114.091101
	[arXiv:1502.01250 [astro-ph.CO]].
	%28 citations counted in INSPIRE as of 12 Jul 2020

	%\cite{Press:1973iz}
	\bibitem{Press:1973iz}
	W.~H.~Press and P.~Schechter,
	\emph{Formation of galaxies and clusters of galaxies by selfsimilar gravitational condensation},
	Astrophys. J. \textbf{187}, 425-438 (1974)
	%doi:10.1086/152650
	%3235 citations counted in INSPIRE as of 04 Aug 2020




	\bibitem{Akrami:2018odb}
	Y.~Akrami {\it et al.} [Planck Collaboration],
	\emph{Planck 2018 results. X. Constraints on inflation},
	arXiv:1807.06211 [astro-ph.CO].

	\bibitem{Sasaki:1986hm}
	M.~Sasaki,
	\emph{Large Scale Quantum Fluctuations in the Inflationary Universe},
	Prog.\ Theor.\ Phys.\  {\bf 76}, 1036 (1986).
	%doi:10.1143/PTP.76.1036


	%\cite{Kopp:2010sh}
	\bibitem{Kopp:2010sh}
	M.~Kopp, S.~Hofmann and J.~Weller,
	\emph{Separate Universes Do Not Constrain Primordial Black Hole Formation},
	Phys. Rev. D \textbf{83}, 124025 (2011)
	%doi:10.1103/PhysRevD.83.124025
	[arXiv:1012.4369 [astro-ph.CO]].
	%44 citations counted in INSPIRE as of 04 Aug 2020


	%\cite{Carr:2014pga}
	\bibitem{Carr:2014pga}
	B.~J.~Carr and T.~Harada,
	\emph{Separate universe problem: 40 years on},
	Phys. Rev. D \textbf{91}, no.8, 084048 (2015)
	%doi:10.1103/PhysRevD.91.084048
	[arXiv:1405.3624 [astro-ph.CO]].
	%12 citations counted in INSPIRE as of 04 Aug 2020




	\bibitem{Musco:2012au}
	I.~Musco and J.~C.~Miller,
	\emph{Primordial black hole formation in the early universe: critical behaviour and self-similarity},
	Class. Quant. Grav. \textbf{30}, 145009 (2013)
	%doi:10.1088/0264-9381/30/14/145009
	[arXiv:1201.2379 [gr-qc]].
	%102 citations counted in INSPIRE as of 10 Jul 2020

	\bibitem{Harada:2013epa}
	T.~Harada, C.~M.~Yoo and K.~Kohri,
	\emph{Threshold of primordial black hole formation},
	Phys. Rev. D \textbf{88}, no.8, 084051 (2013)
	%doi:10.1103/PhysRevD.88.084051
	[arXiv:1309.4201 [astro-ph.CO]].
	%134 citations counted in INSPIRE as of 10 Jul 2020

	%\cite{Escriva:2020tak}
\bibitem{Escriva:2020tak}
A.~Escriv\`a, C.~Germani and R.~K.~Sheth,
``Analytical thresholds for black hole formation in general cosmological backgrounds,''
[arXiv:2007.05564 [gr-qc]].
%9 citations counted in INSPIRE as of 22 Jan 2021

	%\cite{Escriva:2019phb}
\bibitem{Escriva:2019phb}
A.~Escriv\`a, C.~Germani and R.~K.~Sheth,
``Universal threshold for primordial black hole formation,''
Phys. Rev. D \textbf{101}, no.4, 044022 (2020)
%doi:10.1103/PhysRevD.101.044022
[arXiv:1907.13311 [gr-qc]].
%43 citations counted in INSPIRE as of 22 Jan 2021

	\bibitem{Jedamzik:1999am}
	K.~Jedamzik and J.~C.~Niemeyer,
	\emph{Primordial black hole formation during first order phase transitions},
	Phys. Rev. D \textbf{59}, 124014 (1999)
	%doi:10.1103/PhysRevD.59.124014
	[arXiv:astro-ph/9901293 [astro-ph]].
	%113 citations counted in INSPIRE as of 10 Jul 2020

	%\cite{Pi:2017gih}
\bibitem{Pi:2017gih}
S.~Pi, Y.~l.~Zhang, Q.~G.~Huang and M.~Sasaki,
%``Scalaron from $R^2$-gravity as a heavy field,''
JCAP \textbf{05}, 042 (2018)
%doi:10.1088/1475-7516/2018/05/042
[arXiv:1712.09896 [astro-ph.CO]].
%70 citations counted in INSPIRE as of 22 Jan 2021

	%\cite{Cai:2018tuh}
\bibitem{Cai:2018tuh}
Y.~F.~Cai, X.~Tong, D.~G.~Wang and S.~F.~Yan,
%``Primordial Black Holes from Sound Speed Resonance during Inflation,''
Phys. Rev. Lett. \textbf{121}, no.8, 081306 (2018)
%doi:10.1103/PhysRevLett.121.081306
[arXiv:1805.03639 [astro-ph.CO]].
%48 citations counted in INSPIRE as of 22 Jan 2021


	\bibitem{Chen:2019zza}
	C.~Chen and Y.~F.~Cai,
	\emph{Primordial black holes from sound speed resonance in the inflaton-curvaton mixed scenario},
	JCAP \textbf{10}, 068 (2019)
	%doi:10.1088/1475-7516/2019/10/068
	[arXiv:1908.03942 [astro-ph.CO]].
	%10 citations counted in INSPIRE as of 09 Jul 2020


	\bibitem{Chen:2020uhe}
	C.~Chen, X.~H.~Ma and Y.~F.~Cai,
	``Dirac-Born-Infeld realization of sound speed resonance mechanism for primordial black holes,''
	Phys. Rev. D \textbf{102}, no.6, 063526 (2020)
	doi:10.1103/PhysRevD.102.063526
	[arXiv:2003.03821 [astro-ph.CO]].


	\bibitem{Kuhnel:2015vtw}
	F.~Kühnel, C.~Rampf and M.~Sandstad,
	\emph{Effects of Critical Collapse on Primordial Black-Hole Mass Spectra},
	Eur. Phys. J. C \textbf{76}, no.2, 93 (2016)
	%doi:10.1140/epjc/s10052-016-3945-8
	[arXiv:1512.00488 [astro-ph.CO]].
	%28 citations counted in INSPIRE as of 10 Jul 2020


	%\cite{Niemeyer:1997mt}
	\bibitem{Niemeyer:1997mt}
	J.~C.~Niemeyer and K.~Jedamzik,
	\emph{Near-critical gravitational collapse and the initial mass function of primordial black holes},
	Phys. Rev. Lett. \textbf{80}, 5481-5484 (1998)
	%doi:10.1103/PhysRevLett.80.5481
	[arXiv:astro-ph/9709072 [astro-ph]].
	%190 citations counted in INSPIRE as of 04 Aug 2020

	%\cite{Neilsen:1998qc}
	\bibitem{Neilsen:1998qc}
	D.~W.~Neilsen and M.~W.~Choptuik,
	\emph{Critical phenomena in perfect fluids},
	Class. Quant. Grav. \textbf{17}, 761-782 (2000)
	%doi:10.1088/0264-9381/17/4/303
	[arXiv:gr-qc/9812053 [gr-qc]].
	%65 citations counted in INSPIRE as of 04 Aug 2020

	%\cite{Musco:2008hv}
	\bibitem{Musco:2008hv}
	I.~Musco, J.~C.~Miller and A.~G.~Polnarev,
	\emph{Primordial black hole formation in the radiative era: Investigation of the critical nature of the collapse},
	Class. Quant. Grav. \textbf{26}, 235001 (2009)
	%doi:10.1088/0264-9381/26/23/235001
	[arXiv:0811.1452 [gr-qc]].
	%96 citations counted in INSPIRE as of 04 Aug 2020

	%\cite{Musco:2004ak}
	\bibitem{Musco:2004ak}
	I.~Musco, J.~C.~Miller and L.~Rezzolla,
	\emph{Computations of primordial black hole formation},
	Class. Quant. Grav. \textbf{22}, 1405-1424 (2005)
	%doi:10.1088/0264-9381/22/7/013
	[arXiv:gr-qc/0412063 [gr-qc]].
	%142 citations counted in INSPIRE as of 04 Aug 2020

	\bibitem{Carr:2016drx}
	B.~Carr, F.~Kuhnel and M.~Sandstad,
	\emph{Primordial Black Holes as Dark Matter},
	Phys.\ Rev.\ D {\bf 94}, no. 8, 083504 (2016)
	%doi:10.1103/PhysRevD.94.083504
	[arXiv:1607.06077 [astro-ph.CO]].


	%\cite{Carr:2016hva}
	\bibitem{Carr:2016hva}
	B.~J.~Carr, K.~Kohri, Y.~Sendouda and J.~Yokoyama,
	\emph{Constraints on primordial black holes from the Galactic gamma-ray background},
	Phys. Rev. D \textbf{94}, no.4, 044029 (2016)
	%doi:10.1103/PhysRevD.94.044029
	[arXiv:1604.05349 [astro-ph.CO]].
	%57 citations counted in INSPIRE as of 04 Aug 2020



	%\cite{Carr:2017jsz}
	\bibitem{Carr:2017jsz}
	B.~Carr, M.~Raidal, T.~Tenkanen, V.~Vaskonen and H.~Veermäe,
	\emph{Primordial black hole constraints for extended mass functions},
	Phys. Rev. D \textbf{96}, no.2, 023514 (2017)
	%doi:10.1103/PhysRevD.96.023514
	[arXiv:1705.05567 [astro-ph.CO]].
	%203 citations counted in INSPIRE as of 04 Aug 2020




	%\cite{Carr:2018poi}
	\bibitem{Carr:2018poi}
	B.~Carr and F.~Kuhnel,
	\emph{Primordial black holes with multimodal mass spectra},
	Phys. Rev. D \textbf{99}, no.10, 103535 (2019)
	%doi:10.1103/PhysRevD.99.103535
	[arXiv:1811.06532 [astro-ph.CO]].
	%12 citations counted in INSPIRE as of 04 Aug 2020


	%\cite{Hawking:1974sw}
	\bibitem{Hawking:1974sw}
	S.~W.~Hawking,
	\emph{Particle Creation by Black Holes},
	Commun. Math. Phys. \textbf{43}, 199-220 (1975)
	%doi:10.1007/BF02345020
	%8385 citations counted in INSPIRE as of 04 Aug 2020


	%\cite{Page:1976df}
	\bibitem{Page:1976df}
	D.~N.~Page,
	\emph{Particle Emission Rates from a Black Hole: Massless Particles from an Uncharged, Nonrotating Hole},
	Phys. Rev. D \textbf{13}, 198-206 (1976)
	%doi:10.1103/PhysRevD.13.198
	%792 citations counted in INSPIRE as of 04 Aug 2020

	%\cite{Page:1977um}
	\bibitem{Page:1977um}
	D.~N.~Page,
	\emph{Particle Emission Rates from a Black Hole. 3. Charged Leptons from a Nonrotating Hole},
	Phys. Rev. D \textbf{16}, 2402-2411 (1977)
	%doi:10.1103/PhysRevD.16.2402
	%188 citations counted in INSPIRE as of 04 Aug 2020

	%\cite{Page:1976ki}
	\bibitem{Page:1976ki}
	D.~N.~Page,
	\emph{Particle Emission Rates from a Black Hole. 2. Massless Particles from a Rotating Hole},
	Phys. Rev. D \textbf{14}, 3260-3273 (1976)
	%doi:10.1103/PhysRevD.14.3260
	%317 citations counted in INSPIRE as of 04 Aug 2020


	%\cite{MacGibbon:1987my}
	\bibitem{MacGibbon:1987my}
	J.~H.~MacGibbon,
	\emph{Can Planck-mass relics of evaporating black holes close the universe?},
	Nature \textbf{329}, 308-309 (1987)
	%doi:10.1038/329308a0
	%142 citations counted in INSPIRE as of 04 Aug 2020

	%\cite{Arbey:2019mbc}
	\bibitem{Arbey:2019mbc}
	A.~Arbey and J.~Auffinger,
	\emph{BlackHawk: A public code for calculating the Hawking evaporation spectra of any black hole distribution},
	Eur. Phys. J. C \textbf{79}, no.8, 693 (2019)
	%doi:10.1140/epjc/s10052-019-7161-1
	[arXiv:1905.04268 [gr-qc]].
	%12 citations counted in INSPIRE as of 04 Aug 2020


	%\cite{MacGibbon:1991tj}
	\bibitem{MacGibbon:1991tj}
	J.~H.~MacGibbon,
	\emph{Quark and gluon jet emission from primordial black holes. 2. The Lifetime emission},
	Phys. Rev. D \textbf{44}, 376-392 (1991)
	%doi:10.1103/PhysRevD.44.376
	%117 citations counted in INSPIRE as of 04 Aug 2020

	%\cite{Kawasaki:2017bqm}
	\bibitem{Kawasaki:2017bqm}
	M.~Kawasaki, K.~Kohri, T.~Moroi and Y.~Takaesu,
	\emph{Revisiting Big-Bang Nucleosynthesis Constraints on Long-Lived Decaying Particles},
	Phys. Rev. D \textbf{97}, no.2, 023502 (2018)
	%doi:10.1103/PhysRevD.97.023502
	%[arXiv:1709.01211 [hep-ph]].
	%106 citations counted in INSPIRE as of 21 Jan 2021


	%\cite{Protheroe:1994dt}
	\bibitem{1995PhRvD..51.4134P}
	R.~J.~Protheroe, T.~Stanev and V.~S.~Berezinsky,
	\emph{Electromagnetic cascades and cascade nucleosynthesis in the early universe},
	Phys. Rev. D \textbf{51}, 4134-4144 (1995)
	%doi:10.1103/PhysRevD.51.4134
	[arXiv:astro-ph/9409004 [astro-ph]].
	%52 citations counted in INSPIRE as of 12 Jul 2020

	%\cite{Montmerle:1977vc}
    \bibitem{Montmerle:1977vc}
    T.~Montmerle,
    \emph{On the Possible Existence of Cosmological Cosmic Rays. 1. The framework for light-element and gamma-ray production},
    Astrophys. J. \textbf{216}, 177 (1977)
    %doi:10.1086/155679
    %0 citations counted in INSPIRE as of 12 Sep 2020

	%\cite{Ginzburg:1990sk}
	\bibitem{Berezinskii}
	V.~S.~Berezinsky, S.~V.~Bulanov, V.~A.~Dogiel, V.~L.~Ginzburg and V.~S.~Ptuskin,
	\emph{Astrophysics of cosmic rays},
	(North-Holland, New York, 1990)
	%132 citations counted in INSPIRE as of 12 Jul 2020

	%\cite{Svensson:1990pfo}
	\bibitem{1990ApJ...349..415S}
	R.~Svensson and A.~A.~Zdziarski,
	\emph{Photon-photon scattering of gamma rays at cosmological distances},
	Astrophys. J. \textbf{349}, 415-428 (1990)
	%doi:10.1086/168325
	%48 citations counted in INSPIRE as of 12 Jul 2020

	\bibitem{Maximon}
	L. Maximon,
	\emph{Simple analytic expressions for the total Born approximation cross section for pair production in a Coulomb field},
	Journal of Research of the National Bureau of Standards-B: Mathematical Sciences. \textbf{72B}, no.1, 79
	%	doi: 10.6028/jres.072B.011.






	\bibitem{Abramovich}
	S.~N. Abramovich, Y.~B. Guzhovskij, V.~A. Zherebtsov and A.~G. Zvenigorodskij,
	\emph{NUCLEAR PHYSICS CONSTANTS FOR THERMONUCLEAR FUSION. A Reference Handbook},
	INDC(CCP)-326/L+F (1989).


    \bibitem{Wang2017}
	M.~Wang \textit{et al.},
	\emph{The AME2016 atomic mass evaluation (II). Tables, graphs and references},
	Chin. Phys. C \textbf{41}, no.3, 030003 (2017)



	%\cite{Huang:1976jyl}
	\bibitem{Huang:1976jyl}
	K.~N.~Huang, M.~Aoyagi, M.~H.~Chen, B.~Crasemann and H.~Mark,
	\emph{Neutral-atom electron binding energies from relaxed-orbital relativistic Hartree-Fock-Slater calculations 2 $\leq$ Z $\leq$ 106},
	Atom. Data Nucl. Data Tabl. \textbf{18}, 243-291 (1976)
	%doi:10.1016/0092-640X(76)90027-9
	%86 citations counted in INSPIRE as of 08 Aug 2020



	%\cite{He:2013ica}
	\bibitem{He:2013ica}
	J.~J.~He, S.~Z.~Chen, C.~E.~Rolfs, S.~W.~Xu, J.~Hu, X.~W.~Ma, M.~Wiescher, R.~J.~Deboer, T.~Kajino, M.~Kusakabe, L.~Y.~Zhang, S.~Q.~Hou, X.~Q.~Yu, N.~T.~Zhang, G.~Lian, Y.~H.~Zhang, X.~H.~Zhou, H.~S.~Xu, G.~Q.~Xiao and W.~L.~Zhan,
	\emph{A drop in the $^{6}Li(p,\gamma)^{7}Be$ reaction at low energies},
	Phys. Lett. B \textbf{725}, 287-291 (2013)
	%doi:10.1016/j.physletb.2013.07.044
	%23 citations counted in INSPIRE as of 30 Jul 2020




	\bibitem{Kawano1992}
	L.~Kawano,
	\emph{Let's go: Early universe. 2. Primordial nucleosynthesis: The Computer way},
	FERMILAB-PUB-92-004-A (1992).
	%31 citations counted in INSPIRE as of 03 Aug 2020


	\bibitem{Smith:1992yy}
	M. Smith, L. Kawano and R. Malaney,
	\emph{Experimental, Computational, and Observational Analysis of Primordial Nucleosynthesis},
	Astrophys.\ J.\ Suppl.\  {\bf 85}, 219 (1993)
	%doi: 10.1086/191763


	\bibitem{Cyburt2010}
	R.Cyburt {\it et al.},
	\emph{The JINA REACLIB Database: Its Recent Updates and Impact on Type-I X-ray Bursts},
	Astrophys. J. Suppl. Ser. {\bf 189}, 240 (2010)
	%doi:10.1088/0067-0049/189/1/240

	%\cite{Coc:2015bhi}
	\bibitem{Coc2015}
	A.~Coc, P.~Petitjean, J.~P.~Uzan, E.~Vangioni, P.~Descouvemont, C.~Iliadis and R.~Longland,
	\emph{New reaction rates for improved primordial D/H calculation and the cosmic evolution of deuterium},
	Phys. Rev. D \textbf{92}, no.12, 123526 (2015)
	%doi:10.1103/PhysRevD.92.123526
	[arXiv:1511.03843 [astro-ph.CO]].
	%50 citations counted in INSPIRE as of 03 Aug 2020


	%\cite{Descouvemont:2004cw}
	\bibitem{Descov2004}
	P.~Descouvemont, A.~Adahchour, C.~Angulo, A.~Coc and E.~Vangioni-Flam,
	\emph{Compilation and R-matrix analysis of Big Bang nuclear reaction rates},
	Atom. Data Nucl. Data Tabl. \textbf{88}, 203-236 (2004)
	%doi:10.1016/j.adt.2004.08.001
	[arXiv:astro-ph/0407101 [astro-ph]].
	%191 citations counted in INSPIRE as of 03 Aug 2020


	%\cite{Patrignani:2016xqp}
	\bibitem{Patrignani}
	C.~Patrignani \textit{et al.} [Particle Data Group],
	\emph{Review of Particle Physics},
	Chin. Phys. C \textbf{40}, no.10, 100001 (2016)
	%doi:10.1088/1674-1137/40/10/100001
	%5861 citations counted in INSPIRE as of 03 Aug 2020


	%\cite{Aghanim:2018eyx}
    \bibitem{Aghanim:2018eyx}
    N.~Aghanim \textit{et al.} [Planck],
    \emph{Planck 2018 results. VI. Cosmological parameters},
    [arXiv:1807.06209 [astro-ph.CO]].
    %3276 citations counted in INSPIRE as of 17 Aug 2020



	%\cite{Khatri:2010ed}
    \bibitem{Khatri:2010ed}
    R.~Khatri and R.~A.~Sunyaev,
    \emph{Time of primordial Be-7 conversion into Li-7, energy release and doublet of narrow cosmological neutrino lines},
    Astron. Lett. \textbf{37}, 367 (2011)
    %doi:10.1134/S1063773711060041
    [arXiv:1009.3932 [astro-ph.CO]].
    %15 citations counted in INSPIRE as of 18 Aug 2020





	%\cite{Aver:2015iza}
	\bibitem{Aver2015}
	E.~Aver, K.~A.~Olive and E.~D.~Skillman,
	\emph{The effects of He I $\lambda$10830 on helium abundance determinations},
	JCAP \textbf{07}, 011 (2015)
	%doi:10.1088/1475-7516/2015/07/011
	[arXiv:1503.08146 [astro-ph.CO]].
	%143 citations counted in INSPIRE as of 15 Jul 2020

	%\cite{Cooke:2017cwo}
	\bibitem{Cooke2018}
	R.~J.~Cooke, M.~Pettini and C.~C.~Steidel,
	\emph{One Percent Determination of the Primordial Deuterium Abundance},
	Astrophys. J. \textbf{855}, no.2, 102 (2018)
	%doi:10.3847/1538-4357/aaab53
	[arXiv:1710.11129 [astro-ph.CO]].
	%107 citations counted in INSPIRE as of 03 Aug 2020

	%\cite{VangioniFlam:2002sa}
	\bibitem{VangioniFlam:2002sa}
	E.~Vangioni-Flam, K.~A.~Olive, B.~D.~Fields and M.~Casse,
	\emph{On the baryometric status of He-3},
	Astrophys. J. \textbf{585}, 611-616 (2003)
	%doi:10.1086/346232
	[arXiv:astro-ph/0207583 [astro-ph]].
	%63 citations counted in INSPIRE as of 15 Jul 2020



	\bibitem{2003SSRv..106....3G}
	J. Geiss and G. Gloeckler,
	\emph{Isotopic Composition of H, He and Ne in the Protosolar Cloud},
	Spa. Sci. Rev. {\bf{106}}, 3 (2003)
	%doi:10.1023/A:1024651232758


	%\cite{Bania:2002yj}
	\bibitem{Bania:2002yj}
	T.~M.~Bania, R.~T.~Rood and D.~S.~Balser,
	\emph{The cosmological density of baryons from observations of 3He+ in the Milky Way},
	Nature \textbf{415}, 54-57 (2002)
	%doi:10.1038/415054a
	%197 citations counted in INSPIRE as of 15 Jul 2020

	%\cite{Bellomo:2017zsr}
	\bibitem{Bellomo:2017zsr}
	N.~Bellomo, J.~L.~Bernal, A.~Raccanelli and L.~Verde,
	\emph{Primordial Black Holes as Dark Matter: Converting Constraints from Monochromatic to Extended Mass Distributions},
	JCAP \textbf{01}, 004 (2018)
	%doi:10.1088/1475-7516/2018/01/004
	[arXiv:1709.07467 [astro-ph.CO]].
	%46 citations counted in INSPIRE as of 04 Aug 2020



%\cite{Bugaev:2008gw}
\bibitem{Bugaev:2008gw}
E.~Bugaev and P.~Klimai,
%``Constraints on amplitudes of curvature perturbations from primordial black holes,''
Phys. Rev. D \textbf{79}, 103511 (2009)
doi:10.1103/PhysRevD.79.103511
[arXiv:0812.4247 [astro-ph]].
%46 citations counted in INSPIRE as of 19 Mar 2021



%\cite{Kribs:1999bs}
\bibitem{Kribs:1999bs}
G.~D.~Kribs, A.~K.~Leibovich and I.~Z.~Rothstein,
%``Bounds from primordial black holes with a near critical collapse initial mass function,''
Phys. Rev. D \textbf{60}, 103510 (1999)
doi:10.1103/PhysRevD.60.103510
[arXiv:astro-ph/9904021 [astro-ph]].
%25 citations counted in INSPIRE as of 19 Mar 2021


	%\cite{Bugaev:2009kq}
    \bibitem{Bugaev:2009kq}
    E.~V.~Bugaev and P.~A.~Klimai,
    %``Bound on induced gravitational wave background from primordial black holes,''
    JETP Lett. \textbf{91}, 1-5 (2010)
    doi:10.1134/S0021364010010017
    [arXiv:0911.0611 [astro-ph.CO]].
%6 citations counted in INSPIRE as of 19 Mar 2021

%\cite{Coc:2011az}
	\bibitem{Coc:2011az}
	A.~Coc, S.~Goriely, Y.~Xu, M.~Saimpert and E.~Vangioni,
	\emph{Standard Big-Bang Nucleosynthesis up to CNO with an improved extended nuclear network},
	Astrophys. J. \textbf{744}, 158 (2012)
	%doi:10.1088/0004-637X/744/2/158
	[arXiv:1107.1117 [astro-ph.CO]].
	%108 citations counted in INSPIRE as of 05 Aug 2020

	%\cite{Pitrou:2018cgg}
	\bibitem{Pitrou:2018cgg}
	C.~Pitrou, A.~Coc, J.~P.~Uzan and E.~Vangioni,
	\emph{Precision big bang nucleosynthesis with improved Helium-4 predictions},
	Phys. Rept. \textbf{754}, 1-66 (2018)
	%doi:10.1016/j.physrep.2018.04.005
	[arXiv:1801.08023 [astro-ph.CO]].
	%95 citations counted in INSPIRE as of 05 Aug 2020


    %\cite{Coc:2014oia}
    \bibitem{Coc:2014oia}
    A.~Coc, J.~P.~Uzan and E.~Vangioni,
    \emph{Standard big bang nucleosynthesis and primordial CNO Abundances after Planck},
     JCAP {\bf 1410}, 050 (2014)
    %  doi:10.1088/1475-7516/2014/10/050
    [arXiv:1403.6694 [astro-ph.CO]].
    %%CITATION = doi:10.1088/1475-7516/2014/10/050;%%
    %67 citations counted in INSPIRE as of 22 Aug 2020



    %\cite{Suzuki:2002qa}
    \bibitem{Suzuki:2002qa}
    T.~K.~Suzuki and S.~Inoue,
    \emph{Cosmic ray production of lithium-6 by structure formation shocks in the early milky way: a fossil record of dissipative processes during galaxy formation},
    Astrophys.\ J.\  {\bf 573}, 168 (2002)
    %  doi:10.1086/340487
    [astro-ph/0201190].
     %%CITATION = doi:10.1086/340487;%%
    %42 citations counted in INSPIRE as of 22 Aug 2020

    %\cite{Rollinde:2004kz}
    \bibitem{Rollinde:2004kz}
    E.~Rollinde, E.~Vangioni-Flam and K.~A.~Olive,
    \emph{Cosmological cosmic rays and the observed Li-6 plateau in metal poor halo stars},
    Astrophys.\ J.\  {\bf 627}, 666 (2005)
    %  doi:10.1086/430401
    [astro-ph/0412426].
    %%CITATION = doi:10.1086/430401;%%
    %39 citations counted in INSPIRE as of 22 Aug 2020

    %\cite{Tatischeff:2006tw}
    \bibitem{Tatischeff:2006tw}
    V.~Tatischeff and J.-P.~Thibaud,
    \emph{Is Li-6 in metal-poor halo stars produced in situ by solar-like flares ?},
    Astron.\ Astrophys.\  {\bf 469}, 265 (2007)
    %  doi:10.1051/0004-6361:20066635
    [astro-ph/0610756].
    %%CITATION = doi:10.1051/0004-6361:20066635;%%
    %27 citations counted in INSPIRE as of 22 Aug 2020

	%\cite{Prantzos:2012wt}
	\bibitem{Prantzos:2012wt}
	N.~Prantzos,
	\emph{Production and evolution of Li, Be and B isotopes in the Galaxy},
	Astron. Astrophys. \textbf{542}, A67 (2012)
	%doi:10.1051/0004-6361/201219043
	[arXiv:1203.5662 [astro-ph.GA]].
	%35 citations counted in INSPIRE as of 05 Aug 2020


	%\cite{Asplund:2005yt}
	\bibitem{Asplund:2005yt}
	M.~Asplund, D.~L.~Lambert, P.~E.~Nissen, F.~Primas and V.~V.~Smith,
	\emph{Lithium isotopic abundances in metal-poor halo stars},
	Astrophys. J. \textbf{644}, 229-259 (2006)
	%doi:10.1086/503538
	[arXiv:astro-ph/0510636 [astro-ph]].
	%327 citations counted in INSPIRE as of 05 Aug 2020

    %\cite{Cayrel:2007te}
    \bibitem{Cayrel:2007te}
    R.~Cayrel {\it et al.},
    \emph{Line shift, line asymmetry, and the 6Li/7Li isotopic ratio determination},
    Astron.\ Astrophys.\  {\bf 473}, L37 (2007)
    %  doi:10.1051/0004-6361:20078342
    [arXiv:0708.3819 [astro-ph]].
    %%CITATION = doi:10.1051/0004-6361:20078342;%%
    %72 citations counted in INSPIRE as of 22 Aug 2020


	%\cite{Lind:2013iza}
	\bibitem{Lind:2013iza}
	K.~Lind, J.~Melendez, M.~Asplund, R.~Collet and Z.~Magic,
	\emph{The lithium isotopic ratio in very metal-poor stars},
	Astron. Astrophys. \textbf{554}, A96 (2013)
	%doi:10.1051/0004-6361/201321406
	[arXiv:1305.6564 [astro-ph.SR]].
	%48 citations counted in INSPIRE as of 05 Aug 2020

	%\cite{Richard:2001qp}
\bibitem{Richard:2001qp}
O.~Richard, G.~Michaud, J.~Richer, S.~Turcotte, S.~Turck-Chieze and D.~A.~VandenBerg,
\emph{Models of metal poor stars with gravitational settling and radiative accelerations: I. evolution and abundance anomalies},
Astrophys. J. \textbf{580}, 1100-1117 (2002)
%doi:10.1086/343733
[arXiv:astro-ph/0112113 [astro-ph]].
%40 citations counted in INSPIRE as of 05 Nov 2020

%\cite{Richard:2004pj}
\bibitem{Richard:2004pj}
O.~Richard, G.~Michaud and J.~Richer,
\emph{Implications of WMAP observations on Li abundance and stellar evolution models},
Astrophys. J. \textbf{619}, 538-548 (2005)
%doi:10.1086/426470
[arXiv:astro-ph/0409672 [astro-ph]].
%157 citations counted in INSPIRE as of 05 Nov 2020

	\bibitem{Korn2006}
	A.~J.~Korn, F.~Grundahl, O.~Richard, P~.S.~Barklem, L.~Mashonkina, R. Collet, N.~Piskunov and B.~Gustafsson,
	\emph{A probable stellar solution to the cosmological lithium discrepancy},
	Nature \textbf{442}, 7103 (2006).
	%doi:10.1038/nature05011

	\bibitem{Pinsonneault2002}
	M. H. Pinsonneault, G. Steigman, T. P. Walker, and V. K. Narayanan,
	\emph{Stellar Mixing and the Primordial Lithium Abundance},
	Astrophys. J. \textbf{574}, 398 (2002).
	%doi:10.1086/340119




    \bibitem{1972A&AS....7..417M}
    J. P. Meyer,
    \emph{Deuterons and He3 formation and destruction in proton induced spallation of light nuclei (Z <= 8)}
    Astro. \& Astrophys. \textbf{7}, 417 (1972)

\bibitem{1969NCimA..63..529B} Blair, I.~M., Taylor, A.~E., Chapman, W.~S., et al.\ 1969, Nuovo Cimento A Serie, 63, 529. %doi:10.1007/BF02756231

%\cite{Mcgill:1984jq}
\bibitem{Mcgill:1984jq}
J.~A.~Mcgill, G.~W.~Hoffmann, M.~L.~Barlett, R.~W.~Fergerson, E.~C.~Milner, R.~E.~Chrien, R.~J.~Sutter, T.~Kozlowski and R.~L.~Stearns,
\emph{PROTON + NUCLEUS INCLUSIVE (P, P') SCATTERING AT 800-MeV}
Phys. Rev. C \textbf{29}, 204-208 (1984)
%doi:10.1103/PhysRevC.29.204
%20 citations counted in INSPIRE as of 26 Jan 2021

%\cite{Nagae:1986wu}
\bibitem{Nagae:1986wu}
T.~Nagae, S.~Sasaki, K.~Tokushuku, H.~Sano, M.~Sekimoto, I.~Arai, A.~Manabe, H.~Nunokawa, H.~Sakamoto and K.~Aoki, \textit{et al.}
\emph{Quasifree Production of Delta0 Isobars in Proton - Nucleus Reactions at 3.9-GeV/c}
Phys. Lett. B \textbf{191}, 31-35 (1987)
%doi:10.1016/0370-2693(87)91316-5
%19 citations counted in INSPIRE as of 26 Jan 2021

\bibitem{1991PhRvL..67.1982C} Chiba, J., Kobayashi, T., Nagae, T., et al.\ 1991, Phys. Rev. Lett., 67, 1982. %doi:10.1103/PhysRevLett.67.1982

%\cite{Engel:1996ic}
\bibitem{Engel:1996ic}
A.~Engel, A.~K.~Dutt-Mazumder, R.~Shyam and U.~Mosel,
\emph{Pion production in proton proton collisions in a covariant one boson exchange model}
Nucl. Phys. A \textbf{603}, 387-414 (1996)
%doi:10.1016/0375-9474(96)80008-F
[arXiv:nucl-th/9601026 [nucl-th]].
%45 citations counted in INSPIRE as of 26 Jan 2021

%\cite{AbdelBary:2004qb}
\bibitem{AbdelBary:2004qb}
M.~Abdel-Bary \textit{et al.} [GEM],
\emph{Detailed comparison of the pp ---\ensuremath{>} $\pi$+pn and pp ---\ensuremath{>} $\pi$+d reactions at 951-MeV}
Phys. Lett. B \textbf{610}, 31-36 (2005)
%doi:10.1016/j.physletb.2005.01.100
[arXiv:nucl-ex/0412002 [nucl-ex]].
%9 citations counted in INSPIRE as of 26 Jan 2021

%\cite{Kurbatov:2007zv}
\bibitem{Kurbatov:2007zv}
V.~Kurbatov, M.~Buscher, S.~Dymov, D.~Gusev, M.~Hartmann, A.~Kacharava, A.~Khoukaz, V.~Komarov, A.~Kulikov and G.~Macharashvili, \textit{et al.}
\emph{Energy dependence of forward S(0)-1 diproton production in the pp ---\ensuremath{>} pp $\pi_0$ reaction}
Phys. Lett. B \textbf{661}, 22-27 (2008)
%doi:10.1016/j.physletb.2008.01.051
[arXiv:0712.1186 [nucl-ex]].
%20 citations counted in INSPIRE as of 26 Jan 2021

%\cite{Maeda:2008gv}
\bibitem{Maeda:2008gv}
Y.~Maeda, M.~Segawa, T.~Ishida, A.~Kacharava, M.~Nomachi, Y.~Shimbara, Y.~Sugaya, K.~Tamura, T.~Yagita and K.~Yasuda, \textit{et al.}
\emph{Differential cross section and analyzing power of the vec- p p ---\ensuremath{>} pp $\pi_0$ reaction at a beam energy of 390-MeV}
Phys. Rev. C \textbf{77}, 044004 (2008)
%doi:10.1103/PhysRevC.77.044004
[arXiv:0802.2331 [nucl-ex]].
%2 citations counted in INSPIRE as of 26 Jan 2021

%\cite{Reno:1987qw}
\bibitem{Reno:1987qw}
M.~H.~Reno and D.~Seckel,
\emph{Primordial Nucleosynthesis: The Effects of Injecting Hadrons}
Phys. Rev. D \textbf{37}, 3441 (1988)
%doi:10.1103/PhysRevD.37.3441
%217 citations counted in INSPIRE as of 26 Jan 2021



\end{thebibliography}
\end{document}